\documentclass[twocolumn, tighten]{aastex631}
\usepackage[utf8]{inputenc}
\usepackage{natbib,graphics,
graphicx,amsmath,amssymb,threeparttable,xcolor}

\newcommand{\ff}{\ensuremath{f_{\rm F444W}}}
\newcommand{\fo}{\ensuremath{f_{\rm F150W}}}
\newcommand{\fsm}{\ensuremath{f_{{\rm 850\, \mu m}}}}

\shortauthors{McKay et al.}

\begin{document}

\title{The Physical Properties and Morphologies of Faint Dusty Star-forming Galaxies Identified with JWST}

\correspondingauthor{Stephen J. McKay}
\email{sjmckay3@wisc.edu}

\author[0000-0003-4248-6128]{S.~J.~McKay}
\affiliation{Department of Physics, University of Wisconsin--Madison, 1150 University Avenue,
Madison, WI 53706, USA}

\author[0000-0002-3306-1606]{A.~J.~Barger}
\affiliation{Department of Astronomy, University of Wisconsin--Madison, 475 N. Charter Street, Madison, WI 53706, USA}
\affiliation{Department of Physics and Astronomy, University of Hawaii, 2505 Correa Road, Honolulu, HI 96822, USA}
\affiliation{Institute for Astronomy, University of Hawaii, 2680 Woodlawn Drive, Honolulu, HI 96822, USA}

\author[0000-0002-6319-1575]{L.~L.~Cowie}
\affiliation{Institute for Astronomy, University of Hawaii, 2680 Woodlawn Drive, Honolulu, HI 96822, USA}

\author[0000-0003-3910-6446]{M.~J.~Nicandro~Rosenthal}
\affiliation{Department of Astronomy, University of Wisconsin--Madison, 475 N. Charter Street, Madison, WI 53706, USA}

\begin{abstract}

We identify a sample of 234 dusty star-forming galaxies (DSFGs) in the A2744 and GOODS-S fields
using JWST/NIRCam-selected galaxies as priors for SCUBA-2 measurements.
This method provides a large number of galaxies both above an 850$\,\mathrm{\mu}$m flux of 2~mJy (47 bright DSFGs) and 
below (187 faint DSFGs), representing the largest sample of individually identified (i.e., not stacked) faint DSFGs to date.
We observe that $\ff$ and $\fo$ are tightly correlated with redshift, with fainter sources lying at higher redshifts, suggesting that the observed 
near-infrared flux may be an effective way of selecting high-redshift DSFGs.
We study the physical properties and morphologies of the DSFGs through spectral energy distribution fitting and parametric surface brightness profile modeling. 
Other than the lower star formation rates (SFRs) and total infrared luminosities in the faint DSFGs (SFR~$= 80^{+81}_{-45}$; $L_{\rm IR} = 11.82^{+0.30}_{-0.36}$) compared to the bright DSFGs (SFR~$= 254^{+135}_{-131}$; $L_{\rm IR} = 12.32^{+0.20}_{-0.30}$),
the populations have similar properties.
The stellar masses do not appear to be strongly dependent on either the SFRs or the submillimeter flux.
These results suggest that the faint DSFGs are drawn from the same population of galaxies as the bright DSFGs.
Based on visual inspections, we find a lower merger fraction ($\sim21$\%) relative to previous HST-based studies, suggesting that dust attenuation may have impacted previous estimates of DSFG merger rates.
\end{abstract}

\section{Introduction}

It is well established that a large fraction of the cosmic star formation history (SFH) is shrouded by dust,
with estimates that dust-obscured star formation is more prevalent than unobscured star formation out to redshifts of at least $z\sim4$--5 \citep[e.g.,][]{bouwens20, zavala21, sun24c}. 
The most powerful individual contributors to the dust-obscured portion of the cosmic SFH are dusty star-forming galaxies (DSFGs), originally discovered with surveys at 850~$\mu$m \citep{smail97,barger98,hughes98,eales99}. 

While thousands of DSFGs have been detected through wide-field surveys with bolometer cameras on single-dish telescopes, the majority of these are brighter than $\sim$2~mJy at 850~$\mu$m, i.e., above the confusion limits of single-dish submillimeter telescopes ($\sim1.6$~mJy for SCUBA-2 on the 15-m JCMT at 850~$\mu$m; \citealt{cowie17}); thus, historically, these were called submillimeter galaxies (SMGs).
\footnote{``DSFG'' refers to any object with a far-infrared (FIR) to millimeter detection. Here we use the terms ``bright" ($\fsm>2$~mJy) and ``faint'' ($\fsm<2$~mJy) DSFGs to compare different submillimeter flux ranges. We generally only use ``SMGs'' to refer to submillimeter-selected samples with $\fsm \gtrsim 2$~mJy, corresponding to the classical definition.}
Despite their extreme star formation rates (SFRs), which can exceed $\sim$1000~${\rm M}_\odot\,$yr$^{-1}$, DSFGs with $\fsm > 2$~mJy (``bright DSFGs'') contribute only $\sim$20--30\% of the total star formation rate density (SFRD) above $z\sim 2$ due to their rarity \citep[e.g.,][]{barger12,barger14,cowie17}.
On the other hand, DSFGs with $\fsm < 2$~mJy (``faint DSFGs''), which make up most of the dust-obscured SFH, have been much more difficult to study in large numbers. 

Single-dish confusion noise prevents faint DSFGs from being directly selected for efficient follow-up with interferometers such as the Atacama Large Millimeter/submillimeter Array (ALMA) \citep[e.g.,][]{cowie18}.
They must therefore be selected using either direct interferometric surveys, which are costly and yield relatively small numbers of sources in limited areas \citep[e.g.,][]{walter16,aravena16}, or alternative techniques to probe beneath the confusion limit.
These techniques include taking advantage of the magnification from massive lensing clusters \citep[e.g.,][]{smail97,cowie02,cowie22,knudsen08,chen13,hsu16,fujimoto25,fujimoto23a}, 
or using predetermined optical to near-infrared (NIR) color selections as priors for making measurements in submillimeter maps \citep[e.g.,][]{wang12,caputi12,chen16, wang16,wang19,barger23,mckay24}.
As a result, there are only a handful of analyses of faint DSFG samples that have contrasted them with bright DSFG samples \citep[e.g.,][]{aravena16,chen16,cowie22,suzuki24,uematsu24}, and these have been limited by their small sample sizes.

Added to these difficulties, studies of faint DSFGs suffer from the same limitations as those of bright DSFGs. Most critically, their optical/NIR counterparts are challenging to identify due to the $\sim$7$''$--30$''$ resolution of single-dish submillimeter/millimeter telescopes. This hampers reliable estimates of their stellar masses and photometric redshifts. 
Moreover, even when their precise positions are known,
some DSFGs show little to no optical emission \citep[e.g.,][]{chen15, dacunha15, cowie18, stach19} at the typical depths of HST or ground-based surveys.

Finally, before the launch of JWST, high-resolution imaging redward of rest-frame $\sim$0.5~$\mu$m was not available for DSFGs above $z\sim2$. 
Since the rest-frame optical/NIR emission traces the majority of the stellar mass, and stellar morphologies can help distinguish between merger-driven or secular star-formation triggering mechanisms, 
this prevented a full understanding of both DSFGs' evolutionary pathways and the processes driving their SFRs. While deep HST observations at 1.6~$\mu$m suggested that
DSFGs resembled large, massive disks, with high rates of mergers/irregular morphologies and spatial offsets between the UV/optical and submillimeter emission \citep{swinbank10, targett13, chen15},
there was evidence that structured dust obscuration could be impacting these rest-frame UV/optical morphologies.
Now, JWST studies are placing stronger constraints on the stellar masses and morphologies of bright DSFGs \citep[e.g.,][]{chen22, gillman23,gillman24, hodge24}, including detecting counterparts too faint for HST \citep[e.g.,][]{perez-gonzalez23, xiao23, williams23, gottumukkala23, sun24a}; however, little is known about these properties for faint DSFGs. 

Rather than simply analyzing the JWST data for known DSFGs, we can invert the process and use the JWST data to identify DSFGs.
\citet{barger23} showed that red JWST/NIRCam F444W/F150W colors reliably identify NIR counterparts to SCUBA-2 850~$\mu$m-selected SMGs, building on past selections with HST and/or Spitzer \citep[e.g.,][]{wang12,caputi12,chen16, wang16,wang19}. They further showed that these red galaxies could be used as priors to identify SCUBA-2 850~$\mu$m sources below the confusion limit. In \citet{mckay24}, we verified that this selection is highly accurate by demonstrating that $\sim95$\% of the red NIRCam galaxies with SCUBA-2 detections are also associated with $>3\sigma$ ALMA detections.
This procedure provides large, uniform samples of both bright and faint DSFGs with accurate JWST counterparts, making it possible to study how the two populations compare.

We apply the \citet{barger23} selection across the A2744 and GOODS-S fields to develop a large sample of 850~$\mu$m and/or 1.2~mm-detected DSFGs, most of which are fainter than 2~mJy at 850~$\mu$m. In this paper, we focus on the stellar properties of these galaxies, which we measure from both integrated spectral energy distribution (SED) fits and surface brightness modeling.

In Section~2, we summarize the multiwavelength data in the A2744 and GOODS-S fields. In Section~3, we describe the NIRCam selection that we use to identify DSFGs, and we detail some initial characteristics of our sample. In Section~4, we fit the full optical-to-millimeter SEDs of the sample to constrain their stellar properties. 
We then describe the basic properties of the DSFG sample derived from our SED fitting in order to 
understand the nature of the galaxies identified by our NIRCam selection. 
In particular, we compare the properties of the faint DSFGs with
those of the well-studied bright DSFGs.
In Section~5, we analyze the surface brightness profiles of the DSFGs and investigate how many of them 
exhibit major merger signatures. 
Finally, in Section~6, we summarize our results.

Throughout the paper, we assume a flat concordance $\Lambda$CDM cosmology with $\Omega_m= 0.3$, $\Omega_\Lambda = 0.7$, and $H_0 = 70.0$~km~s$^{-1}$~Mpc$^{-1}$. We assume a \citet{kroupa01} initial mass function (IMF). All magnitudes are quoted in the AB system \citep{oke83}.

\section{Data}
\label{data}

\subsection{A2744}
The A2744 field has JWST observations from several surveys, including the GLASS-A2744 \citep{treu22} and UNCOVER \citep{bezanson22} programs.  We use the catalog of isophotal photometry from \citet{paris23} for consistency with \citet{barger23}. This includes eight bands of HST imaging from 0.4--1.6~$\mu$m (ACS: F435W, F606W, F775W, and F814W; WFC3: F105W, F125W, F140W, and F160W) and eight bands of JWST/NIRCam imaging from 0.9--4.4~$\mu$m (F090W, F115W, F150W, F200W, F277W, F356W, F410M, and F444W). We also make use of the additional data products provided for some sources by the UNCOVER team, such as lensing magnifications and photometric and spectroscopic redshifts (DR2; \citealt{weaver24}). The lensing magnifications are derived from the lensing model of \citet{furtak23b}, which is highly constrained using both HST and JWST multiply imaged sources. 

The SCUBA-2 data in A2744 were presented in \citet{cowie22} and discussed further in \citet{barger23} and \citet{mckay24}. The SCUBA-2 images are centered on the inner cluster region and have minimum 1$\sigma$ rms noise values at 450~$\mu$m and 850~$\mu$m of 2.8~mJy and 0.26~mJy, respectively.

In addition to the previous ALMA coverage of the central A2744 cluster region from the 1.2~mm ALMA Lensing Cluster Survey (ALCS; \citealt{fujimoto23a}) and the 1.1~mm ALMA Frontier Fields Survey (AFFS; \citealt{gonzalez-lopez17, munoz-arancibia23}) surveys, \citet{fujimoto25} recently published the Deep UNCOVER-ALMA Legacy High-Z (DUALZ) survey data and catalog, which covered a 24~arcmin$^2$ region corresponding to the UNCOVER 
NIRCam mosaic with 1.2~mm ALMA imaging down to a minimum rms value of 32.7~$\mu$Jy. 
We use the data products from that survey, since they include the previous ALMA data in the field.

\subsection{GOODS-S}

We include the GOODS-S field in our study to take advantage of its extremely deep JWST and submillimeter/millimeter observations, along with unparalleled multiwavelength coverage from the X-ray to the radio.

We primarily use the JWST data from the JADES survey \citep{eisenstein23}. We use the imaging and photometric catalog from the second data release (JADES DR2; \citealt{eisenstein23b}), which includes NIRCam photometry from the JEMS \citep{williams23} and FRESCO \citep{oesch23} surveys, for a total of 14 NIRCam bands covering 0.9--4.8~$\mu$m (F090W, F115W, F150W, F182M F200W, F210M, F277W, F335M, F356W, F410M, F430M, F444W, F460M, and F480M) in some areas. We use the photometry measured in the largest Kron apertures (Kron parameter = 2.5) to recover the total fluxes. The JADES DR2 catalog also includes the 0.4--1.6~$\mu$m HST photometry (ACS: F435W, F606W, F775W, F814W, and F850LP; WFC3: F105W, F125W, F140W, and F160W) measured from the Hubble Legacy Field project \citep{illingworth16, whitaker19}.

The GOODS-S has extremely deep SCUBA-2 450~$\mu$m and 850~$\mu$m data, which have been extensively analyzed in \citet{cowie18}, \citet{barger19}, and \citet{barger22}. The minimum 1$\sigma$ rms noise values of the current 450~$\mu$m and 850~$\mu$m maps are 1.67~mJy and 0.18~mJy, respectively. 

Portions of the GOODS-S have been observed by numerous ALMA programs covering a range of depths, frequencies, and beam sizes. Rather than re-reduce the data, we make use of published images or submillimeter/millimeter source catalogs, where possible. We use the ALMA 1.2~mm catalog from v2.0 of the GOODS-ALMA survey \citep{gomez-guijarro22} and the ALMA 1.2~mm catalog from the ASAGAO survey \citep{hatsukade18}. We also use the ALMA 870~$\mu$m, 1.2~mm, 2~mm, and 3~mm fluxes from \citet{cowie18} and \citet{mckay23}. 

To search for fainter ALMA sources, we use the 26~arcmin$^2$ combined 1.2~mm continuum image (with a $uv$-taper of 250~k$\lambda$ and primary beam correction applied) released by \citet{hatsukade18}, who combined data from \citet{dunlop17} and \citet{franco18} to obtain a depth of $\sim$30~$\mu$Jy. The resulting synthesized beam size of this image is 0$\farcs$59$\times$~0$\farcs$53. 

Finally, when performing our SED fits, we use the Spitzer/MIPS 24~$\mu$m photometry from \citet{elbaz11} and the Herschel/PACS 100 and 160~$\mu$m and Herschel/SPIRE 250 and 350~$\mu$m photometry from the HerMES survey \citep{oliver12}. We do not use the SPIRE 500~$\mu$m data due to the low spatial resolution and large uncertainties, and because we have the deep SCUBA-2 450~$\mu$m data.

\begin{figure}[!t]
    \centering
    \includegraphics[width=\linewidth]{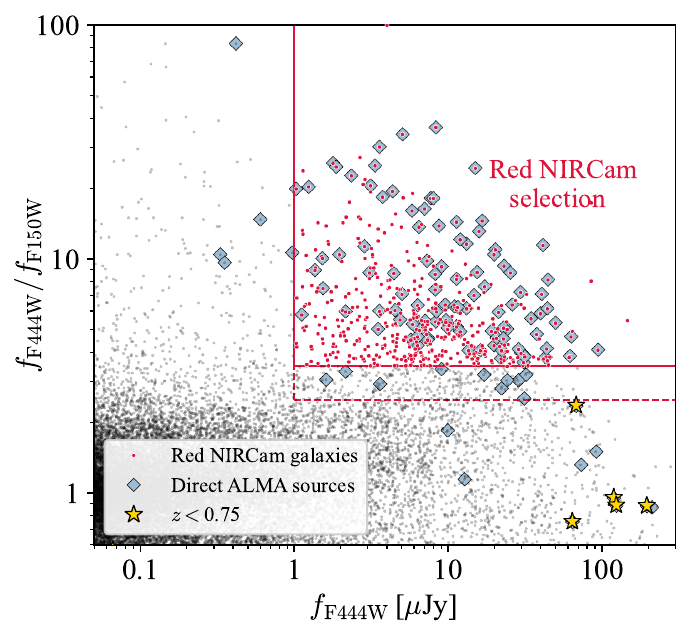}
    \caption{The red NIRCam galaxy selection (Equation~\ref{redselection}; red solid lines) shown in $\ff/\fo$ color and $\ff$ flux space. The sources from the \citet{paris23} and \citet{eisenstein23b} catalogs that satisfy this selection are shown as red points, while the remainder of the catalog sources are shown as black points. The ALMA 870~$\mu$m and/or 1.1/1.2~mm sources from \citet{hatsukade18}, \citet{cowie18}, \citet{gomez-guijarro22},  and \citet{fujimoto25} that have NIRCam counterparts are shown as blue diamonds (note that four sources lie off the lower-left corner of the plot and may be spurious). ALMA sources with $z < 0.75$, including A2744 cluster members, are denoted by yellow stars. The red dotted lines illustrate an alternative selection with the color cut relaxed to $\ff/\fo > 2.5$. Note that the F444W fluxes for sources in A2744 have not been corrected for lensing.}
    \label{fig:red_color_select}
\end{figure}

\section{Sample Selection}

\subsection{Red NIRCam Galaxies}
\label{redpriors}

The photometric selection outlined in \citet{barger23} for identifying DSFG counterparts is as follows:
\begin{equation}
\label{redselection}
\begin{cases}
\ff/\fo > 3.5, \\
\ff > 1~{\rm \mu Jy}.
\end{cases}
\end{equation}
We illustrate this red color selection in $\ff/\fo$ color and $\ff$ flux space as the red solid lines in Figure~\ref{fig:red_color_select}. We plot all sources from the JADES DR2 or the GLASS-A2744 photometric catalogs (black points), marking those which satisfy Equation~\ref{redselection} (red points). We also show the ALMA 870~$\mu$m and/or 1.1/1.2~mm sources from \citet{cowie18}, \citet{hatsukade18}, \citet{gomez-guijarro22}, and \citet{fujimoto25} (blue diamonds).
There are 132 unique ALMA sources from these surveys that lie on the F150W and F444W imaging.  Two of these appear to match to red galaxies that are not deblended from brighter foreground objects in the JADES catalog, so we exclude them. Of the remaining 130 ALMA sources, 125 have clear ($\ff > 0.1$~$\mu$Jy) NIRCam counterparts at the ALMA positions. 

For the other five ALMA sources, one has no NIRCam counterpart within 1$\farcs$5, and four match to very faint ($\ff \lesssim 0.05$~$\mu$Jy), blue NIRCam sources; i.e., they lie off the lower-left side of Figure~\ref{fig:red_color_select}. All five of these objects are near the ALMA detection limits and are consistent with being spurious based on negative peak analyses in the original ALMA surveys, so we mark them as questionable and do not consider them further in our analysis. 

We find that 100 out of 125 ALMA sources with robust NIRCam counterparts satisfy the criteria of Equation~\ref{redselection}. Almost all of the 25 ALMA sources that do not satisfy the selection lie close to the selection boundary, being slightly fainter in F444W or slightly bluer in $\ff/\fo$. The ALMA detections that fall in the lower-right corner of Figure~\ref{fig:red_color_select} mostly correspond to sources with $z < 0.75$ (yellow stars), including several A2744 cluster members.

It is clear from Figure~\ref{fig:red_color_select} that although the red color selection is very effective at capturing the known ALMA sources, it is not complete. Excluding the five sources at $z<0.75$, this selection captures $\sim$83\% of the known ALMA sources with robust NIRCam counterparts.
If we relaxed our selection to $\ff/\fo>2.5$, $\ff>1~\mu$Jy (red dotted lines in Figure~\ref{fig:red_color_select}), we would recover an additional 10 ALMA sources (110, or 92\% of the total). However, in total, $\sim$800 galaxies satisfy this relaxed selection, compared to $\sim$450 galaxies that satisfy the original selection, meaning that the efficiency decreases drastically as we go to lower $\ff/\fo$ ratios. 
Thus, for this paper, as a good trade-off between achieving high completeness and high purity while also providing an efficient sample to use as priors in selecting faint DSFGs, we use the original selection from \citet{barger23}. 

To define our sample of red NIRCam galaxies, we begin by selecting all sources in either the GLASS-A2744 or the JADES DR2 photometric catalogs that satisfy Equation~\ref{redselection}.

In A2744, the SCUBA-2 coverage does not extend all the way to the edge of the NIRCam mosaic, so we restrict our selection to the overlapping region. We also remove sources that have {\tt FLAG~=~1} in the \citet{paris23} catalog and combine multiple sources that correspond to a single galaxy. After making these changes, we find 158 total red NIRCam galaxies in A2744. The effective NIRCam+SCUBA-2 area from which these 158 galaxies are selected is 41.3~arcmin$^2$, giving a surface density of 3.82~sources~arcmin$^{-2}$.

In the GOODS-S, the JADES data include several regions where the NIRCam F150W imaging does not cover the entire F444W region. As a result, some sources in the catalog lie only partially on the F150W image and have inflated $\ff/\fo$ ratios due to the lower F150W fluxes. We visually inspect each source to make sure we are not including sources that fall off the edge of the F150W coverage. This leaves 295 sources that satisfy the red color-selection criteria and are not flagged. However, two sources correspond to the same galaxy, so we combine them, giving a total of 294 galaxies in the GOODS-S. Since the NIRCam imaging we use covers an effective area of 63.4~arcmin$^2$, this corresponds to a surface density of 4.64~sources~arcmin$^{-2}$.

After combining the samples from both fields, we have a final catalog of 452 red NIRCam galaxies.

\subsection{Submillimeter/millimeter Properties of Red NIRCam Galaxies}
\label{submmprops}

We next determine how many of the red NIRCam galaxies are associated with significant SCUBA-2 850~$\mu$m or ALMA continuum detections. The method we use for obtaining SCUBA-2 fluxes based on prior positions follows that of \citet{barger23}, who demonstrated that it finds the directly detected (i.e., $>5\sigma$) sources and identifies marginal ($>3\sigma$) sources. Thus, rather than match to existing SCUBA-2 catalogs, we begin with the JWST prior positions. We first measure the SCUBA-2 850~$\mu$m flux at the position of each red NIRCam galaxy position. Then, working down from brighter to fainter 850~$\mu$m fluxes, we extract the peak flux within 4$''$ of each prior. We clean the source from the SCUBA-2 image using the matched-filter point spread function (PSF), in order to limit contamination of fainter sources by the wings of brighter sources. We then measure the rms noise at the position of each source.

We consider a galaxy to be detected in the SCUBA-2 850~$\mu$m data if it is associated with a $>3\sigma$ SCUBA-2 source. 
As we showed in \citet{mckay24}, this method identifies the correct counterpart to SCUBA-2 850~$\mu$m sources with an accuracy of up to 95\%, which is comparable to or better than a combined radio\,+\,machine learning method used prior to the launch of JWST \citep[][]{an18}. 

Across both the A2744 and GOODS-S fields, there are 184 red NIRCam galaxies with $>3\sigma$ SCUBA-2 850~$\mu$m detections. Of these, 33 have known ALMA 870~$\mu$m fluxes from \citet{cowie18}, which we use in place of the SCUBA-2 850~$\mu$m fluxes for the rest of our analysis (hereafter, we will refer to both as 850~$\mu$m fluxes). 
We also include in our sample four red NIRCam galaxies with ALMA 870~$\mu$m fluxes but $< 3 \sigma$ SCUBA-2 850~$\mu$m detections.\footnote{We note that there are two ALMA 870~$\mu$m sources from \citet{cowie18} (sources \#1 and \#29 in their Table~4) that do fall on the JADES F444W and F150W NIRCam imaging but are not included in our sample. Both objects are clearly red galaxies associated with the ALMA 870~$\mu$m emission but are blended with foreground galaxies in the JADES catalog. We exclude these objects to ensure a uniform selection process.} To check for biases between sources that have ALMA 870~$\mu$m detections and those that only have SCUBA-2 850~$\mu$m data, we compare the SCUBA-2 850~$\mu$m to ALMA 870~$\mu$m fluxes for sources which have both, finding a median ratio of $f_{\text{SCUBA-2}}/f_{\rm ALMA}=1.01$.

By running the same detection procedure on 1000 randomly selected positions in the SCUBA-2 850~$\mu$m images, we constrain the false-positive fraction to be $\sim$10\%, meaning that roughly 18 of our $>3\sigma$ SCUBA-2 sources could be false positives. If we were to lower the SCUBA-2 detection threshold to $2\sigma$, then we would find 257 red NIRCam galaxies with 850~$\mu$m detections; however, in this scenario, the false-positive fraction increases to $\sim$20\%.

In \citet{mckay24}, we used the red NIRCam galaxies as priors to recover faint ALMA 1.2~mm sources that were not bright enough to be directly detected in the ALMA images. We do the same here for sources that fall within either the DUALZ 1.2~mm {\it Wide} map \citep{fujimoto25} in A2744 or the combined 1.2~mm image from \citet{hatsukade18} in the GOODS-S. We use the publicly available primary beam-corrected mosaics for both fields. 
For each prior position lying in one of these ALMA mosaics, we measure the 1.2~mm flux by finding the largest local 
peak within $1\farcs0$ of the NIRCam position. 
We estimate the rms noise by measuring the standard deviation in 5000 random pixels within 12$''$ of the source, after masking out any $>4\sigma$ peaks in the region. We consider a galaxy detected if the peak flux is more than 3$\times$ the estimated rms noise; we adopt these peak fluxes and rms errors as our 1.2~mm fluxes and errors. However, for sources matching to the 1.1/1.2~mm ALMA catalogs in Section~\ref{redpriors}, we use the published fluxes.

In total, there are 234 red NIRCam galaxies with at least a $>3\sigma$ detection at 850~$\mu$m and/or at 1.1/1.2~mm. We hereafter refer to these galaxies as the \textit{red DSFG sample}. This is the sample on which we focus for the remainder of the paper. We summarize the red DSFG sample and the parent sample of red NIRCam galaxies in Table~\ref{tab:samples}. 

\begin{figure}
    \centering
    \includegraphics[width=\linewidth]{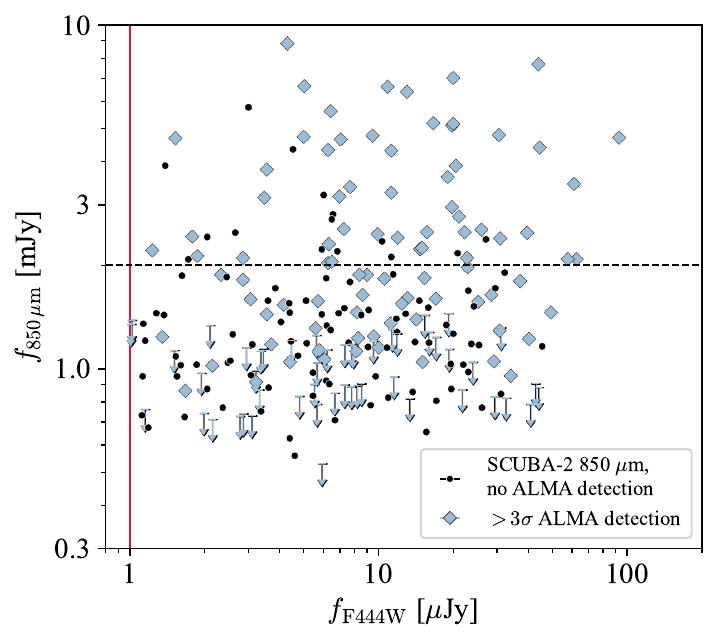}
    \caption{The 850~$\mu$m flux vs. F444W flux for the red DSFG sample, distinguishing between sources with at least a $>3\sigma$ ALMA detection (light blue diamonds) and those with $>3\sigma$ SCUBA-2 850~$\mu$m detections but either no ALMA coverage or no ALMA detection (black points). For sources with $<3\sigma$ SCUBA-2 850~$\mu$m detections (note that these have $>3\sigma$ ALMA 1.1/1.2~mm detections), we show the $3\sigma$ upper limits (downward facing arrows). We mark our selection criterion of $\ff > 1~\mu$Jy (red solid line). We also show the $\fsm = 2$~mJy threshold between bright and faint DSFGs (black dashed line). Note that fluxes for sources in the A2744 field have not been corrected for lensing magnification.}
    \label{fig:f444w_scuba2}
\end{figure}

We emphasize that $\sim$60\% of the red DSFG sample comprises new DSFGs revealed by the JWST selection, rather than previously known ALMA sources. In Figure~\ref{fig:f444w_scuba2}, we show 850~$\mu$m flux versus F444W flux for all 234 galaxies. Of these, 47 galaxies (20\%) have 850~$\mu$m fluxes brighter than 2~mJy, while 187 (80\%) have 850~$\mu$m fluxes or $3\sigma$ upper limits less than 2~mJy (we discuss the relative properties of these two subsets in Sections~\ref{seds} and \ref{morph}). The median 850~$\mu$m flux of the entire red DSFG sample is $\langle\fsm\rangle = 1.15$~mJy. 

In Table~\ref{tab:catalog}, we provide the full catalog of galaxies in the red DSFG sample, including source IDs, coordinates, redshifts, 850~$\mu$m and 1.2~mm fluxes and errors, and various properties that we estimate from our SED fits and morphological analysis.

\begin{deluxetable}{cccc}
\tablewidth{\textwidth}
\tablecaption{\label{tab:samples}
Sample Summary}
\tablehead{
& \colhead{GOODS-S} & \colhead{A2744} & \colhead{Total} }
\startdata
Red NIRCam Galaxies&294&158&452\\
Red DSFG sample\tablenotemark{\scriptsize a}  & 144 & 90 & 234 \\
&&&\\
\multicolumn{4}{c}{Red DSFG Sample}\\
\hline
$f_{\rm850\,\mu m}>2$~mJy &35& 12& 47 \\
$f_{\rm850\,\mu m}<2$~mJy &109& 78&187 \\
Spectroscopic\tablenotemark{\scriptsize b} redshifts &77&47&124\\
\enddata
\tablenotetext{ a}{The red DSFG sample refers to red NIRCam galaxies with $>3\sigma$ 850~$\mu$m or 1.1/1.2~mm detections.}
\tablenotetext{ b}{This includes the HST and JWST grism redshifts (see Section~\ref{redshifts}).}
\end{deluxetable}

\begin{figure}
    \centering
    \includegraphics[width=\linewidth]{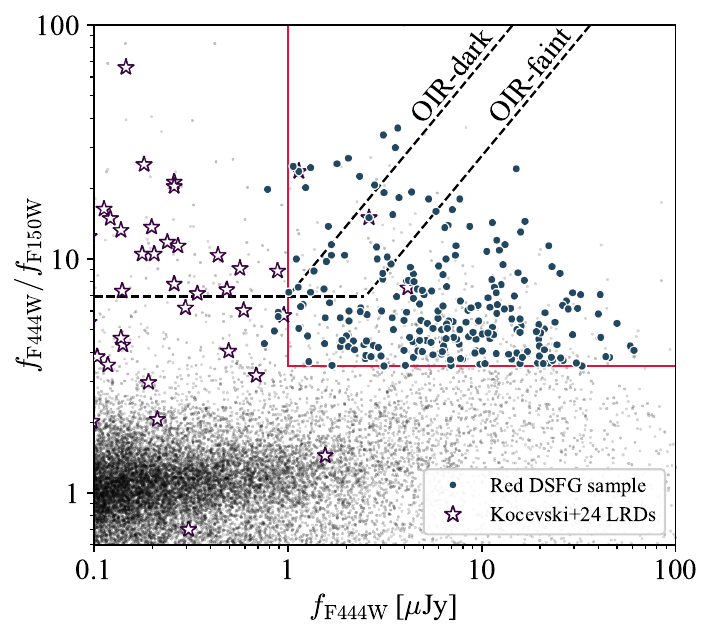}
    \caption{The $\ff/\fo$ colors and $\ff$ fluxes of the red DSFG sample (dark-blue points). We show the color selections for OIR-faint and OIR-dark galaxies (black dashed lines; \citealt{gottumukkala23}). ``Little red dots'' (LRDs) from \citet{kocevski24} in the A2744 and GOODS-S fields are shown as purple stars. 
    We show the remaining sources in the \citet{paris23} and \citet{eisenstein23b} NIRCam catalogs as black points. Note that here, unlike Figure~\ref{fig:red_color_select}, the $\ff$ fluxes for sources in A2744 have been corrected for lensing.
   }
    \label{fig:redcolors_comparison}
\end{figure}

\subsection{Comparison with Other Red Galaxy Selections}
\label{other_selections}

A growing number of studies using HST, Spitzer, and/or JWST data have indicated that massive galaxies appearing faint or undetected at optical/NIR wavelengths (sometimes called ``HST-dark'' or ``OIR-dark'', with varying definitions) may contribute substantially to the stellar mass density and SFRD at $z>3$ \citep[e.g.,][]{wang12, wang16, wang19, alcalde-pampliega19, manning22, barrufet23, gottumukkala23, williams23, gibson24, xiao23a, xiao23, barrufet24}.
These galaxies frequently exhibit reddened optical slopes and ongoing star formation, as evidenced by H$\alpha$ detections \citep[e.g.,][]{wang19, yamaguchi19, perez-gonzalez23, xiao23a, barrufet24}. However, only a subset of them have high-enough SFRs to be detected at current submillimeter/millimeter survey limits and $\sim$10--20\% have been identified as massive quiescent galaxies \citep{perez-gonzalez23,barrufet24}. 

As the sample of DSFGs studied in this paper are also selected through red NIR colors, it is worth understanding how our sample compares with these populations. For simplicity, we adopt the selections used in \citet{gottumukkala23} (see also, e.g., \citealt{perez-gonzalez23}) and define {\it OIR-faint} to refer to galaxies with $\ff/\fo > 6.9$ and $\fo < 0.36$~$\mu$Jy (25~mag), and we use {\it OIR-dark} to refer to the subset with $\ff/\fo > 6.9$ and $\fo < 0.14$~$\mu$Jy (26~mag). 

In Figure~\ref{fig:redcolors_comparison}, we contrast the red NIRCam galaxy selection criteria
shown in Figure~\ref{fig:red_color_select} (red solid lines; Equation~\ref{redselection}) 
with the OIR-faint and OIR-dark selection criteria (black dashed lines). 
We show the red DSFG sample with dark-blue circles.
Of the 234 sources in the red DSFG sample, a total of 35 (15\%) are OIR-faint, with 16 of these being OIR-dark. 

The 35 OIR-faint DSFGs comprise 53\% of the total number of red NIRCam galaxies that are OIR-faint. 
This is similar to the fraction of DSFGs in the total red NIRCam selected galaxy population.

However, below $\ff = 1$~$\mu$Jy, there are very few DSFGs, as was first shown by 
\citet{barger23} for $f_{\rm F444W} \geq 0.05 \,\mu$Jy.
This is consistent with the picture put forth by, e.g., \citet{gottumukkala23}, in which the majority of the OIR-dark galaxies are ``normal'' star-forming galaxies with significant dust obscuration, whose SFRs would not be high enough to be detected via their cold dust emission by typical submillimeter/millimeter surveys.

A second red population that we can compare with our selection is ``little red dots'' (LRDs), a class of objects revealed by JWST that typically display compact morphologies and characteristic V-shaped SEDs in the rest-frame optical (though the exact definition varies; e.g., \citealt{furtak23a, kokorev23, matthee24, kocevski24, greene24, taylor24,labbe23}). Some of these seem to host moderate-luminosity broad-line active galactic nuclei (AGNs; for which hot dusty torus emission can produce a steep red rest-frame optical/NIR slope), despite showing only very marginal X-ray emission in deep stacked data \citep[e.g.,][]{tasnim24}. 


\begin{longrotatetable}
\begin{deluxetable*}{cccccccccccccccc}
\tablewidth{0pt}
\tabletypesize{\scriptsize}
\tablecaption{\label{tab:catalog}
Red DSFG Sample}
\tablehead{
\colhead{ID}&
\colhead{R.A.}&
\colhead{Decl.}&
\colhead{$z$}&
\colhead{Type}&
\colhead{$\fsm$}&
\colhead{$f_{\rm 1.2\,mm}$}&
\colhead{SFR}&
\colhead{$\log(M_*)$}&
\colhead{$A_V$}&
\colhead{$\log(L_{\rm IR})$}&
\colhead{$\log(L_{\rm UV})$}&
\colhead{$n_{\rm F444W}$}&
\colhead{$n_{\rm F150W}$}&
\colhead{$R_{e,\,{\rm F444W}}$}&
\colhead{$R_{e,\,{\rm F150W}}$}\\
\colhead{}&
\colhead{(J2000)}&
\colhead{(J2000)}&
\colhead{}&
\colhead{}&
\colhead{(mJy)}&
\colhead{(mJy)}&
\colhead{(${\rm M}_\odot\,{\rm yr}^{-1}$)}&
\colhead{}&
\colhead{}&
\colhead{}&
\colhead{}&
\colhead{}&
\colhead{}&
\colhead{(kpc)}&
\colhead{(kpc)}
}
\startdata
GS-1&53.1027942&$-27.8928805$&1.920&2&$3.25\pm0.14$&$0.793\pm0.06$&368$^{+24}_{-23}$&10.12$^{+0.04}_{-0.05}$&2.10$^{+0.06}_{-0.06}$&12.49$^{+0.03}_{-0.03}$&9.93$^{+0.02}_{-0.02}$&0.77&1.14&1.89&2.62\\
GS-2&53.1015140&$-27.8699588$&3.10$^{+0.35}_{-0.09}$&0&$0.17\pm0.24$&\nodata&160$^{+22}_{-18}$&10.40$^{+0.05}_{-0.05}$&2.66$^{+0.07}_{-0.05}$&12.13$^{+0.06}_{-0.05}$&9.40$^{+0.03}_{-0.03}$&3.44&0.74&0.6&2.2\\
GS-3&53.1261982&$-27.7996239$&1.886&1&$0.88\pm0.19$&$0.146\pm0.07$&34$^{+10}_{-9}$&10.08$^{+0.04}_{-0.04}$&2.24$^{+0.09}_{-0.09}$&11.46$^{+0.11}_{-0.14}$&9.05$^{+0.02}_{-0.02}$&0.76&1.49&1.08&1.26\\
GS-4&53.1159040&$-27.9133627$&2.68$^{+0.05}_{-0.61}$&0&$1.87\pm0.39$&\nodata&79$^{+21}_{-15}$&9.74$^{+0.08}_{-0.10}$&1.94$^{+0.08}_{-0.09}$&11.82$^{+0.10}_{-0.09}$&9.50$^{+0.05}_{-0.05}$&3.11&0.9&1.15&2.49\\
GS-5&53.1115530&$-27.9093844$&2.21$^{+0.54}_{-0.09}$&0&$1.85\pm0.36$&\nodata&169$^{+35}_{-34}$&9.94$^{+0.08}_{-0.09}$&1.61$^{+0.09}_{-0.07}$&12.15$^{+0.08}_{-0.10}$&10.00$^{+0.03}_{-0.03}$&1.09&2.33&2.94&4.48\\
\enddata
\tablecomments{Coordinates are the JWST/NIRCam coordinates listed in the catalogs of \citet{eisenstein23b} and \citet{paris23}, respectively. For sources in the A2744 field, fluxes and derived quantities have been corrected for lensing magnification. The $z$ Type values are as follows: 0 = photometric redshift, 1 = spectroscopic redshift, 2 = HST grism redshift, and 3 = JWST grism redshift. For the SFR, $M_*$, $A_V$, $L_{\rm IR}$, and $L_{\rm UV}$ columns, the values listed are the median values of the posterior distributions, with the errors referring to the 16th--84th percentile ranges of the posteriors. We do not list errors for the morphological parameters because the uncertainties reported by {\tt GALFIT} are known to underestimate the true errors.}
\tablerefs{Redshift references: 
\citet{bacon23}, \citet{barger19}, \citet{barrufet24}, \citet{cowie23}, \citet{deugenio24}, \citet{garilli21}, \citet{gonzalez-lopez19}, \citet{inami17}, \citet{kriek15}, \citet{kurk12}, \citet{mclure18}, \citet{mckay23}, \citet{mignoli05}, \citet{momcheva16}, \citet{munoz-arancibia23}, \citet{naidu24}, \citet{pentericci18}, \citet{price24}, \citet{sun22}, and \citet{vanzella08}.}
\tablecomments{Table \ref{tab:catalog} is published in its entirety in the electronic edition of the {\it Astrophysical Journal}.  A portion is shown here for guidance regarding its form and content.}
\end{deluxetable*}
\end{longrotatetable}

Nearly all of the DSFGs are spatially extended in $\ff$, so we do not expect there to be any substantial overlap 
with the LRD samples.
To illustrate this, we use the catalog of LRDs compiled by \citet{kocevski24}, who selected compact JWST sources with red optical and blue UV slopes from several legacy field LRD samples. There are 69 sources in their sample across the GOODS-S and A2744 fields, which we plot in Figure~\ref{fig:redcolors_comparison} (purple stars). The LRDs generally have similar $\ff/\fo$ colors but fainter F444W fluxes than the red DSFG sample. This is not by construction, but reflects a real distinction between the populations. LRDs also tend to lie at $z>3$, while the red DSFG sample (and the DSFG population in general) peaks around $z\sim 2.5$ (see Section~\ref{redshifts}).

Cross-matching the red DSFG sample to the \citet{kocevski24} catalog, we find that 
just two of our red DSFGs are also selected as LRDs. One of these matched DSFGs is also the only X-ray source among the LRD sample in the GOODS-S ($f_{\rm 0.5-7 \,keV} = 1.1\times10^{-15}$~erg\,s$^{-1}$\,cm$^{-2}$; see \citealt{kocevski24} for an extended discussion of this source). In the mid- to far-infrared (FIR), this source is detected in Spitzer/MIPS and SCUBA-2 850~$\mu$m data ($\fsm=1.26$~mJy, $\sim3.1\sigma$). 

A stacking of the remaining LRD positions in the SCUBA-2 850~$\mu$m data results in a stacked flux of $\fsm = -0.056 \pm 0.043$~mJy. Thus, the majority of LRDs are not DSFGs, even down to faint submillimeter fluxes, in agreement with the stacking analysis of \citet{labbe23} of 20 sources with ALMA 1.2~mm coverage in A2744.

\subsection{Redshifts}
\label{redshifts}

We compile spectroscopic redshifts from the literature for the sources in the red DSFG sample, where available.  

In the GOODS-S, these redshifts are taken from a variety of surveys and previously compiled catalogs \citep{ mignoli05, vanzella08, kurk12, lefevre15,kriek15,inami17, cowie18, mclure18, pentericci18,gonzalez-lopez19, garilli21, bacon23,cowie23,mckay23}, as well as JWST/NIRSpec programs \citep{bunker23, deugenio24}. 
For two sources, we adopt spectroscopic redshifts obtained from our own Keck/MOSFIRE observations. 
We also include HST grism redshifts from the 3D-HST program  \citep{momcheva16}, though we note that these tend to be more uncertain than other spectroscopic or JWST grism redshifts. 

In A2744, the spectroscopic redshifts come mainly from the ALT JWST/NIRCam grism program of \citet{naidu24} and the UNCOVER JWST/NIRSpec program of \citet{price24}.
We limit the \citet{price24} redshifts to those with quality flag $q = 3$. We also include literature redshifts from the compilations of \citet{munoz-arancibia23} and \citet{sun22} (the redshifts for these three sources come from ALMA program \#2017.1.01219.S; PI: F. Bauer).  Finally, we include six spectroscopic redshifts obtained from our own Keck/MOSFIRE observations. 

In total, 124 galaxies (53\%) in the red DSFG sample have spectroscopic or grism redshifts; of these, 32 are HST grism redshifts. For all but four of the remaining 110 sources without spectroscopic or grism redshifts, we adopt photometric redshifts from the {\tt EAZY} code, published in the \citet{eisenstein23b} and \citet{weaver24}  catalogs. These redshifts are measured using just the HST and JWST photometry in each field. In Section~\ref{testing_photzs}, we discuss the four cases where there is no {\tt EAZY} redshift or where we adopt a different redshift. 

\begin{figure}
    \centering
    \includegraphics[width=0.98\linewidth]{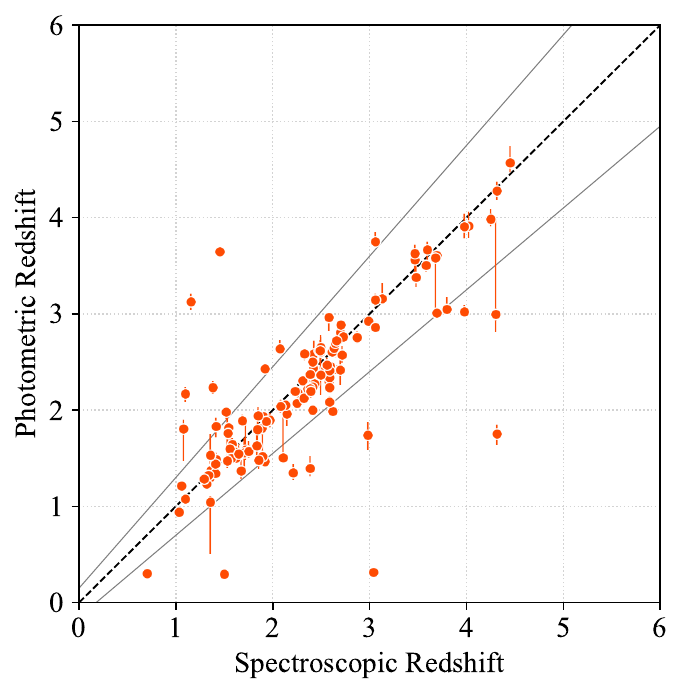}
    \caption{Photometric redshift from {\tt EAZY} (error bars are the 16th--84th percentile ranges of the posteriors)
    vs. spectroscopic redshift (including HST grism)
    for sources in the red DSFG sample with spectroscopic redshifts (orange points). We also show the one-to-one correlation (black dashed line) and the $|z_{\rm spec} - z_{\rm phot}|/(1 + z_{\rm spec}) = 0.15$ threshold (gray solid lines). Note that 87\% of the sources fall within this threshold.
    }
    \label{fig:eazy_redshifts}
\end{figure}

In Figure~\ref{fig:eazy_redshifts}, we compare the photometric redshifts to the spectroscopic redshifts for the sources that have both. The accuracy of the photometric redshifts in predicting the true redshifts is very high, with 87\% of the sources having $|z_{\rm spec} - z_{\rm phot}|/(1 + z_{\rm spec}) < 0.15$ and 91\% having $|z_{\rm spec} - z_{\rm phot}|/(1 + z_{\rm spec}) < 0.20$.

\begin{figure}[!t]
    \centering
    \includegraphics[width=\linewidth]{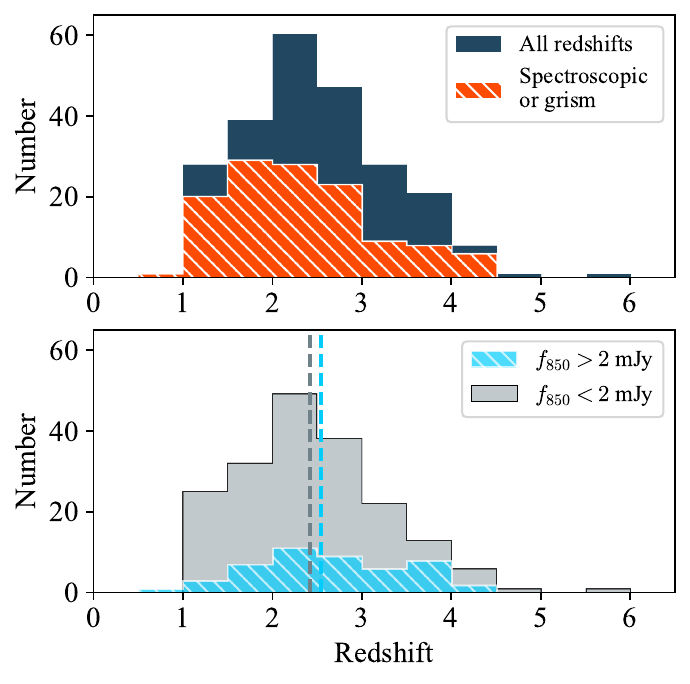}
    \caption{Top panel: Distribution of the best available redshifts for the red DSFG sample (dark blue), including adopted photometric redshifts, compared with the distribution of only spectroscopic/grism redshifts (hatched orange). 
    Bottom panel: Comparison of redshift distributions for the $\fsm < 2$~mJy (gray histogram) and $\fsm > 2$~mJy (hatched light blue histogram) subsets. We also show the median redshifts of $\langle z \rangle = 2.42$ for the $\fsm < 2$~mJy subset (gray dashed line) and $\langle z \rangle = 2.54$ for the $\fsm > 2$~mJy subset (light blue dashed line).}
    \label{fig:redshifts}
\end{figure}

\begin{figure*}
    \centering
    \includegraphics[width=\linewidth]{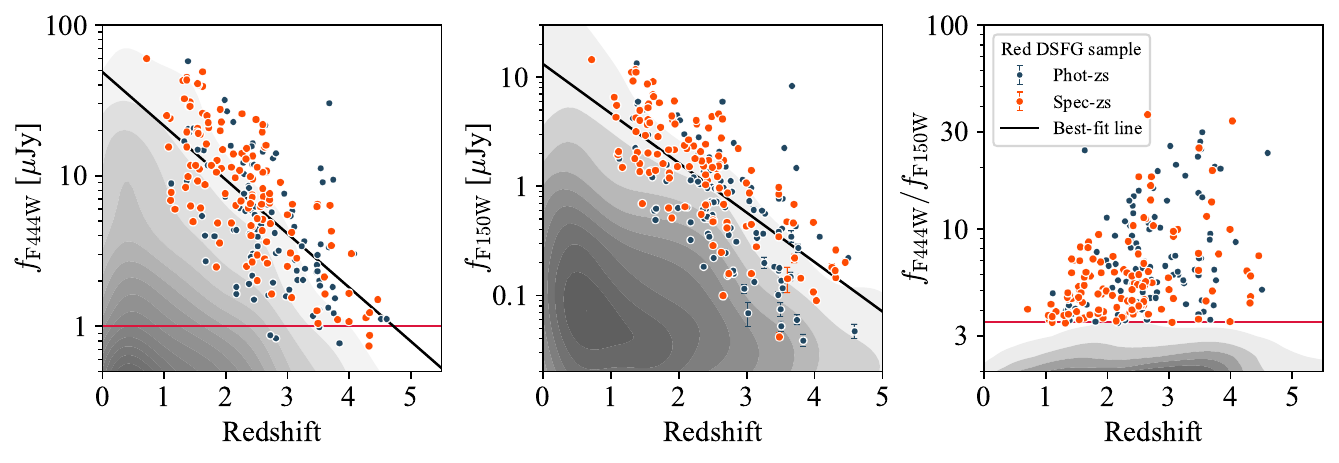}
    \caption{For sources in the red DSFG sample, $\ff$ (left), $\fo$ (center), and $\ff/\fo$ (right) vs. redshift, distinguishing between photometric redshifts (dark-blue points) and spectroscopic redshifts (orange points). For most of the DSFGs, the y-axis error bars are smaller than the data points. In the left and center panels, we show the best-fit linear relationship (black line). In the left and right panels, we show the red color selection of Equation~\ref{redselection} (red solid line). For comparison, we show the distribution of galaxies from the \citet{paris23} JWST catalog (gray contours), using photometric redshifts from \citet{weaver24}.  Fluxes for sources in A2744 have been corrected for lensing.
    }
    \label{fig:redshift_flux_ratio}
\end{figure*}

In Figure~\ref{fig:redshifts}, we show the final redshift distribution of the red DSFG sample. In the top panel, we show the breakdown of photometric versus spectroscopic/grism redshifts, while in the bottom panel, we show the redshift distributions for the galaxies with $\fsm>2$~mJy and $\fsm < 2$~mJy, respectively. 

The median redshift of the full red DSFG sample is $\langle z \rangle = 2.43$, with a 16th--84th percentile range of $z = 1.57$--3.46. This generally agrees with other samples of 850~$\mu$m-selected DSFGs \citep[e.g.,][]{chapman05,danielson17, cowie18, dudzeviciute20}. Interestingly, the faint DSFGs show
no sign of having a different redshift distribution than the bright DSFGs, which stands in contrast to the trend of increasing redshift with increasing submillimeter flux noted by other DSFG studies \citep[e.g.,][]{ivison02, pope05, smolcic12,cowie17,stach19,birkin21,chen22b}.

In Figure~\ref{fig:redshift_flux_ratio}, we show how the F444W flux, F150W flux, and F444W/F150W flux ratios of the red DSFG sample vary with redshift. For comparison, we show the distributions for the JWST catalog of \citet{paris23} in A2744, using the {\tt EAZY} photometric redshifts from \citet{weaver24}. This figure demonstrates that for the DSFGs, both $\ff$ and $\fo$ have a strong negative correlation with redshift. The reason for the relationship is that we are seeing the tip of the luminosity function at any given redshift (\citealt{naidu24}; A. Barger et al., in preparation), with the DSFGs clearly having brighter $\ff$ and $\fo$ than the vast majority of JWST sources. By selection, they also exhibit by far the reddest $\ff/\fo$ colors, corresponding to their dusty nature. 
We note that we also see a correlation between the $\ff/\fo$ color and redshift (right panel), though there is a larger degree of scatter. This would suggest that the small percentage of bluer DSFGs missed by our color selection---most of which also have $\ff \gtrsim 10\,\mu$Jy (see Figure~\ref{fig:red_color_select})---may lie at lower redshifts on average.

Our finding of a correlation between $f_{\rm F444W}$ and redshift agrees with the correlation found in \citet{barger22} using Spitzer/IRAC 4.5~$\mu$m fluxes for 450~$\mu$m-selected DSFGs. 
The best-fit line for our $f_{\rm F444W}$--$z$ correlation is
\begin{equation}
    \log(f_{\rm F444W}) = (-0.36 \pm 0.02) (1+z)  +(2.05 \pm 0.09),
\end{equation}
with $\ff$ in $\mu$Jy. The best-fit line for our $f_{\rm F150W}$--$z$ correlation is
\begin{equation}
    \log(f_{\rm F150W})  = (-0.45 \pm 0.03) (1+z) +(1.57 \pm 0.11).
\end{equation}

The tightness of these correlations suggests that it may be possible to select high-redshift DSFGs from submillimeter samples based on their NIRCam 1.5~$\mu$m or 4.4~$\mu$m fluxes. It also suggests that rare $z>5$ DSFGs may be inherently missed by our selection due to having fainter $\ff$.

\begin{figure*}[!t]
    \centering
    \includegraphics[width=0.495\linewidth]{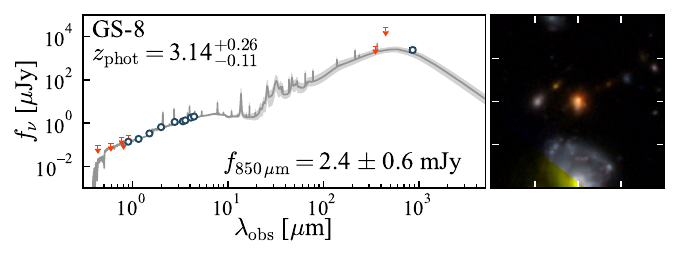}
    \includegraphics[width=0.495\linewidth]{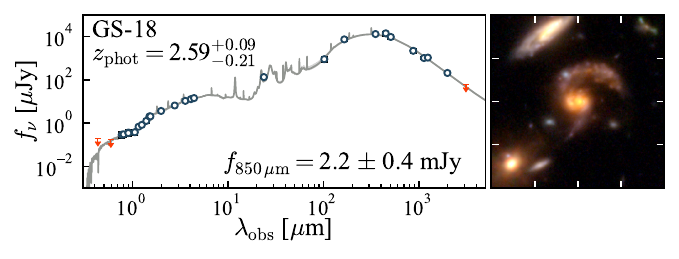}
    \includegraphics[width=0.495\linewidth]{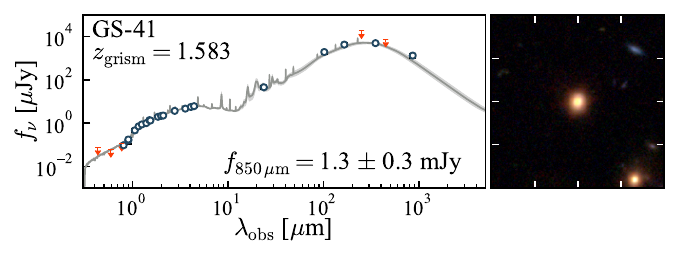}
    \includegraphics[width=0.495\linewidth]{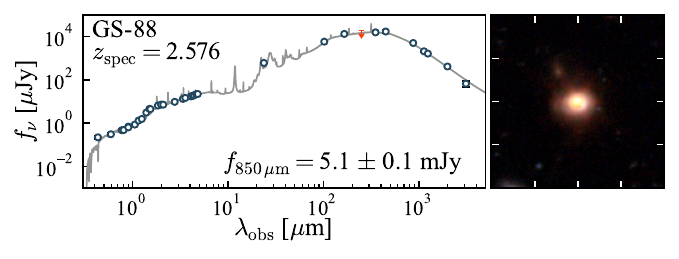}
    \includegraphics[width=0.495\linewidth]{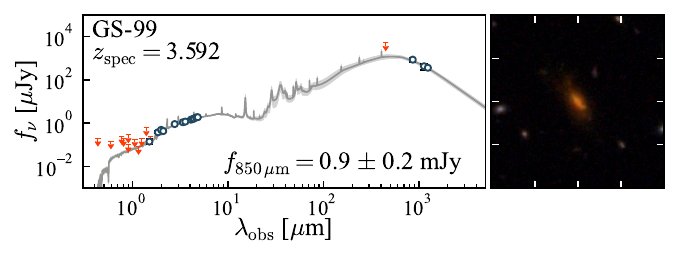}
    \includegraphics[width=0.495\linewidth]{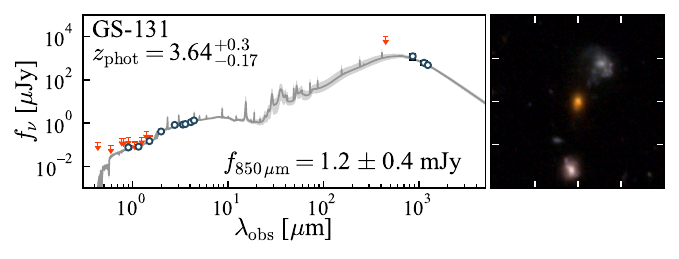}
    \includegraphics[width=0.495\linewidth]{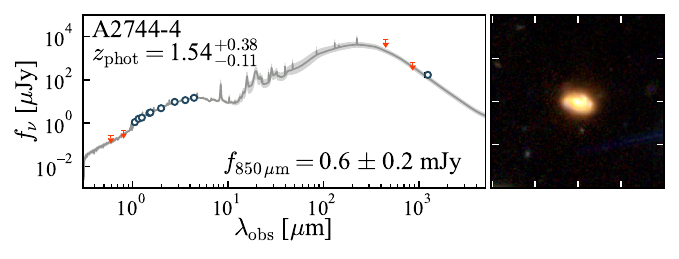}
    \includegraphics[width=0.495\linewidth]{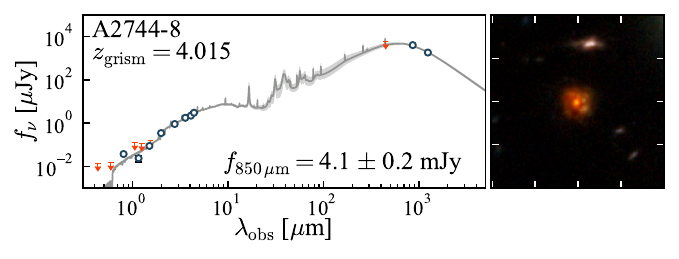}
    \includegraphics[width=0.495\linewidth]{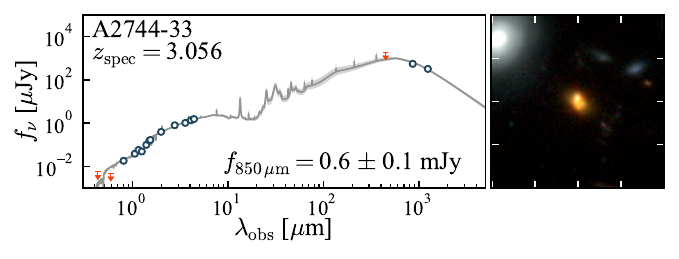}
    \includegraphics[width=0.495\linewidth]{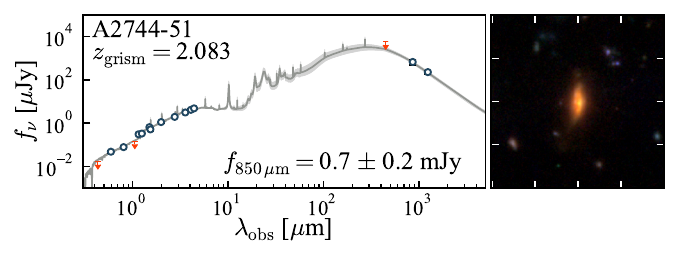}
    \includegraphics[width=0.495\linewidth]{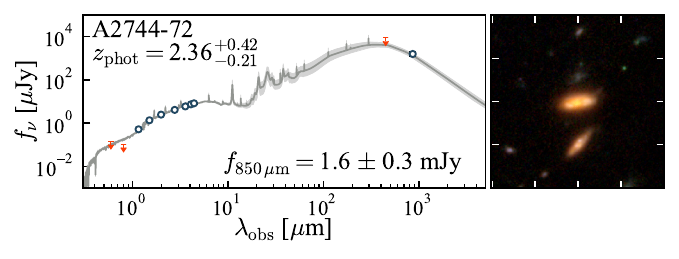}
    \includegraphics[width=0.495\linewidth]{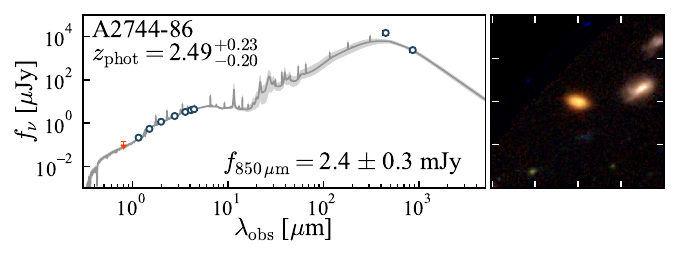}
    \caption{SED fits from {\tt BAGPIPES} and RGB cutouts (R: F444W, G: F277W, B: F150W) for 12 example sources from the red DSFG sample. The cutouts are 8$''$ on a side. We show the observed photometry on the SED panels (dark-blue circles if $>3\sigma$ detection; otherwise, red arrow for 3$\sigma$ upper limit). The adopted redshifts (either spectroscopic, HST or JWST grism, or photometric) and 850~$\mu$m fluxes (delensed for A2744) are marked on the SED panels.
    }
    \label{fig:sedpanels}
\end{figure*}
\vskip 2cm

\section{Physical Properties from SED Fitting}
\label{seds}

\subsection{SED Fitting with {\tt BAGPIPES}}
\label{pipes}

We fit the SEDs of the red DSFG sample with the {\tt BAGPIPES} code, which uses a fully Bayesian framework to constrain the posterior likelihood distributions of the model parameters.
When performing the SED fits, we consider all the available HST, JWST, Spitzer/MIPS, Herschel/PACS and SPIRE, SCUBA-2, and ALMA photometry, as described in Sections~\ref{data} and \ref{submmprops}. Before fitting the SEDs, we correct the observed A2744 photometry for lensing using the lensing magnifications published in \citet{weaver24}. We also include a 5\% error floor to ensure that the fits are not overly constrained by individual data points with very low rms errors.

For our SED models, we assume a delayed exponential SFH with ages between 1 Myr and 14 Gyr (in practice, the code restricts the model ages to be less than the age of the Universe at a given redshift) and the e-folding time, $\tau$, between 100~Myr and 9~Gyr. Variants of this form have often been used to model the SFHs of dusty galaxies \citep[e.g.,][]{wang19, barrufet23, barrufet24, yan24, perez-gonzalez24, sun24} as it can account for both recent bursts and older generations of star formation.

We use a \citet{calzetti00} dust attenuation curve, which is a common choice for high-redshift studies \citep[e.g.,][]{lim20,barrufet23,xiao23a,uematsu24,sun24}. The magnitude of dust attenuation is parameterized by the $V$-band attenuation, $A_V$, which we allow to vary between 0 and 6; furthermore, we assume a factor of 2 higher attenuation in the stellar birth clouds. {\tt BAGPIPES} uses the \citet{bruzual03} stellar population models and the dust emission models of \citet{draine07}. The stellar metallicity is allowed to range between 0.2 and 1.2 times solar metallicity. Finally, we include nebular emission based on the {\tt CLOUDY} photoionization code \citep{ferland17} with the ionization parameter $\log U$ fixed to $-2.0$.

In the main {\tt BAGPIPES} fits, we fix the model redshift to the best available redshift, prioritizing first spectroscopic and/or JWST grism redshifts, then HST grism redshifts, and, finally, photometric redshifts, as discussed in Section~\ref{redshifts}. Given the high accuracy of the {\tt EAZY} redshift estimates (see Figure~\ref{fig:eazy_redshifts}), we simply adopt these redshifts as the best photometric redshift estimate, except in four cases where we estimate the redshift with {\tt BAGPIPES} (Section~\ref{testing_photzs}).

In Figure~\ref{fig:sedpanels}, we show examples of the best-fit SEDs for a representative subset of the red DSFG sample. In each panel, we also show a color image of the galaxy (R: F444W, G: F277W, B: F150W).
We have chosen the examples to span a range of morphologies (major mergers, as well as isolated disks or spheroids), redshifts, and rest-frame optical SED shapes. Some of the sources are detected in up to ten FIR to millimeter bands, while others either are not covered by or are undetected in any FIR bands other than 850~$\mu$m, and, thus, would not otherwise have been identified as DSFGs. 

We measure total SFRs for the red DSFG sample by converting their rest-frame UV and integrated IR luminosities from the best-fit {\tt BAGPIPES} SEDs to SFRs following the prescription of \citet{kennicutt98} (see also \citealt{bell05}), normalized to a \citet{kroupa01} IMF. The total SFR is given by

\begin{multline}
        {\rm SFR} \, [{\rm M}_\odot~{\rm yr}^{-1}] = 1.16 \times 10^{-10} L_{\rm IR} \\+ 3.35\times10^{-10} \, L_{\rm UV},
\end{multline}

\noindent where both $L_{\rm IR}$ and $L_{\rm UV}$ are in units of ${\rm L}_\odot$. $L_{\rm IR}$ is integrated between rest-frame 8 and 1000~$\mu$m, and $L_{\rm UV}$ is computed as the monochromatic luminosity, $\nu L_\nu$, at rest-frame 2800~\AA. We measure the flux at rest-frame 2800~\AA~using a top-hat filter with a width of 350~\AA~\citep[e.g.,][]{xiao23a} convolved with the best-fit {\tt BAGPIPES} spectrum.

We do not correct $L_{\rm UV}$ for dust attenuation, since we use $L_{\rm IR}$ to trace the obscured star formation directly. Note that in our sample of dusty galaxies, the UV contribution to the SFR is typically dwarfed by the IR contribution ($L_{\rm IR}/L_{\rm UV} \sim 100$; see Table~\ref{tab:results}).

\begin{deluxetable*}{cccc}
\tablewidth{\textwidth}
\tablecaption{\label{tab:results}
Median Properties of the Red DSFG Sample}
\tablehead{
& \colhead{$f_{\rm 850 \mu m}<2$~mJy} & \colhead{$f_{\rm 850 \mu m}>2$~mJy} & \colhead{All} }
\startdata
Number&187&47&234\\
$z$&2.42$^{+0.88}_{-0.87}$&2.54$^{+1.06}_{-0.88}$&2.43$^{+1.03}_{-0.86}$\\
$f_{\rm 850\,\mu m}$ [mJy]&0.95$^{+0.54}_{-0.54}$&3.16$^{+1.61}_{-0.86}$&1.15$^{+1.24}_{-0.63}$\\
$\log(M_*/{\rm M}_\odot)$&10.29$^{+0.48}_{-0.39}$&10.51$^{+0.38}_{-0.46}$&10.35$^{+0.45}_{-0.42}$\\
SFR [${\rm M}_\odot\,$yr$^{-1}$] &80$^{+81}_{-45}$&254$^{+135}_{-131}$&96$^{+142}_{-56}$\\
sSFR [Gyr$^{-1}$] &0.53$^{+0.49}_{-0.41}$&0.94$^{+0.56}_{-0.53}$&0.57$^{+0.63}_{-0.39}$\\
$\log(L_{\rm IR}/{\rm L}_\odot) $&11.82$^{+0.30}_{-0.36}$&12.32$^{+0.20}_{-0.30}$&11.89$^{+0.40}_{-0.36}$\\
$\log(L_{\rm UV}/{\rm L}_\odot)$&9.89$^{+0.38}_{-0.58}$&9.93$^{+0.59}_{-0.34}$&9.89$^{+0.41}_{-0.58}$\\
$\log(L_{\rm IR}/L_{\rm UV}) $&1.90$^{+0.55}_{-0.29}$&2.40$^{+0.34}_{-0.57}$&2.00$^{+0.54}_{-0.35}$\\
$A_V$ [mag]&1.55$^{+0.70}_{-0.37}$&2.01$^{+0.59}_{-0.65}$&1.61$^{+0.70}_{-0.42}$\\
$n_{\,{\rm F444W}}$&1.72$^{+1.97}_{-0.81}$&1.76$^{+2.01}_{-0.81}$&1.72$^{+2.02}_{-0.81}$\\
$n_{\,{\rm F150W}}$&1.14$^{+2.14}_{-0.67}$&0.98$^{+1.98}_{-0.78}$&1.13$^{+2.03}_{-0.73}$\\
$R_{e,\,{\rm F444W}}$ [kpc]&1.63$^{+1.60}_{-0.98}$&2.00$^{+0.83}_{-0.92}$&1.78$^{+1.42}_{-1.09}$\\
$R_{e,\,{\rm F150W}}$ [kpc]&3.20$^{+2.81}_{-1.78}$&3.16$^{+3.29}_{-1.39}$&3.16$^{+2.88}_{-1.72}$\\
$(b/a)_{\,{\rm F444W}}$&0.53$^{+0.24}_{-0.25}$&0.65$^{+0.19}_{-0.22}$&0.56$^{+0.23}_{-0.26}$\\
$(b/a)_{\,{\rm F150W}}$&0.51$^{+0.28}_{-0.25}$&0.59$^{+0.26}_{-0.30}$&0.53$^{+0.27}_{-0.27}$\\
Secure Merger \%&$19\%\pm3$\%&$32\%\pm8$\%&$21\%\pm3$\%\\
\enddata
\tablecomments{Median values and the 16th--84th percentile ranges (i.e., not the errors on the median), except for the secure merger fraction, where the errors are the Poisson errors. For sources in the A2744 field, all properties are corrected for lensing magnification. For the S\'ersic index, effective radius, and axis ratio, we exclude sources flagged for poor {\tt GALFIT} fits (see Section~\ref{galfit}).
}
\end{deluxetable*}

\subsection{Testing Photometric Redshifts \label{testing_photzs}}

The {\tt EAZY} photometric redshifts from both \citet{eisenstein23b} and \cite{weaver24} are based only on HST and JWST data, so it is interesting to see whether the addition of the submillimeter/millimeter photometry helps to constrain the photometric redshifts. We therefore perform a second round of {\tt BAGPIPES} fits in which we allow the redshift to vary from $z=0$ to $z=14$ with a uniform prior. The constraints on the rest of the model parameters are treated identically to the SED fits in Section~\ref{pipes}. We take the median of the posterior distribution as the {\tt BAGPIPES} photometric redshift for each source.

\begin{figure}
    \centering
    \includegraphics[width=\linewidth]{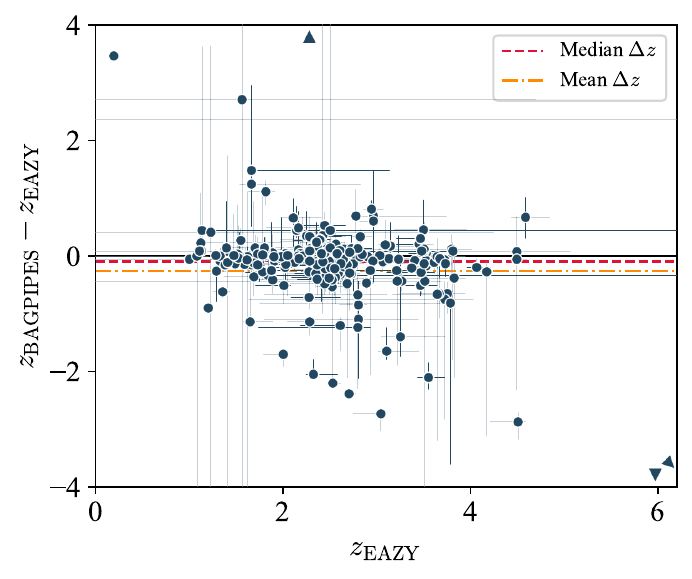}
    \caption{Difference between {\tt BAGPIPES} and {\tt EAZY} photometric redshifts for the red DSFG sample (dark-blue points). The median (red dashed line) and mean (orange dashed--dotted line) are shown for reference against the zero line (black solid line). Error bars refer to the 16th--84th percentile ranges on the individual redshift fits; the y-axis error bars incorporate both the {\tt BAGPIPES} and {\tt EAZY} errors. For three sources lying outside the plot range, we plot representative triangles indicating their relative positions.
    }
    \label{fig:pipes_eazy_comp}
\end{figure}

In Figure~\ref{fig:pipes_eazy_comp}, we show the difference between the {\tt BAGPIPES} and {\tt EAZY} photometric redshifts for the red DSFG sample. We also show the median and mean differences, both of which are close to zero. This implies that both codes are in agreement for most of our sample. There are clearly a number of outliers, for which the addition of the submillimeter data may be strongly influencing the final redshift estimate, but we do not observe an overall trend towards lower photometric redshifts in the {\tt BAGPIPES} fits.

Although these {\tt BAGPIPES} fits generally perform well in recovering the spectroscopic redshifts for sources that have both, the fraction with $|z_{\rm spec} - z_{\rm phot}|/(1 + z_{\rm spec}) < 0.15$ is 80\%, marginally lower than the {\tt EAZY} results (87\%). This may be due in part to the choice of a uniform redshift prior (the {\tt EAZY} fits use a more complex prior distribution). We also run a set of SED fits in which we use a Gaussian distribution centered on the {\tt EAZY} redshift as a prior for {\tt BAGPIPES}, but even in this case, the fraction with $|z_{\rm spec} - z_{\rm phot}|/(1 + z_{\rm spec}) < 0.15$ is just 82\%. 

Therefore, for nearly all galaxies in our sample, we keep the {\tt EAZY} redshift as our best photometric redshift estimate.
There are just four sources for which we adopt the {\tt BAGPIPES} redshifts instead: 

First, for one source (A2744-66) with no published {\tt EAZY} redshift, we use the {\tt BAGPIPES} redshift of $z_{\rm phot} =1.37^{+0.62}_{-0.10}$. 

Second, the {\tt EAZY} redshift for A2744-55 (a $\fsm = 4.8$~mJy galaxy) is $z_{\rm phot} = 0.197^{+0.004}_{-0.004}$, but the {\tt BAGPIPES} best-fit redshift is $z_{\rm phot}=3.66_{-0.08}^{+0.06}$. This is also consistent with the appearance of the source in the NIRCam image.

Third, there is a bright quasar (GS-59) with an X-ray detection ($f_{\rm 0.5-7 \,keV} = 5.7\times10^{-16}$~erg\,s$^{-1}$\,cm$^{-2}$, implying $L_X \gtrsim 10^{43}$~erg\,s$^{-1}$ for $z\gtrsim1.5$). The {\tt EAZY} redshift is $z_{\rm phot}=4.50^{+0.09}_{-0.30}$. This source is also detected by ALMA at 870~$\mu$m, 1.1~mm, 1.2~mm, and 2~mm. When fixed to the {\tt EAZY} redshift, {\tt BAGPIPES} suggests that this source is a $\log(L_{\rm IR}/{\rm L}_\odot) \sim 13$ galaxy with a stellar mass of $\log(M_*/{\rm M}_\odot) \sim 11.8$. By contrast, when the redshift is allowed to vary, {\tt BAGPIPES} converges on a much lower $z_{\rm phot} = 1.63^{+0.13}_{-0.05}$ and does not find a secondary redshift solution at $z>3$. 

\begin{figure}
    \centering
    \includegraphics[width=\linewidth]{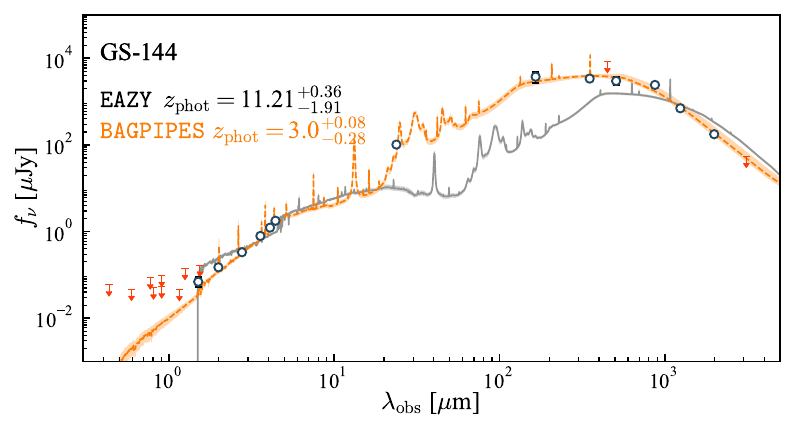}
    \caption{{\tt BAGPIPES} SED fits for GS-144, the source with the highest {\tt EAZY} photometric redshift in the red DSFG sample ($z_{\rm phot} = 11.21$, measured using only the HST+JWST data), but which is also detected by ALMA at 870~$\mu$m, 1.2~mm, and 2~mm. We show the {\tt BAGPIPES} fits with the redshift fixed to the {\tt EAZY} $z_{\rm phot} = 11.21$ (gray solid line) and with the redshift allowed to vary (orange dashed line), resulting in $z_{\rm phot} = 3.01$. 
    }
    \label{fig:sed_comp}
\end{figure}

Finally, the source with the highest {\tt EAZY} redshift in the sample is a GOODS-S galaxy (GS-144) at $z_{\rm phot} =11.21^{+0.36}_{-1.91}$. This source is coincident with a 2.4~mJy ALMA 870~$\mu$m source from \citet{cowie18} that is also detected in ALMA 1.2~mm and 2~mm imaging \citep{mckay23}. The {\tt BAGPIPES} photometric redshift is $z_{\rm phot} = 3.01^{+0.08}_{-0.28}$. This is an example of a DSFG whose strong dust attenuation ($A_V \sim 3.1$) causes it to mimic the photometric colors of an extremely high-redshift galaxy when only the optical to NIR photometry are considered \citep[e.g.,][]{naidu22,zavala23,meyer24}.  In Figure~\ref{fig:sed_comp}, we show the two {\tt BAGPIPES} fits for this galaxy: the first with the redshift fixed to the {\tt EAZY} value, and the second where the redshift is allowed to vary.

We emphasize that adopting the {\tt BAGPIPES} redshifts for these four sources has no significant impact on the median properties of our sample.

\begin{figure*}
    \centering
    \includegraphics[width=\linewidth]{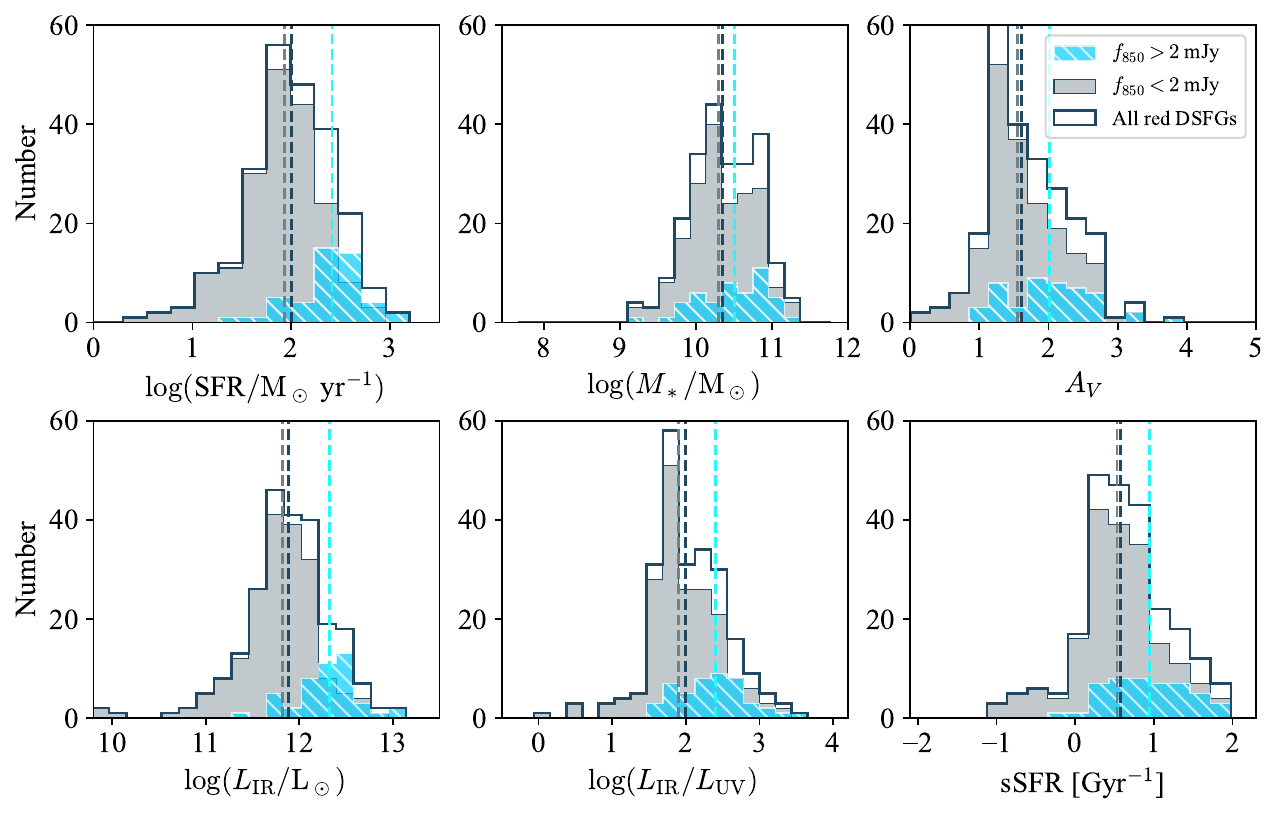}
    \caption{Distributions of the main physical parameters inferred from our SED fits for the red DSFG sample. We show the distributions for the entire sample (open), as well as those for the subsets with $\fsm < 2$~mJy (gray) and $\fsm > 2$~mJy (hatched blue) separately. We show the median values as dashed lines.
    }
    \label{fig:sample_properties}
\end{figure*}

\subsection{SED Results}
\label{sed_results}

In Figure~\ref{fig:sample_properties}, we plot histograms of the SFR, specific SFR
(sSFR~$\equiv {\rm SFR}/M_*$), $L_{\rm IR}$, and $L_{\rm IR}/L_{\rm UV}$
for the full red DSFG sample (open), as well as for the sample separated into $\fsm > 2$~mJy 
 (hatched blue) and $\fsm < 2$~mJy (gray). 
We give the median values (dashed lines) in
Table~\ref{tab:results}, along with the 16th--84th percentile ranges of the stellar masses, 
SFRs, dust attenuations, and IR and UV luminosities. 

In general, we find that the galaxies in the full sample are best fit with a dusty SED model.
For our assumption of a \citet{calzetti00} attenuation curve, we find
$A_V = 1$--3 for almost all sources (median $A_V = 1.61$).
The galaxies tend to be massive (median $ \log (M_*/{\rm M}_\odot) = 10.35$),
and they lie between $1\lesssim z \lesssim 5$. They have median $\log(L_{\rm IR} /{\rm L}_\odot) = 11.9$ and 
median SFR~$= 96\, {\rm M}_\odot\,{\rm yr}^{-1}$. 
The sample spans over 2~dex in SFR and $M_*$, encompassing both moderately star-forming galaxies with SFRs as low as 5--50~${\rm M}_\odot\,{\rm yr}^{-1}$ and extreme star-forming galaxies with SFRs up to $\sim$1000~${\rm M}_\odot\,{\rm yr}^{-1}$.

As expected, the sources with $\fsm < 2$~mJy (80\% of the sample) have substantially ($\sim0.5$~dex) 
lower SFRs and $L_{\rm IR}$ than those with $\fsm > 2$~mJy.  
However, the stellar masses are not significantly different. A Mann--Whitney test indicates that the stellar
mass distributions are only $\sim2.1\sigma$ deviant from the hypothesis that they are drawn from
the same parent distribution.

In terms of their SFRs alone, the faintest DSFGs overlap with the bright end of the extinction-corrected,
UV-selected Lyman break galaxy (LBG) population \citep{cowie17}.
However, we note that all but 14 (i.e., 94\%) of the red DSFG 
sample have $L_{\rm IR}/L_{\rm UV} > 20$ (see Figure~\ref{fig:sample_properties}); i.e., their SFRs 
are completely dominated by dust-obscured star formation (traced by $L_{\rm IR}$) rather than by unobscured star formation (traced by $L_{\rm UV}$).
We perform a quick test to see if the dust-corrected SFRs recover the total SFRs based on both the IR and UV luminosities (i.e., the SFRs measured in Section~\ref{pipes}): Using the measured $A_V$, we correct the observed UV luminosities for dust and convert these to SFRs using the relation in \citet{kennicutt98}. We find that the SFR$_{\rm UV,\, corr}$ underpredicts the SFR$_{\rm IR+UV}$ by a factor of 2.0, on average; furthermore, we do not see a significant difference between the bright and faint DSFGs. This discrepancy could be explained if these galaxies contain extremely dust-obscured regions from which no UV light escapes and hence are not accounted for in the dust correction (in some DSFGs the FIR and UV emission appear spatially offset; e.g., \citealt{barro16}, \citealt{hodge16}, \citealt{elbaz17}). However, a full analysis that takes into account systematic uncertainties to quantify the effect on estimates of the cosmic SFRD from LBG samples is outside the scope of this paper. 

\begin{figure*}
    \centering
    \includegraphics[width=\linewidth]{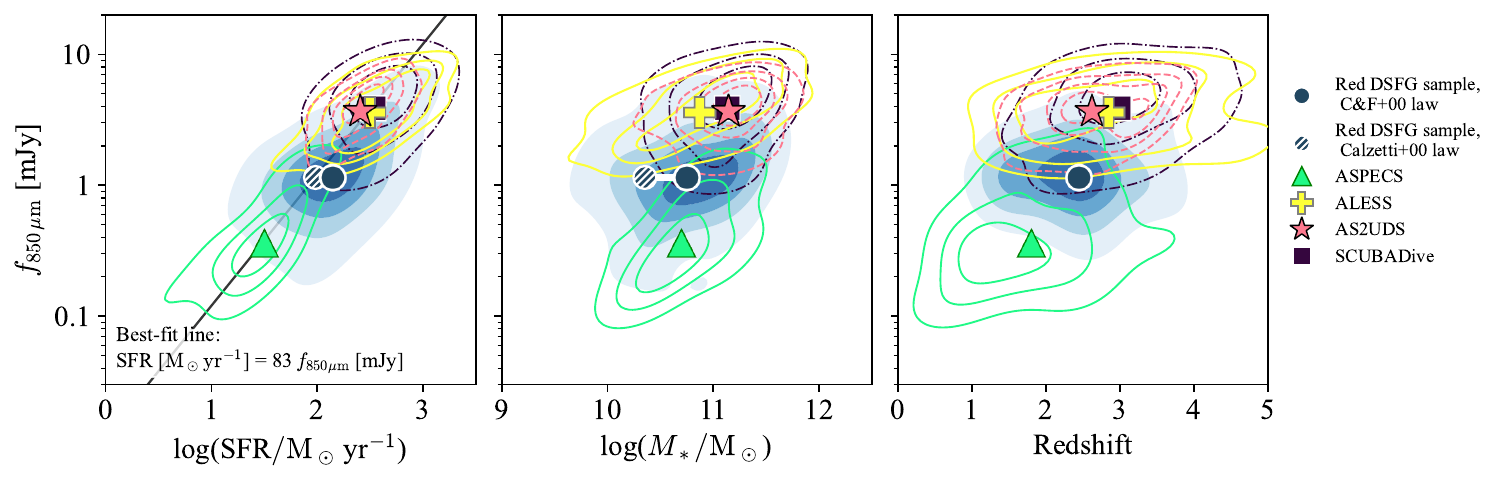}
    \caption{Comparison of derived properties assuming a \citet{charlot00} attenuation curve for the red DSFG sample (blue circle and filled contours) with those of several literature samples: the ASPECS main sample \citep[green triangle and solid contours;][]{aravena20}, ALESS SMGs \citep[yellow cross and solid contours;][]{dacunha15}, AS2UDS SMGs \citep[pink star and dashed contours;][]{dudzeviciute20}, and SCUBADive \citep[purple square and dashed--dotted contours;][]{mckinney24}. For each sample, we plot the distribution of $\fsm$ vs. SFR (left), stellar mass (center), and redshift (right) as Gaussian kernel density estimations, and we mark the median values (large symbols). The errors on the medians are generally smaller than the symbols, so we do not plot them. For sources with no direct $\fsm$ measurement, we convert $f_{\rm 1.2~mm}$ to $\fsm$ assuming a modified blackbody SED (see text). In the left panel, we show the best-fit SFR--$\fsm$ relation for $\fsm$ between 0.1 and 10~mJy. In the left and center panels, we also show the shift in median properties for our sample under our original assumption of a \citet{calzetti00} attenuation curve (hatched blue and white circle). }
    \label{fig:comp_properties}
\end{figure*}

We next compare the properties of our sample with those of several submillimeter/millimeter-selected DSFG samples for which SFRs and stellar masses have been published (we convert these to a \citealt{kroupa01} IMF). In Figure~\ref{fig:comp_properties}, we plot $\fsm$ versus SFR (left), stellar mass (center), and redshift (right). The comparison samples are the 99 sources in the main ALESS sample \citep{dacunha15}, the 707 sources in the AS2UDS sample \citep{dudzeviciute20}, the 289 sources in the SCUBADive sample \citep[][]{mckinney24}, and the 35 sources in the main ASPECS sample \citep{aravena20}. 

These samples span a wide range in submillimeter flux: ALESS, AS2UDS, and SCUBADive are based primarily on ALMA follow-up of single-dish submillimeter surveys with flux limits of 2--4~mJy. Note that due to various factors, including multiple ALMA sources contributing to the SCUBA-2 flux in some cases, roughly 15\% of the ALMA sources from these surveys end up having $\fsm<2$~mJy, so there is some overlap with our sample in terms of $\fsm$ alone. However, these samples all have median $\fsm\sim3.7$~mJy, compared to the median $\fsm\sim1.2$~mJy for our sample.
Compared to the other surveys, ASPECS probes a much deeper flux regime ($\sigma_{\rm 1.2~mm} = 9.3\,\mu$Jy/beam$^{-1}$) over a much smaller area ($\sim5$~arcmin$^{-2}$).
Since ASPECS does not have ALMA data at 870~$\mu$m, we convert their $f_{\rm 1.2~mm}$ to $\fsm$ assuming a modified blackbody SED with $T=35$~K and $\beta=1.8$, a frequent choice in the literature. The resulting median $\fsm$ for ASPECS is 0.36~mJy. Different assumptions on $T$ and $\beta$ within a range of reasonable values ($T = 20$--55~K and $\beta = 1.5$--2.2) produce less than a factor of 2 difference in the individual $\fsm$ values.

All of the SED-derived properties for the literature samples were determined using a variant of the two-component \citet{charlot00} attenuation curve, implemented in either the {\tt MAGPHYS} \citep{dacunha08} or {\tt CIGALE} \citep[][]{boquien19} SED-fitting codes.\footnote{A 0.2~dex offset was applied to the SCUBADive masses for consistency with the mass-to-light ratio in \citet{dacunha15}; see \citet{mckinney24}} 
However, it has been shown that the choice of the \citet{charlot00} attenuation curve can increase stellar mass measurements for SMGs by $\sim0.4$~dex relative to the \citet{calzetti00} curve (e.g., \citealt{uematsu24}; see also \citealt{lofaro17}). 

For consistency, since {\tt BAGPIPES} supports the use of dust attenuation curves, we reran our SED fits using the \citet{charlot00} attenuation curve with a power-law slope of $\delta = -0.7$, in agreement with \citet{mckinney24}. In Figure~\ref{fig:comp_properties}, we show the properties derived from these fits, though we also plot the median values derived with the \citet{calzetti00} curve for comparison.
The main effect of the \citet{charlot00} model is to increase the measured stellar mass and $A_V$, since the shallower attenuation curve requires higher attenuation values to match the observed photometry. However, since {\tt BAGPIPES} assumes energy balance between the UV/optical absorption and the IR emission, this also produces a $\sim$0.15~dex increase in the $L_{\rm IR}$, and therefore the SFR, on average. This highlights the need for caution when comparing properties derived from SED fits with different modeling assumptions \citep[e.g.,][]{hainline11, michalowski14, lofaro17, pacifici23, uematsu24}.

We see that the overall median SFR of the red DSFG sample is a factor of 2--3 lower than that of ALESS, AS2UDS, and SCUBADive (which all have median SFRs~$\sim$~300~${\rm M}_\odot$\,yr$^{-1}$) but above that of ASPECS (median SFR~$\sim 30~{\rm M}_\odot$\,yr$^{-1}$). This reflects the fact that the median $\fsm$ of our sample sits in between ASPECS---which is deep enough to probe the dust emission of ``typical'' star-forming galaxies---and classic SMG samples such as ALESS/AS2UDS/SCUBADive, though, as noted above, there is substantial overlap between the $\fsm$ distributions.

As has been seen previously for the bright DSFGs (e.g., \citealt{barger14,cowie17}),
there is a close linear relation between the SFR and $\fsm$. 
The left panel shows that this relation extends to the faint
DSFGs. By performing a linear fit to all of the DSFG samples, weighted by the individual errors on the data points, we find 

\begin{equation}
    {\rm SFR}~[{\rm M}_\odot~{\rm yr}^{-1}] = (83\pm 11) \times \fsm~{\rm [mJy]}
\end{equation} 

\noindent for the range 0.1--10~mJy (error is derived from bootstrapping the best-fit relation). The results remain unchanged if we restrict to $\fsm>2$~mJy.
Our relation is slightly lower than the results of \citet{barger14}, who found SFR~$=134\times \fsm$ (when converted to a \citealt{kroupa01} IMF), and of \citet{cowie17}, who found SFR~$=143 \times \fsm$. However, there are a variety of systematics involved that may be responsible for driving this difference; e.g., comparing SFRs derived from SED codes with those measured directly from IR luminosities. If we consider only the red DSFG sample, then the best-fit relation is ${\rm SFR} = (132\pm 18) \times \fsm$, in agreement with the results of \citet{barger14} and \citet{cowie17} (all three of these studies used direct IR luminosity conversions to measure their SFRs). Note that \citet{dudzeviciute20} investigated this relationship for AS2UDS alone and suggested that $\fsm$ was more strongly correlated with cold dust mass, which we do not explore here. However, from Figure~\ref{fig:comp_properties}, which includes the entire AS2UDS sample (pink contours), we can see that the linear $\fsm$--SFR relationship provides a good fit across a wide range of submillimeter fluxes.

We find a median stellar mass of $\log(M_*/{\rm M}_\odot) = 10.74$ for the red DSFG sample. This is very close to the ASPECS result of $\log(M_*/{\rm M}_\odot) \sim 10.65$. For the $\fsm > 2$~mJy sources, we find a median $\log(M_*/{\rm M}_\odot) = 10.94$, which is similar to those measured for ALESS ($\log(M_*/{\rm M}_\odot) \sim 10.9$), SCUBADive ($\log(M_*/{\rm M}_\odot) \sim 11.1$), and AS2UDS ($\log(M_*/{\rm M}_\odot) \sim 11.1$). The bootstrapped errors on the medians for these surveys range from 0.02--0.1~dex, and the statistical uncertainty associated with different models for the parametric SFH is $\sim0.1$--0.2~dex \citep[e.g.,][]{hainline11,michalowski14, pacifici23}. 
Under our original choice of the \citet{calzetti00} curve, we measure a median stellar mass that is $\sim$0.4~dex lower, but which is in good agreement with other color-based or NIR-dropout selections that employ similar modeling assumptions \citep[e.g.,][]{wang19,xiao23a,perez-gonzalez23,gottumukkala23}.

We note that the JWST data produce much lower statistical uncertainties on the stellar masses compared to previous studies that relied on HST and Spitzer/IRAC data. We find typical 1$\sigma$ uncertainties of $\sim$0.07~dex from the {\tt BAGPIPES} fits, 
which are insignificant compared to the systematic modeling uncertainties discussed above. These can be compared to the ALESS 0.4--0.5~dex statistical uncertainties on the stellar mass \citep[][]{dacunha15}, which are based on UV to NIR photometry from ground-based and Spitzer/IRAC observations. 

Finally, although the red DSFG sample alone does not exhibit significant variation in redshift with submillimeter flux (Section~\ref{redshifts}), in the right panel of Figure~\ref{fig:comp_properties} we see that the median redshifts of the samples do increase as we go from the faintest to the brightest sources, in agreement with the trends noted by other studies \citep[e.g.,][]{ivison02, pope05, smolcic12,cowie17,stach19,birkin21,chen22b}. 
However, the trend is not steep, and there is substantial scatter over two orders of magnitude in $\fsm$. A more detailed study of this relationship is beyond the scope of this paper, so we leave it for future work.

\section{Morphologies and Mergers}
\label{morph}

In this section, we use the high resolution of the NIRCam data to characterize the average stellar morphologies of the red DSFG sample. We also investigate whether these sources are undergoing major mergers, which have long been proposed as a primary mechanism for triggering the extreme SFRs of DSFGs \citep[e.g.,][]{sanders96,mihos96,hopkins08,narayanan10}, though some recent simulations and observational studies have suggested that their merger rates may not differ substantially from those of field galaxies \citep[e.g.,][]{mcalpine19, gillman24}. We focus on the morphologies in the F150W and F444W bands. For the majority of our sample, these roughly correspond to the rest-frame optical and NIR, which, respectively, trace the young and old stellar populations.

\subsection{Surface Brightness Profile Fitting}
\label{galfit}

To measure the morphologies of the red DSFG sample, we use the {\tt GALFIT} tool \citep{peng02,peng10} to fit a two-dimensional (2D) parametric model to the images of each galaxy. {\tt GALFIT} uses a nonlinear least-squares fitting process (built on the Levenberg--Marquardt algorithm) to determine the combination of parameters that minimizes the fit residuals, given a predetermined set of parametric models. It incorporates the instrumental PSF by convolving it with the provided model before fitting the data.

For our fits, we choose a single S\'ersic profile, which is commonly used to model the light profiles of star-forming and passive galaxies in both the local Universe and at high redshifts \citep[e.g.,][]{wuyts11, vanderwel12, kartaltepe23}. Although more complex models (e.g., multiple S\'ersic profiles, or a S\'ersic + point-source profile) are sometimes necessary to capture the detailed morphologies of star-forming galaxies, they can also overfit the data in some cases \citep[e.g.,][]{gillman23}.

We perform the fits on 8$''$$\,\times\,$8$''$ cutouts of the F444W and F150W images centered on the NIRCam position of each galaxy. We perform simultaneous fits of any sources brighter than 26~mag and within 3$''$ of the target galaxy \citep[e.g.,][]{nelson23}, but we use the segmentation maps provided by the respective JWST surveys to mask out other sources. The choice of F150W and F444W corresponds to the rest-frame optical and rest-frame NIR, respectively, for the majority of the red DSFG sample. Since the F444W data have higher signal-to-noise ratio (S/N), we constrain the F150W S\'{e}rsic fits to have the same position angle as the best-fit F444W model. In both bands, we let {\tt GALFIT} compute the noise image (including the sky background and shot noise) from the provided cutouts. 

We produce an empirical PSF by stacking 8$''$$\,\times\,$8$''$ cutouts of bright, unsaturated stars selected from the NIRCam mosaics in each field. This method accounts for the observed broadening of the NIRCam PSF with respect to models \citep[e.g.,][]{nardiello22, zhuang24, sun24a, weaver24}. We identify stars by selecting sources with {\tt flag\_star = 1} (or the equivalent) and $\ff>1$~$\mu$Jy in the respective JWST catalogs. We visually inspect them to ensure that the centers of the stars are not saturated in the relevant NIRCam filters and that the stars do not lie in a crowded field. We then stack the cutouts of the stars to generate the final PSF, which we pass to {\tt GALFIT} to be convolved with the S\'{e}rsic model during the fitting.

To ensure that our choice of cutout size is not biasing our results, we repeated the fits with 6$''$ and 12$''$ image and PSF cutouts and confirmed that the fitted parameters were unchanged (at the $\sim$1\% level).

\begin{figure*}[!t]
    \centering
    \includegraphics[width=0.15\linewidth]{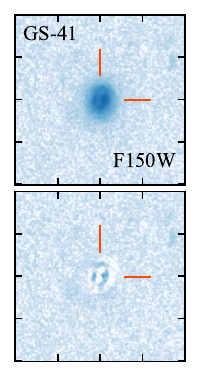}
    \includegraphics[width=0.15\linewidth]{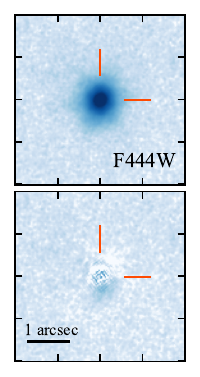}\hspace{4ex}
    \includegraphics[width=0.15\linewidth]{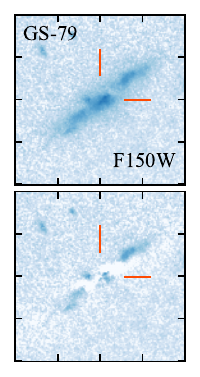}
    \includegraphics[width=0.15\linewidth]{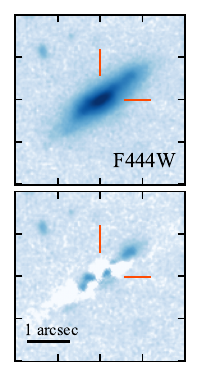}\hspace{4ex}
    \includegraphics[width=0.15\linewidth]{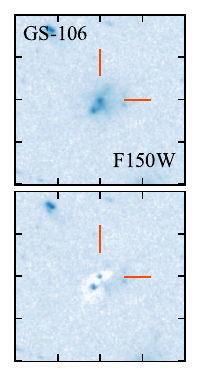}
    \includegraphics[width=0.15\linewidth]{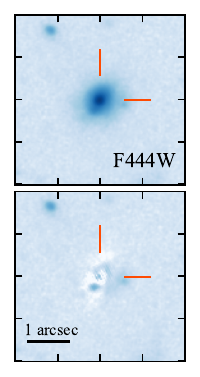}\\
    
    \includegraphics[width=0.15\linewidth]{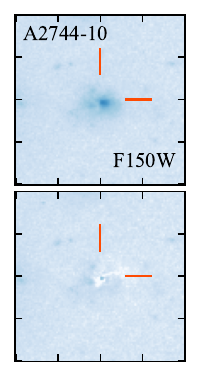}
    \includegraphics[width=0.15\linewidth]{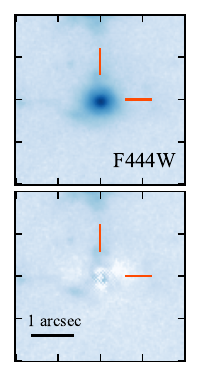}\hspace{4ex}
    \includegraphics[width=0.15\linewidth]{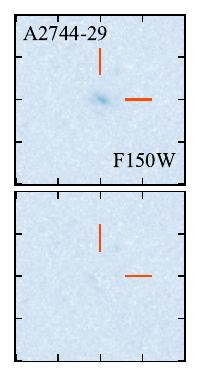}
    \includegraphics[width=0.15\linewidth]{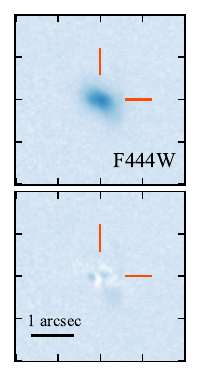}\hspace{4ex}
    \includegraphics[width=0.15\linewidth]{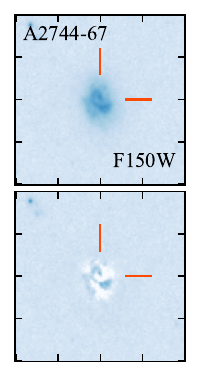}
    \includegraphics[width=0.15\linewidth]{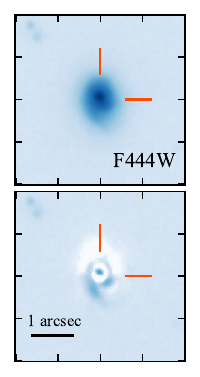}\\
    \includegraphics[width=0.8\linewidth]{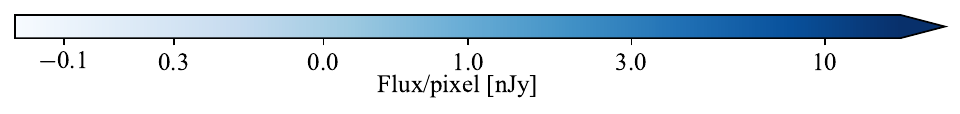}
    
    \caption{Surface brightness profile fitting results from {\tt GALFIT} for six example sources from the red DSFG sample. For each source, we show the results for the F150W (left column) and F444W (right column) bands. For each band, we show the original image (top) and the residual image after subtracting the best-fit S\'ersic profile (bottom). The images are 4$''$ on a side and centered on the NIRCam positions (orange crosshairs). All images are shown on the same flux scale, as indicated by the colorbar}.
    \label{fig:galfitresults}
\end{figure*}

\begin{figure*}
    \centering
    \includegraphics[width=\linewidth]{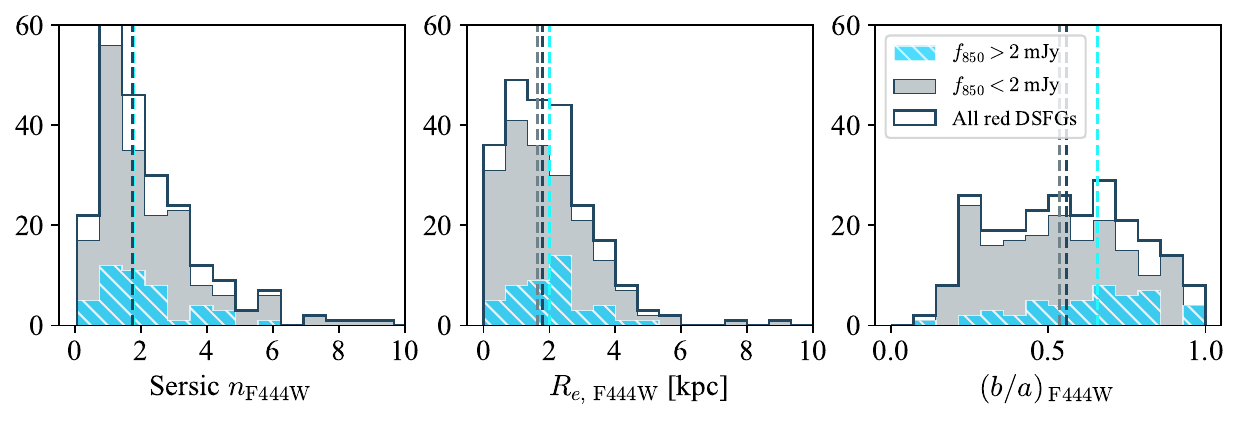}
    \caption{Distributions of the morphological parameters inferred from our surface brightness profile fits for the red DSFG sample. The legend and histogram colors are the same as Figure~\ref{fig:sample_properties}. We exclude sources flagged for poor {\tt GALFIT} fits (see Section~\ref{galfit}).
    }
    \label{fig:morph_distributions}
\end{figure*}

In Figure~\ref{fig:galfitresults}, we present six examples of our {\tt GALFIT} F444W and F150W fits. For galaxies in A2744, we correct all measured effective radii (i.e., half-light radii) for the lensing magnifications by scaling them by $1/\sqrt{\mu}$, where $\mu$ is the magnification factor. In Table~\ref{tab:results}, we give the median values and the 16th--84th percentile ranges of the S\'ersic indices, effective radii, and axis ratios that we measure in F444W and F150W for the red DSFG sample. We show the distributions of the measured morphological parameters in Figure~\ref{fig:morph_distributions}.

In many cases, the {\tt GALFIT} residuals show substructure, such as spiral arms, clumps, or tidal features, that become more apparent once the S\'{e}rsic model has been subtracted from the image. This has also been noted by other recent studies of DSFGs \citep{lebail24, polletta24, hodge24, price23}. Indeed, \citet{jain24} recently reported that one of the objects in our sample (A2744-53) is a grand design spiral at $z\sim4$.
The red DSFG sample also tends to appear clumpier at 1.5~$\mu$m than at 4.4~$\mu$m, suggestive of structured dust obscuration that may drastically alter their morphologies between the rest-frame optical and NIR \citep[e.g.,][]{polletta24, gillman24,hodge24}. 

We flag sources with measured $R_e > 50$~kpc, which is typically produced by low-S/N detections in F150W or by nearby bright sources/mergers that influence the fit. 
The fraction of sources in the red DSFG sample with good fits that are not flagged in F444W is 98\%, while in F150W it is 92\% due to the fact that some sources have no F150W detection and hence cannot be modeled. We exclude these flagged objects from the median results quoted in Table~\ref{tab:results} and from the histograms in Figure~\ref{fig:morph_distributions}.

\subsection{Identifying Mergers}

Many previous studies indicated that DSFGs tend to exhibit disturbed morphologies and/or be engaged in some stage of a major merger.
For example, in their sample of 48 ALMA-detected DSFGs, \citet{chen15} found a disturbed/merger incidence of $\sim$80\% and offsets between the rest-frame optical and FIR, from which they inferred that most DSFGs with $\fsm > 2$~mJy are early/mid-stage mergers. Similarly, \citet{franco18} determined that $\sim60$\% of their sample of 18 DSFGs from the GOODS-ALMA survey had irregular morphologies or merger signatures. 

However, these studies relied on HST data; thus, for $z>2$, they were unable to probe the stellar light beyond rest-frame wavelengths of $\sim0.5$~$\mu$m. 
With the substantial increase in sensitivity and wavelength coverage afforded by JWST, we can investigate the merger fraction of the red DSFG sample at longer rest-frame wavelengths with less susceptibility to the effects of dust. 

\begin{figure}[!t]
    \centering
    \includegraphics[width=0.32\linewidth]{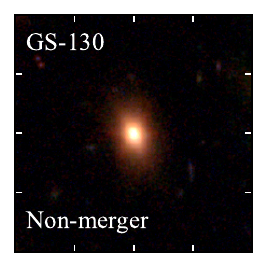}
    \includegraphics[width=0.32\linewidth]{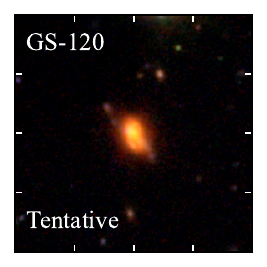}
    \includegraphics[width=0.32\linewidth]{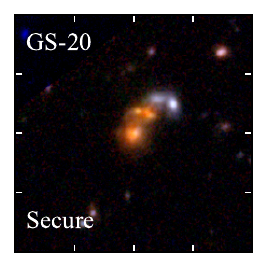} \\
    \includegraphics[width=0.32\linewidth]{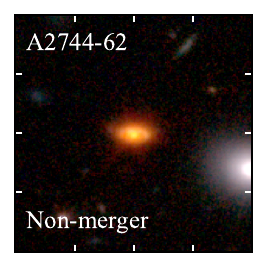}
    \includegraphics[width=0.32\linewidth]{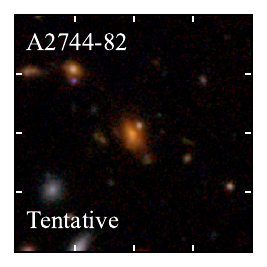}
    \includegraphics[width=0.32\linewidth]{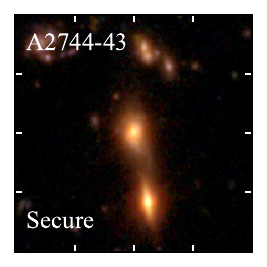}
    \caption{RGB cutouts (R: F444W, G: F277W, B: F150W) and merger classifications for six example sources from the red DSFG sample. The cutouts are 8$''$ on a side.
    }
    \label{fig:merger_ex}
\end{figure}

To identify mergers, we perform visual inspections of each galaxy using 8$''$$\,\times\,$8$''$ RGB image cutouts along with the residual images from our {\tt GALFIT} fits.
We denote as ``secure'' (or major) mergers galaxies that exhibit clear signs of interactions, such as tidal tails, or that have close companions of similar colors. We denote as ``tentative'' mergers those
that are, for example, located in a crowded field, or that have smaller companions (see, e.g., \citealt{gillman24}). 
In Figure~\ref{fig:merger_ex}, we show examples of each type of merger classification. 
These visual classifications are inherently subjective; thus, while we should be able to identify major mergers with high reliability, there may be some minor mergers/interactions that we miss.

Among our sample, there are two separate pairs of red DSFGs where the galaxies appear to be merging with one another (we mark all four as secure mergers). There are a further eight red DSFGs that appear to be interacting with other galaxies in the red NIRCam galaxy sample that were not selected as red DSFGs
(i.e., the second galaxy does not have a submillimeter detection).

In total, we classify 50 ($21\%\pm3$\%) of the red DSFGs as secure mergers, and an additional 33 ($14\%\pm2$\%) as tentative mergers. These results are in line with some recent results from JWST-based studies. For example, \citet{cheng23} reported that for their sample of 19 DSFGs with $\fsm \sim 1$--2~mJy, most galaxies exhibited nondisturbed, disk-dominated morphologies in the NIRCam F444W or F360M bands. Other NIRCam studies have reinforced this picture, with major merger fractions of $\sim$20\% and undisturbed disk morphologies comprising nearly half of DSFG samples---consistent with the percentages in non-DSFG field samples \citep[e.g.,][]{gillman24}. 
The implication seems to be that previous HST-based studies were more affected by dust attenuation, which can produce clumpy/disturbed-looking morphologies in sources that resemble smooth disks at longer wavelengths \citep[e.g.,][]{boogaard24}.

We note that \citet{hodge24} found that even in NIRCam data, $\sim 54\% \pm23$\% of their sample of 12 DSFGs from ALESS displayed evidence for mergers/interactions, and only $\sim23\%\pm8$\% were classified as undisturbed disks. We argue that there are two reasons for this: 
First, the methods used to determine merger fractions vary from paper to paper: e.g., many authors combine major mergers and disturbed/irregular morphologies when reporting percentages, so comparing results is not always straightforward. If we combine our ``secure'' merger identifications with those we identified as ``tentative'' (which may serve as a proxy for minor mergers or interactions), then we measure a total merger fraction of $\sim35\%\pm4$\% for the whole red DSFG sample. 

Second, the \citet{hodge24} galaxies are, on average, much brighter DSFGs than ours (median $\fsm = 6.4$~mJy). Within our sample, we see hints that brighter $\fsm$ correlates with higher merger fractions. 
For example, if we restrict to the 19 galaxies in the red DSFG sample with $\fsm > 3.5$~mJy (even these only have a median $\fsm = 4.7$~mJy), then we measure a secure merger fraction of $37\%\pm 14$\%, consistent with \citet{hodge24} within uncertainties. 
If we now combine the secure and tentative merger identifications in these 19 galaxies, then we measure a total merger fraction of $58\%\pm17\%$, in agreement with the \citet{hodge24} result.

\subsection{Bulge Formation and Size Evolution}
\label{optnirmorph}

\begin{figure*}
    \centering
    \includegraphics[width=0.9\linewidth]{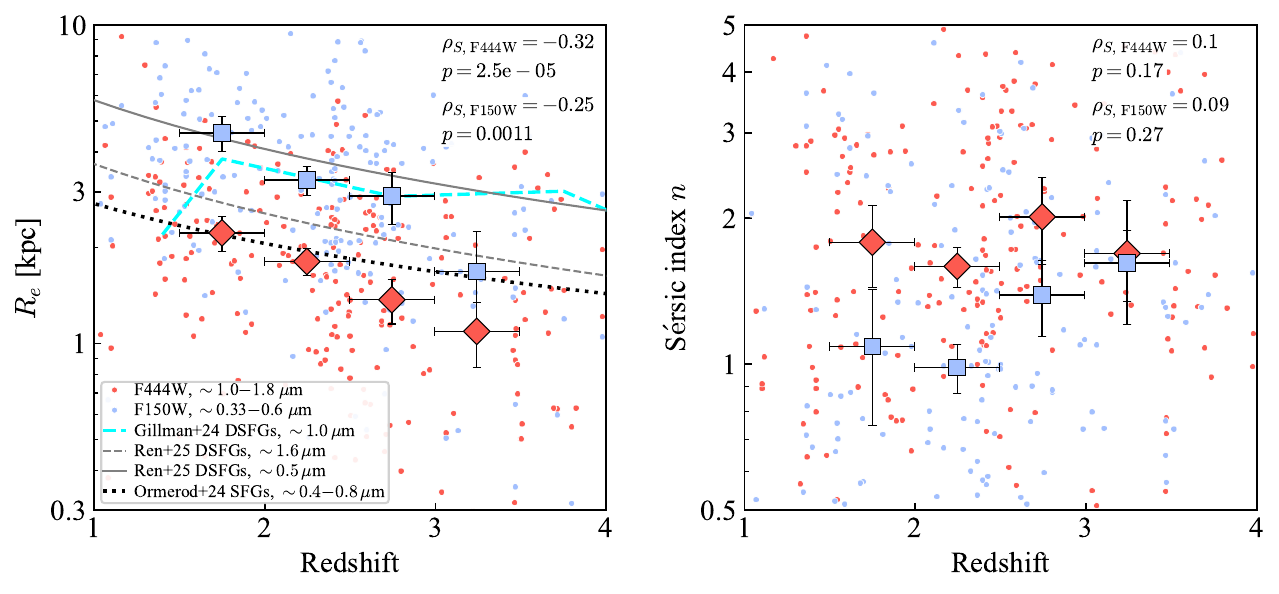}
    \caption{Effective radii (left) and S\'ersic indices (right) vs. redshift for the red DSFG sample. We show measurements made in both the rest-frame optical (F150W; blue points) and rest-frame NIR (F444W; red points); i.e., each galaxy corresponds to two data points in each panel. The medians in four redshift bins are shown for F150W (blue squares) and F444W (red diamonds). For these, the horizontal error bars represent the bin width, and the vertical error bars are the bootstrapped error on the median; we list the corresponding rest-frame wavelengths in the figure legend. In the upper-right corner, we list the Spearman rank correlation coefficients and $p$-values for F150W and F444W, respectively. In the left panel, we show the $R_e$--$z$ relationships found for other samples of DSFGs \citep[][]{gillman24,ren25} and star-forming galaxies \citep[][]{ormerod24} (see figure legend for rest-frame wavelengths). }
    \label{fig:r_e_sersic}
\end{figure*}

As can be seen from Figure~\ref{fig:morph_distributions}, the S\'ersic indices of the red DSFG sample fall mostly between $n_{\rm F444W}=0.5$--4, with median $n_{\rm F444W} = 1.7$, suggestive of disk-dominated morphologies. The distribution of axis ratios that we measure (median $(b/a)_{\rm F444W} = 0.56^{+0.23}_{-0.26}$ with a wide range from $b/a \approx 0.3$--0.9) are also generally consistent with a population of randomly oriented disks \citep[e.g.,][]{padilla08,vanderwel14, tan24}. 
The bright DSFGs are biased towards marginally higher $b/a$ (see Figure~\ref{fig:morph_distributions}), possibly suggesting that their shapes or inclination angles differ slightly from the faint DSFGs, though our current sample is not sufficient to confirm this. 
Overall, we see no significant morphological differences between the $\fsm>2$~mJy and $\fsm<2$~mJy DSFGs.

However, the median effective radius in F150W of $R_{e,\,{\rm F150W}} = 3.16^{+2.88}_{-1.72}$~kpc is substantially larger than the median effective radius in F444W of $R_{e,\,{\rm F444W}}= 1.78^{+1.42}_{-1.09}$~kpc \citep[also noted by, e.g.,][]{chen22, gillman23,  gillman24, boogaard24, hodge24, price23,ren25}. 
The S\'ersic indices in F150W are also somewhat lower (median $n_{\rm F150W} = 1.13^{+2.03}_{-0.73}$) than those in F444W (median $n_{\rm F444W}=1.72^{+2.02}_{-0.81}$). 

The combination of smaller sizes and more compact light profiles at longer wavelengths suggests that the older stellar populations (traced by F444W) are more centrally concentrated than the younger populations (traced by F150W). In other words, we may be seeing an ongoing process of bulge formation in these galaxies, consistent with a picture of inside-out galaxy evolution \citep[e.g.,][]{kamieneski23, gillman23}. These trends may also reflect the impact of dust on the F150W morphologies---e.g., from an obscured central starburst---causing them to be less centrally concentrated and thereby lowering the S\'ersic index and increasing the (measured) effective radius (see, e.g., \citealt{roper22} and \citealt{nedkova24}). 

To look for evolution in the average DSFG morphology, in Figure~\ref{fig:r_e_sersic}, we plot the measured $R_e$ (left) and $n$ (right) in the F150W and F444W bands versus redshift. The large symbols show the binned medians of $R_e$ and $n$ in F150W (blue squares) and F444W (red diamonds) over four redshift bins. We only show the medians for $1.5 \leq z \leq 3.5$, where F150W probes the rest-frame optical (into the UV) and F444W probes the rest-frame NIR. 

In the left panel, we see a clear evolution in the medians from $z=3.5$ to $z=1.5$. We measure the Spearman rank correlation coefficient, $\rho_S$ (a nonparametric indicator of the strength of the association between two variables), for both $R_{e, \, {\rm F150W}}$ and $R_{e, \, {\rm F444W}}$ and redshift, restricting to the redshift range above. The result is $\rho_{S,\, {\rm F150W}} = -0.25$ ($3.9\sigma$) and $\rho_{S,\, {\rm F444W}} = -0.32$ ($4.2\sigma$), indicating a significant negative correlation in each band. This reflects the growth of galaxy size and stellar mass to cosmic noon \citep[e.g.,][]{vanderwel14}, where dusty star formation is known to peak, and indicates that both the young and old stellar populations of $z\gtrsim3$ DSFGs are more compact than those at $z\sim2$. 

For a comparison with other DSFG samples with JWST coverage, we show the relationships found by \citet{ren25} for their sample of $\sim$200 DSFGs at rest-frame 0.6~$\mu$m and 1.6~$\mu$m and the running median from \citet{gillman24} for 80 DSFGs at rest-frame 1.0~$\mu$m (note that there is an overlap of 46 objects between the \citet{ren25} and \citealt{gillman24} samples). For a representative sample of typical star-forming galaxies, we also show the rest-frame 0.4--0.8~$\mu$m best-fit relation from \citet{ormerod24}.

The $R_e$--$z$ trend for our red DSFG sample is generally consistent with the literature within errors, though our sizes tend to be slightly smaller in both the rest-frame NIR and optical. 
These differences may be explained by the slightly different rest-frame wavelengths probed by our choice of filters compared to the literature studies, or to stellar mass distributions due to the different sample selections.
As with the samples of \citet{gillman24} and \citet{ren25}, our sample of DSFGs appears larger in the rest-frame optical than ``normal'' star-forming galaxies at the same redshifts, likely because they represent the most-massive end of the population.

In the right panel of Figure~\ref{fig:r_e_sersic}, we see that there is no statistically significant correlation between $n_{\, {\rm F150W}}$ or $n_{\, {\rm F444W}}$ and redshift for our sample (1.1$\sigma$ and 1.4$\sigma$, respectively), indicating that we do not see a discernible evolution in the shape of DSFGs' stellar light profiles from $z\sim3.5$ to $z\sim1.5$.

\section{Summary}

Using JWST and SCUBA-2,
we identified a large sample of 234 DSFGs covering a wide range of submillimeter fluxes. This allowed us to compare
the properties of faint DSFGs ($\fsm < 2$~mJy) with those of bright DSFGs ($\fsm > 2$~mJy).
We measured their stellar properties using {\tt BAGPIPES} SED fits and their parametric morphologies using {\tt GALFIT}.
Our main results are as follows:

\begin{itemize}

\item Overall, the sample is made up of massive ($\log(M_*/{\rm M}_\odot) \sim 10.3$), dusty ($A_V \sim 1.6$), highly star-forming (SFR~$\sim$~100~${\rm M}_\odot$~yr$^{-1}$) galaxies between $1<z<5$. We found that the SFRs are a linear function of submillimeter flux, extending the results of previous studies to fainter fluxes.
 
\item The JWST data give small statistical errors on the stellar masses of the DSFGs. However, we found that the assumed dust attenuation 
curve has a significant impact on the measured stellar masses; for example, our use of \citet{calzetti00} results in
$\sim$0.4~dex lower stellar masses than \citet{charlot00}. When this is taken into account, our sample has similar stellar masses
to other samples. Remarkably, nearly all of the DSFGs have similar masses, regardless of their SFR or submillimeter flux.

\item 
We identified a tight negative correlation between redshift and both $\ff$ and $\fo$, suggesting that the observed NIR flux may be an effective way to select high-redshift DSFGs from submillimeter samples.

\item Most of the galaxies in our sample appear to be isolated disks, with a moderately low fraction ($\sim21$\%) clearly involved in major mergers. We argued that the high merger fractions quoted for DSFGs prior to JWST may have been due to strong dust attenuation, which caused the images to appear clumpy at rest-frame optical wavelengths.
However, we caution that different methods for identifying mergers can complicate the comparison of results between different studies.

\item
Between the rest-frame optical and rest-frame NIR, 
the sample changes from extended ($R_e \sim 3.2$~kpc), exponential disk-like ($n\sim1.1$) profiles to more compact ($R_e \sim 1.8$~kpc), slightly steeper ($n\sim1.7$) profiles, suggesting the growth of stellar bulges and/or strong central dust attenuation. 

\item
We found a moderate increase of $R_e$ with decreasing redshift in both the rest-frame optical and rest-frame NIR, reflecting the buildup of galaxy size from $z=3.5$ to $z=1.5$. We do not see significant evidence of evolution in the S\'ersic indices over the same redshift range.

\end{itemize}

\vspace{5mm}

\section*{Acknowledgements}
{
We thank the anonymous referee for providing constructive comments that helped to improve the manuscript.
S.~J.~M. would like to thank J. McKinney for providing the derived properties and fluxes for the SCUBADive sample and I. Smail for discussions on the redshift and flux distributions of our sample.

We gratefully acknowledge support for this research from 
the William F. Vilas Estate and from the North American ALMA Science Center through the ALMA Ambassadors program (S.~J.~M.); from
the University of Wisconsin-Madison, Office of the Vice Chancellor for Research with funding from the Wisconsin Alumni Research Foundation (A.~J.~B.);
and from NASA grant 80NSSC22K0483 (L.~L.~C.).

The National Radio Astronomy Observatory is a facility of the National Science Foundation operated under cooperative agreement by Associated Universities, Inc.
This paper makes use of the following ALMA data: 
ADS/JAO.ALMA\#2013.1.00999.S, \\ 
ADS/JAO.ALMA\#2015.1.01425.S, \\ 
ADS/JAO.ALMA\#2015.1.00543.S,\\ 
ADS/JAO.ALMA\#2015.1.00242.S, \\ 
ADS/JAO.ALMA\#2015.1.00098.S, \\ 
ADS/JAO.ALMA\#2017.1.01219.S, \\ 
ADS/JAO.ALMA\#2017.1.00755.S, \\ 
ADS/JAO.ALMA\#2018.1.00035.L, \\ 
ADS/JAO.ALMA\#2021.1.00024.S,\\ 
and ADS/JAO.ALMA\#2022.1.00073.S. \\ 
ALMA is a partnership of ESO (representing its member states), NSF (USA), and NINS (Japan), together with NRC (Canada), MOST and ASIAA (Taiwan), and KASI (Republic of Korea), in cooperation with the Republic of Chile. The Joint ALMA Observatory is operated by 
 ESO, AUI/NRAO, and NAOJ.

The James Clerk Maxwell Telescope is operated by the East Asian Observatory on behalf of The National Astronomical Observatory of Japan, Academia Sinica Institute of Astronomy and Astrophysics, the Korea Astronomy and Space Science Institute, the National Astronomical Observatories of China and the Chinese Academy of Sciences (grant No.~XDB09000000), with additional funding support from the Science and Technology Facilities Council of the United Kingdom and participating universities 
in the United Kingdom and Canada. 

Some of the data presented in this paper were obtained from the Mikulski Archive for Space Telescopes (MAST) at the Space Telescope Science Institute, which is operated by the Association of Universities for Research in Astronomy, Inc., under NASA contract NAS 5-03127 for JWST. The archival JWST data used in this work was from the GLASS-JWST \citep{10.17909/kw3c-n857} and JADES \citep{10.17909/8tdj-8n28} HLSPs plus the UNCOVER first epoch survey data, \dataset[doi: 10.17909/zn4s-0243]{https://doi.org/10.17909/zn4s-0243}.

We wish to recognize and acknowledge 
the very significant cultural role and reverence that the summit of Maunakea has always had within the indigenous Hawaiian community. We are most fortunate to have the opportunity to conduct observations from this mountain.
}

\facilities{ALMA, HST, JCMT, JWST.}
\software{{\tt astropy} \citep{astropy:2013,astropy:2018,astropy:2022}, {\tt BAGPIPES} \citep{carnall18}, {\tt EAZY} \citep{brammer08}, {\tt GALFIT} \citep{peng10}, {\tt scipy} \citep{scipy20}}

\footnotesize
\bibliography{bib.bib}

\begin{thebibliography}{}
\expandafter\ifx\csname natexlab\endcsname\relax\def\natexlab#1{#1}\fi
\providecommand{\url}[1]{\href{#1}{#1}}
\providecommand{\dodoi}[1]{doi:~\href{http://doi.org/#1}{\nolinkurl{#1}}}
\providecommand{\doeprint}[1]{\href{http://ascl.net/#1}{\nolinkurl{http://ascl.net/#1}}}
\providecommand{\doarXiv}[1]{\href{https://arxiv.org/abs/#1}{\nolinkurl{https://arxiv.org/abs/#1}}}

\bibitem[{{Alcalde Pampliega} {et~al.}(2019){Alcalde Pampliega}, {P{\'e}rez-Gonz{\'a}lez}, {Barro}, {Dom{\'\i}nguez S{\'a}nchez}, {Eliche-Moral}, {Cardiel}, {Hern{\'a}n-Caballero}, {Rodriguez-Mu{\~n}oz}, {S{\'a}nchez Bl{\'a}zquez}, \& {Esquej}}]{alcalde-pampliega19}
{Alcalde Pampliega}, B., {P{\'e}rez-Gonz{\'a}lez}, P.~G., {Barro}, G., {et~al.} 2019, \apj, 876, 135

\bibitem[{{An} {et~al.}(2018){An}, {Stach}, {Smail}, {Swinbank}, {Almaini}, {Simpson}, {Hartley}, {Maltby}, {Ivison}, {Arumugam}, {Wardlow}, {Cooke}, {Gullberg}, {Thomson}, {Chen}, {Simpson}, {Geach}, {Scott}, {Dunlop}, {Farrah}, {van der Werf}, {Blain}, {Conselice}, {Micha{\l}owski}, {Chapman}, \& {Coppin}}]{an18}
{An}, F.~X., {Stach}, S.~M., {Smail}, I., {et~al.} 2018, \apj, 862, 101

\bibitem[{{Ananna} {et~al.}(2024){Ananna}, {Bogd{\'a}n}, {Kov{\'a}cs}, {Natarajan}, \& {Hickox}}]{tasnim24}
{Ananna}, T.~T., {Bogd{\'a}n}, {\'A}., {Kov{\'a}cs}, O.~E., {Natarajan}, P., \& {Hickox}, R.~C. 2024, \apjl, 969, L18

\bibitem[{{Aravena} {et~al.}(2016){Aravena}, {Decarli}, {Walter}, {Da Cunha}, {Bauer}, {Carilli}, {Daddi}, {Elbaz}, {Ivison}, {Riechers}, {Smail}, {Swinbank}, {Weiss}, {Anguita}, {Assef}, {Bell}, {Bertoldi}, {Bacon}, {Bouwens}, {Cortes}, {Cox}, {G{\'o}nzalez-L{\'o}pez}, {Hodge}, {Ibar}, {Inami}, {Infante}, {Karim}, {Le Le F{\`e}vre}, {Magnelli}, {Ota}, {Popping}, {Sheth}, {van der Werf}, \& {Wagg}}]{aravena16}
{Aravena}, M., {Decarli}, R., {Walter}, F., {et~al.} 2016, \apj, 833, 68

\bibitem[{{Aravena} {et~al.}(2020){Aravena}, {Boogaard}, {G{\'o}nzalez-L{\'o}pez}, {Decarli}, {Walter}, {Carilli}, {Smail}, {Weiss}, {Assef}, {Bauer}, {Bouwens}, {Cortes}, {Cox}, {da Cunha}, {Daddi}, {D{\'\i}az-Santos}, {Inami}, {Ivison}, {Novak}, {Popping}, {Riechers}, {van der Werf}, \& {Wagg}}]{aravena20}
{Aravena}, M., {Boogaard}, L., {G{\'o}nzalez-L{\'o}pez}, J., {et~al.} 2020, \apj, 901, 79

\bibitem[{{Astropy Collaboration} {et~al.}(2013){Astropy Collaboration}, {Robitaille}, {Tollerud}, {Greenfield}, {Droettboom}, {Bray}, {Aldcroft}, {Davis}, {Ginsburg}, {Price-Whelan}, {Kerzendorf}, {Conley}, {Crighton}, {Barbary}, {Muna}, {Ferguson}, {Grollier}, {Parikh}, {Nair}, {Unther}, {Deil}, {Woillez}, {Conseil}, {Kramer}, {Turner}, {Singer}, {Fox}, {Weaver}, {Zabalza}, {Edwards}, {Azalee Bostroem}, {Burke}, {Casey}, {Crawford}, {Dencheva}, {Ely}, {Jenness}, {Labrie}, {Lim}, {Pierfederici}, {Pontzen}, {Ptak}, {Refsdal}, {Servillat}, \& {Streicher}}]{astropy:2013}
{Astropy Collaboration}, {Robitaille}, T.~P., {Tollerud}, E.~J., {et~al.} 2013, \aap, 558, A33

\bibitem[{{Astropy Collaboration} {et~al.}(2018){Astropy Collaboration}, {Price-Whelan}, {Sip{\H{o}}cz}, {G{\"u}nther}, {Lim}, {Crawford}, {Conseil}, {Shupe}, {Craig}, {Dencheva}, {Ginsburg}, {VanderPlas}, {Bradley}, {P{\'e}rez-Su{\'a}rez}, {de Val-Borro}, {Aldcroft}, {Cruz}, {Robitaille}, {Tollerud}, {Ardelean}, {Babej}, {Bach}, {Bachetti}, {Bakanov}, {Bamford}, {Barentsen}, {Barmby}, {Baumbach}, {Berry}, {Biscani}, {Boquien}, {Bostroem}, {Bouma}, {Brammer}, {Bray}, {Breytenbach}, {Buddelmeijer}, {Burke}, {Calderone}, {Cano Rodr{\'\i}guez}, {Cara}, {Cardoso}, {Cheedella}, {Copin}, {Corrales}, {Crichton}, {D'Avella}, {Deil}, {Depagne}, {Dietrich}, {Donath}, {Droettboom}, {Earl}, {Erben}, {Fabbro}, {Ferreira}, {Finethy}, {Fox}, {Garrison}, {Gibbons}, {Goldstein}, {Gommers}, {Greco}, {Greenfield}, {Groener}, {Grollier}, {Hagen}, {Hirst}, {Homeier}, {Horton}, {Hosseinzadeh}, {Hu}, {Hunkeler}, {Ivezi{\'c}}, {Jain}, {Jenness}, {Kanarek}, {Kendrew}, {Kern}, {Kerzendorf}, {Khvalko}, {King}, {Kirkby}, {Kulkarni},
  {Kumar}, {Lee}, {Lenz}, {Littlefair}, {Ma}, {Macleod}, {Mastropietro}, {McCully}, {Montagnac}, {Morris}, {Mueller}, {Mumford}, {Muna}, {Murphy}, {Nelson}, {Nguyen}, {Ninan}, {N{\"o}the}, {Ogaz}, {Oh}, {Parejko}, {Parley}, {Pascual}, {Patil}, {Patil}, {Plunkett}, {Prochaska}, {Rastogi}, {Reddy Janga}, {Sabater}, {Sakurikar}, {Seifert}, {Sherbert}, {Sherwood-Taylor}, {Shih}, {Sick}, {Silbiger}, {Singanamalla}, {Singer}, {Sladen}, {Sooley}, {Sornarajah}, {Streicher}, {Teuben}, {Thomas}, {Tremblay}, {Turner}, {Terr{\'o}n}, {van Kerkwijk}, {de la Vega}, {Watkins}, {Weaver}, {Whitmore}, {Woillez}, {Zabalza}, \& {Astropy Contributors}}]{astropy:2018}
{Astropy Collaboration}, {Price-Whelan}, A.~M., {Sip{\H{o}}cz}, B.~M., {et~al.} 2018, \aj, 156, 123

\bibitem[{{Astropy Collaboration} {et~al.}(2022){Astropy Collaboration}, {Price-Whelan}, {Lim}, {Earl}, {Starkman}, {Bradley}, {Shupe}, {Patil}, {Corrales}, {Brasseur}, {N{"o}the}, {Donath}, {Tollerud}, {Morris}, {Ginsburg}, {Vaher}, {Weaver}, {Tocknell}, {Jamieson}, {van Kerkwijk}, {Robitaille}, {Merry}, {Bachetti}, {G{"u}nther}, {Aldcroft}, {Alvarado-Montes}, {Archibald}, {B{'o}di}, {Bapat}, {Barentsen}, {Baz{'a}n}, {Biswas}, {Boquien}, {Burke}, {Cara}, {Cara}, {Conroy}, {Conseil}, {Craig}, {Cross}, {Cruz}, {D'Eugenio}, {Dencheva}, {Devillepoix}, {Dietrich}, {Eigenbrot}, {Erben}, {Ferreira}, {Foreman-Mackey}, {Fox}, {Freij}, {Garg}, {Geda}, {Glattly}, {Gondhalekar}, {Gordon}, {Grant}, {Greenfield}, {Groener}, {Guest}, {Gurovich}, {Handberg}, {Hart}, {Hatfield-Dodds}, {Homeier}, {Hosseinzadeh}, {Jenness}, {Jones}, {Joseph}, {Kalmbach}, {Karamehmetoglu}, {Ka{l}uszy{'n}ski}, {Kelley}, {Kern}, {Kerzendorf}, {Koch}, {Kulumani}, {Lee}, {Ly}, {Ma}, {MacBride}, {Maljaars}, {Muna}, {Murphy}, {Norman}, {O'Steen},
  {Oman}, {Pacifici}, {Pascual}, {Pascual-Granado}, {Patil}, {Perren}, {Pickering}, {Rastogi}, {Roulston}, {Ryan}, {Rykoff}, {Sabater}, {Sakurikar}, {Salgado}, {Sanghi}, {Saunders}, {Savchenko}, {Schwardt}, {Seifert-Eckert}, {Shih}, {Jain}, {Shukla}, {Sick}, {Simpson}, {Singanamalla}, {Singer}, {Singhal}, {Sinha}, {Sip{H{o}}cz}, {Spitler}, {Stansby}, {Streicher}, {Sumak}, {Swinbank}, {Taranu}, {Tewary}, {Tremblay}, {Val-Borro}, {Van Kooten}, {Vasovi{'c}}, {Verma}, {de Miranda Cardoso}, {Williams}, {Wilson}, {Winkel}, {Wood-Vasey}, {Xue}, {Yoachim}, {Zhang}, {Zonca}, \& {Astropy Project Contributors}}]{astropy:2022}
{Astropy Collaboration}, {Price-Whelan}, A.~M., {Lim}, P.~L., {et~al.} 2022, \apj, 935, 167

\bibitem[{{Bacon} {et~al.}(2023){Bacon}, {Brinchmann}, {Conseil}, {Maseda}, {Nanayakkara}, {Wendt}, {Bacher}, {Mary}, {Weilbacher}, {Krajnovi{\'c}}, {Boogaard}, {Bouch{\'e}}, {Contini}, {Epinat}, {Feltre}, {Guo}, {Herenz}, {Kollatschny}, {Kusakabe}, {Leclercq}, {Michel-Dansac}, {Pello}, {Richard}, {Roth}, {Salvignol}, {Schaye}, {Steinmetz}, {Tresse}, {Urrutia}, {Verhamme}, {Vitte}, {Wisotzki}, \& {Zoutendijk}}]{bacon23}
{Bacon}, R., {Brinchmann}, J., {Conseil}, S., {et~al.} 2023, \aap, 670, A4

\bibitem[{{Barger} \& {Cowie}(2023)}]{barger23}
{Barger}, A.~J., \& {Cowie}, L.~L. 2023, \apj, 956, 95

\bibitem[{{Barger} {et~al.}(2019){Barger}, {Cowie}, {Bauer}, \& {Gonz{\'a}lez-L{\'o}pez}}]{barger19}
{Barger}, A.~J., {Cowie}, L.~L., {Bauer}, F.~E., \& {Gonz{\'a}lez-L{\'o}pez}, J. 2019, \apj, 887, 23

\bibitem[{{Barger} {et~al.}(2022){Barger}, {Cowie}, {Blair}, \& {Jones}}]{barger22}
{Barger}, A.~J., {Cowie}, L.~L., {Blair}, A.~H., \& {Jones}, L.~H. 2022, \apj, 934, 56

\bibitem[{{Barger} {et~al.}(1998){Barger}, {Cowie}, {Sanders}, {Fulton}, {Taniguchi}, {Sato}, {Kawara}, \& {Okuda}}]{barger98}
{Barger}, A.~J., {Cowie}, L.~L., {Sanders}, D.~B., {et~al.} 1998, \nat, 394, 248

\bibitem[{{Barger} {et~al.}(2012){Barger}, {Wang}, {Cowie}, {Owen}, {Chen}, \& {Williams}}]{barger12}
{Barger}, A.~J., {Wang}, W.~H., {Cowie}, L.~L., {et~al.} 2012, \apj, 761, 89

\bibitem[{{Barger} {et~al.}(2014){Barger}, {Cowie}, {Chen}, {Owen}, {Wang}, {Casey}, {Lee}, {Sanders}, \& {Williams}}]{barger14}
{Barger}, A.~J., {Cowie}, L.~L., {Chen}, C.~C., {et~al.} 2014, \apj, 784, 9

\bibitem[{{Barro} {et~al.}(2016){Barro}, {Kriek}, {P{\'e}rez-Gonz{\'a}lez}, {Trump}, {Koo}, {Faber}, {Dekel}, {Primack}, {Guo}, {Kocevski}, {Mu{\~n}oz-Mateos}, {Rujopakarn}, \& {Seth}}]{barro16}
{Barro}, G., {Kriek}, M., {P{\'e}rez-Gonz{\'a}lez}, P.~G., {et~al.} 2016, \apjl, 827, L32

\bibitem[{{Barrufet} {et~al.}(2023){Barrufet}, {Oesch}, {Weibel}, {Brammer}, {Bezanson}, {Bouwens}, {Fudamoto}, {Gonzalez}, {Gottumukkala}, {Illingworth}, {Heintz}, {Holden}, {Labbe}, {Magee}, {Naidu}, {Nelson}, {Stefanon}, {Smit}, {van Dokkum}, {Weaver}, \& {Williams}}]{barrufet23}
{Barrufet}, L., {Oesch}, P.~A., {Weibel}, A., {et~al.} 2023, \mnras, 522, 449

\bibitem[{{Barrufet} {et~al.}(2025){Barrufet}, {Oesch}, {Marques-Chaves}, {Arellano-Cordova}, {Baggen}, {Carnall}, {Cullen}, {Dunlop}, {Gottumukkala}, {Fudamoto}, {Illingworth}, {Magee}, {McLure}, {McLeod}, {Micha{\l}owski}, {Stefanon}, {van Dokkum}, \& {Weibel}}]{barrufet24}
{Barrufet}, L., {Oesch}, P.~A., {Marques-Chaves}, R., {et~al.} 2025, \mnras, 537, 3453

\bibitem[{{Bell} {et~al.}(2005){Bell}, {Papovich}, {Wolf}, {Le Floc'h}, {Caldwell}, {Barden}, {Egami}, {McIntosh}, {Meisenheimer}, {P{\'e}rez-Gonz{\'a}lez}, {Rieke}, {Rieke}, {Rigby}, \& {Rix}}]{bell05}
{Bell}, E.~F., {Papovich}, C., {Wolf}, C., {et~al.} 2005, \apj, 625, 23

\bibitem[{{Bezanson} {et~al.}(2024){Bezanson}, {Labbe}, {Whitaker}, {Leja}, {Price}, {Franx}, {Brammer}, {Marchesini}, {Zitrin}, {Wang}, {Weaver}, {Furtak}, {Atek}, {Coe}, {Cutler}, {Dayal}, {van Dokkum}, {Feldmann}, {F{\"o}rster Schreiber}, {Fujimoto}, {Geha}, {Glazebrook}, {de Graaff}, {Greene}, {Juneau}, {Kassin}, {Kriek}, {Khullar}, {Maseda}, {Mowla}, {Muzzin}, {Nanayakkara}, {Nelson}, {Oesch}, {Pacifici}, {Pan}, {Papovich}, {Setton}, {Shapley}, {Smit}, {Stefanon}, {Taylor}, \& {Williams}}]{bezanson22}
{Bezanson}, R., {Labbe}, I., {Whitaker}, K.~E., {et~al.} 2024, \apj, 974, 92

\bibitem[{{Birkin} {et~al.}(2021){Birkin}, {Weiss}, {Wardlow}, {Smail}, {Swinbank}, {Dudzevi{\v{c}}i{\={u}}t{\.{e}}}, {An}, {Ao}, {Chapman}, {Chen}, {da Cunha}, {Dannerbauer}, {Gullberg}, {Hodge}, {Ikarashi}, {Ivison}, {Matsuda}, {Stach}, {Walter}, {Wang}, \& {van der Werf}}]{birkin21}
{Birkin}, J.~E., {Weiss}, A., {Wardlow}, J.~L., {et~al.} 2021, \mnras, 501, 3926

\bibitem[{{Boogaard} {et~al.}(2024){Boogaard}, {Gillman}, {Melinder}, {Walter}, {Colina}, {{\"O}stlin}, {Caputi}, {Iani}, {P{\'e}rez-Gonz{\'a}lez}, {van der Werf}, {Greve}, {Wright}, {Alonso-Herrero}, {{\'A}lvarez-M{\'a}rquez}, {Annunziatella}, {Bik}, {Bosman}, {Costantin}, {Crespo G{\'o}mez}, {Dicken}, {Eckart}, {Hjorth}, {Jermann}, {Labiano}, {Langeroodi}, {Meyer}, {Moutard}, {Pei{\ss}ker}, {Pye}, {Rinaldi}, {Tikkanen}, {Topinka}, \& {Henning}}]{boogaard24}
{Boogaard}, L.~A., {Gillman}, S., {Melinder}, J., {et~al.} 2024, \apj, 969, 27

\bibitem[{{Boquien} {et~al.}(2019){Boquien}, {Burgarella}, {Roehlly}, {Buat}, {Ciesla}, {Corre}, {Inoue}, \& {Salas}}]{boquien19}
{Boquien}, M., {Burgarella}, D., {Roehlly}, Y., {et~al.} 2019, \aap, 622, A103

\bibitem[{{Bouwens} {et~al.}(2020){Bouwens}, {Gonz{\'a}lez-L{\'o}pez}, {Aravena}, {Decarli}, {Novak}, {Stefanon}, {Walter}, {Boogaard}, {Carilli}, {Dudzevi{\v{c}}i{\={u}}t{\.{e}}}, {Smail}, {Daddi}, {da Cunha}, {Ivison}, {Nanayakkara}, {Cortes}, {Cox}, {Inami}, {Oesch}, {Popping}, {Riechers}, {van der Werf}, {Weiss}, {Fudamoto}, \& {Wagg}}]{bouwens20}
{Bouwens}, R., {Gonz{\'a}lez-L{\'o}pez}, J., {Aravena}, M., {et~al.} 2020, \apj, 902, 112

\bibitem[{{Brammer} {et~al.}(2008){Brammer}, {van Dokkum}, \& {Coppi}}]{brammer08}
{Brammer}, G.~B., {van Dokkum}, P.~G., \& {Coppi}, P. 2008, \apj, 686, 1503

\bibitem[{{Bruzual} \& {Charlot}(2003)}]{bruzual03}
{Bruzual}, G., \& {Charlot}, S. 2003, \mnras, 344, 1000

\bibitem[{{Bunker} {et~al.}(2024){Bunker}, {Cameron}, {Curtis-Lake}, {Jakobsen}, {Carniani}, {Curti}, {Witstok}, {Maiolino}, {D'Eugenio}, {Looser}, {Willott}, {Bonaventura}, {Hainline}, {{\"U}bler}, {Willmer}, {Saxena}, {Smit}, {Alberts}, {Arribas}, {Baker}, {Baum}, {Bhatawdekar}, {Bowler}, {Boyett}, {Charlot}, {Chen}, {Chevallard}, {Circosta}, {DeCoursey}, {de Graaff}, {Egami}, {Eisenstein}, {Endsley}, {Ferruit}, {Giardino}, {Hausen}, {Helton}, {Hviding}, {Ji}, {Johnson}, {Jones}, {Kumari}, {Laseter}, {L{\"u}tzgendorf}, {Maseda}, {Nelson}, {Parlanti}, {Perna}, {Rauscher}, {Rawle}, {Rix}, {Rieke}, {Robertson}, {Rodr{\'\i}guez Del Pino}, {Sandles}, {Scholtz}, {Sharpe}, {Skarbinski}, {Stark}, {Sun}, {Tacchella}, {Topping}, {Villanueva}, {Wallace}, {Williams}, \& {Woodrum}}]{bunker23}
{Bunker}, A.~J., {Cameron}, A.~J., {Curtis-Lake}, E., {et~al.} 2024, \aap, 690, A288

\bibitem[{{Calzetti} {et~al.}(2000){Calzetti}, {Armus}, {Bohlin}, {Kinney}, {Koornneef}, \& {Storchi-Bergmann}}]{calzetti00}
{Calzetti}, D., {Armus}, L., {Bohlin}, R.~C., {et~al.} 2000, \apj, 533, 682

\bibitem[{{Caputi} {et~al.}(2012){Caputi}, {Dunlop}, {McLure}, {Huang}, {Fazio}, {Ashby}, {Castellano}, {Fontana}, {Cirasuolo}, {Almaini}, {Bell}, {Dickinson}, {Donley}, {Faber}, {Ferguson}, {Giavalisco}, {Grogin}, {Kocevski}, {Koekemoer}, {Koo}, {Lai}, {Newman}, \& {Somerville}}]{caputi12}
{Caputi}, K.~I., {Dunlop}, J.~S., {McLure}, R.~J., {et~al.} 2012, \apjl, 750, L20

\bibitem[{{Carnall} {et~al.}(2018){Carnall}, {McLure}, {Dunlop}, \& {Dav{\'e}}}]{carnall18}
{Carnall}, A.~C., {McLure}, R.~J., {Dunlop}, J.~S., \& {Dav{\'e}}, R. 2018, \mnras, 480, 4379

\bibitem[{{Chapman} {et~al.}(2005){Chapman}, {Blain}, {Smail}, \& {Ivison}}]{chapman05}
{Chapman}, S.~C., {Blain}, A.~W., {Smail}, I., \& {Ivison}, R.~J. 2005, \apj, 622, 772

\bibitem[{{Charlot} \& {Fall}(2000)}]{charlot00}
{Charlot}, S., \& {Fall}, S.~M. 2000, \apj, 539, 718

\bibitem[{{Chen} {et~al.}(2013){Chen}, {Cowie}, {Barger}, {Casey}, {Lee}, {Sanders}, {Wang}, \& {Williams}}]{chen13}
{Chen}, C.-C., {Cowie}, L.~L., {Barger}, A.~J., {et~al.} 2013, \apj, 776, 131

\bibitem[{{Chen} {et~al.}(2015){Chen}, {Smail}, {Swinbank}, {Simpson}, {Ma}, {Alexander}, {Biggs}, {Brandt}, {Chapman}, {Coppin}, {Danielson}, {Dannerbauer}, {Edge}, {Greve}, {Ivison}, {Karim}, {Menten}, {Schinnerer}, {Walter}, {Wardlow}, {Wei{\ss}}, \& {van der Werf}}]{chen15}
{Chen}, C.-C., {Smail}, I., {Swinbank}, A.~M., {et~al.} 2015, \apj, 799, 194

\bibitem[{{Chen} {et~al.}(2016){Chen}, {Smail}, {Swinbank}, {Simpson}, {Almaini}, {Conselice}, {Hartley}, {Mortlock}, {Simpson}, \& {Wilkinson}}]{chen16}
---. 2016, \apj, 831, 91

\bibitem[{{Chen} {et~al.}(2022{\natexlab{a}}){Chen}, {Gao}, {Hsu}, {Liao}, {Ling}, {Lo}, {Smail}, {Wang}, \& {Wang}}]{chen22}
{Chen}, C.-C., {Gao}, Z.-K., {Hsu}, Q.-N., {et~al.} 2022{\natexlab{a}}, \apjl, 939, L7

\bibitem[{{Chen} {et~al.}(2022{\natexlab{b}}){Chen}, {Liao}, {Smail}, {Swinbank}, {Ao}, {Bunker}, {Chapman}, {Hatsukade}, {Ivison}, {Lee}, {Serjeant}, {Umehata}, {Wang}, \& {Zhao}}]{chen22b}
{Chen}, C.-C., {Liao}, C.-L., {Smail}, I., {et~al.} 2022{\natexlab{b}}, \apj, 929, 159

\bibitem[{{Cheng} {et~al.}(2023){Cheng}, {Huang}, {Smail}, {Yan}, {Cohen}, {Jansen}, {Windhorst}, {Ma}, {Koekemoer}, {Willmer}, {Willner}, {Diego}, {Frye}, {Conselice}, {Ferreira}, {Petric}, {Yun}, {Gim}, {Polletta}, {Duncan}, {Holwerda}, {R{\"o}ttgering}, {Honor}, {Hathi}, {Kamieneski}, {Adams}, {Coe}, {Broadhurst}, {Summers}, {Tompkins}, {Driver}, {Grogin}, {Marshall}, {Pirzkal}, {Robotham}, \& {Ryan}}]{cheng23}
{Cheng}, C., {Huang}, J.-S., {Smail}, I., {et~al.} 2023, \apjl, 942, L19

\bibitem[{{Cowie} {et~al.}(2023){Cowie}, {Barger}, \& {Bauer}}]{cowie23}
{Cowie}, L.~L., {Barger}, A.~J., \& {Bauer}, F.~E. 2023, \apj, 952, 28

\bibitem[{{Cowie} {et~al.}(2022){Cowie}, {Barger}, {Bauer}, {Chen}, {Jones}, {Orquera-Rojas}, {Rosenthal}, \& {Taylor}}]{cowie22}
{Cowie}, L.~L., {Barger}, A.~J., {Bauer}, F.~E., {et~al.} 2022, \apj, 939, 5

\bibitem[{{Cowie} {et~al.}(2017){Cowie}, {Barger}, {Hsu}, {Chen}, {Owen}, \& {Wang}}]{cowie17}
{Cowie}, L.~L., {Barger}, A.~J., {Hsu}, L.~Y., {et~al.} 2017, \apj, 837, 139

\bibitem[{{Cowie} {et~al.}(2002){Cowie}, {Barger}, \& {Kneib}}]{cowie02}
{Cowie}, L.~L., {Barger}, A.~J., \& {Kneib}, J.~P. 2002, \aj, 123, 2197

\bibitem[{{Cowie} {et~al.}(2018){Cowie}, {Gonz{\'a}lez-L{\'o}pez}, {Barger}, {Bauer}, {Hsu}, \& {Wang}}]{cowie18}
{Cowie}, L.~L., {Gonz{\'a}lez-L{\'o}pez}, J., {Barger}, A.~J., {et~al.} 2018, \apj, 865, 106

\bibitem[{{da Cunha} {et~al.}(2008){da Cunha}, {Charlot}, \& {Elbaz}}]{dacunha08}
{da Cunha}, E., {Charlot}, S., \& {Elbaz}, D. 2008, \mnras, 388, 1595

\bibitem[{{da Cunha} {et~al.}(2015){da Cunha}, {Walter}, {Smail}, {Swinbank}, {Simpson}, {Decarli}, {Hodge}, {Weiss}, {van der Werf}, {Bertoldi}, {Chapman}, {Cox}, {Danielson}, {Dannerbauer}, {Greve}, {Ivison}, {Karim}, \& {Thomson}}]{dacunha15}
{da Cunha}, E., {Walter}, F., {Smail}, I.~R., {et~al.} 2015, \apj, 806, 110

\bibitem[{{Danielson} {et~al.}(2017){Danielson}, {Swinbank}, {Smail}, {Simpson}, {Casey}, {Chapman}, {da Cunha}, {Hodge}, {Walter}, {Wardlow}, {Alexander}, {Brandt}, {de Breuck}, {Coppin}, {Dannerbauer}, {Dickinson}, {Edge}, {Gawiser}, {Ivison}, {Karim}, {Kovacs}, {Lutz}, {Menten}, {Schinnerer}, {Wei{\ss}}, \& {van der Werf}}]{danielson17}
{Danielson}, A.~L.~R., {Swinbank}, A.~M., {Smail}, I., {et~al.} 2017, \apj, 840, 78

\bibitem[{{D'Eugenio} {et~al.}(2025){D'Eugenio}, {Cameron}, {Scholtz}, {Carniani}, {Willott}, {Curtis-Lake}, {Bunker}, {Parlanti}, {Maiolino}, {Willmer}, {Jakobsen}, {Robertson}, {Johnson}, {Tacchella}, {Cargile}, {Rawle}, {Arribas}, {Chevallard}, {Curti}, {Egami}, {Eisenstein}, {Kumari}, {Looser}, {Rieke}, {Rodr{\'\i}guez Del Pino}, {Saxena}, {{\"U}bler}, {Venturi}, {Witstok}, {Baker}, {Bhatawdekar}, {Bonaventura}, {Boyett}, {Charlot}, {Danhaive}, {Hainline}, {Hausen}, {Helton}, {Ji}, {Ji}, {Jones}, {Juod{\v{z}}balis}, {Maseda}, {P{\'e}rez-Gonz{\'a}lez}, {Perna}, {Pusk{\'a}s}, {Shivaei}, {Silcock}, {Simmonds}, {Smit}, {Sun}, {Villanueva}, {Williams}, \& {Zhu}}]{deugenio24}
{D'Eugenio}, F., {Cameron}, A.~J., {Scholtz}, J., {et~al.} 2025, \apjs, 277, 4

\bibitem[{{Draine} \& {Li}(2007)}]{draine07}
{Draine}, B.~T., \& {Li}, A. 2007, \apj, 657, 810

\bibitem[{{Dudzevi{\v{c}}i{\={u}}t{\.{e}}} {et~al.}(2020){Dudzevi{\v{c}}i{\={u}}t{\.{e}}}, {Smail}, {Swinbank}, {Stach}, {Almaini}, {da Cunha}, {An}, {Arumugam}, {Birkin}, {Blain}, {Chapman}, {Chen}, {Conselice}, {Coppin}, {Dunlop}, {Farrah}, {Geach}, {Gullberg}, {Hartley}, {Hodge}, {Ivison}, {Maltby}, {Scott}, {Simpson}, {Simpson}, {Thomson}, {Walter}, {Wardlow}, {Weiss}, \& {van der Werf}}]{dudzeviciute20}
{Dudzevi{\v{c}}i{\={u}}t{\.{e}}}, U., {Smail}, I., {Swinbank}, A.~M., {et~al.} 2020, \mnras, 494, 3828

\bibitem[{{Dunlop} {et~al.}(2017){Dunlop}, {McLure}, {Biggs}, {Geach}, {Micha{\l}owski}, {Ivison}, {Rujopakarn}, {van Kampen}, {Kirkpatrick}, {Pope}, {Scott}, {Swinbank}, {Targett}, {Aretxaga}, {Austermann}, {Best}, {Bruce}, {Chapin}, {Charlot}, {Cirasuolo}, {Coppin}, {Ellis}, {Finkelstein}, {Hayward}, {Hughes}, {Ibar}, {Jagannathan}, {Khochfar}, {Koprowski}, {Narayanan}, {Nyland}, {Papovich}, {Peacock}, {Rieke}, {Robertson}, {Vernstrom}, {Werf}, {Wilson}, \& {Yun}}]{dunlop17}
{Dunlop}, J.~S., {McLure}, R.~J., {Biggs}, A.~D., {et~al.} 2017, \mnras, 466, 861

\bibitem[{{Eales} {et~al.}(1999){Eales}, {Lilly}, {Gear}, {Dunne}, {Bond}, {Hammer}, {Le F{\`e}vre}, \& {Crampton}}]{eales99}
{Eales}, S., {Lilly}, S., {Gear}, W., {et~al.} 1999, \apj, 515, 518

\bibitem[{{Eisenstein} {et~al.}(2023{\natexlab{a}}){Eisenstein}, {Willott}, {Alberts}, {Arribas}, {Bonaventura}, {Bunker}, {Cameron}, {Carniani}, {Charlot}, {Curtis-Lake}, {D'Eugenio}, {Endsley}, {Ferruit}, {Giardino}, {Hainline}, {Hausen}, {Jakobsen}, {Johnson}, {Maiolino}, {Rieke}, {Rieke}, {Rix}, {Robertson}, {Stark}, {Tacchella}, {Williams}, {Willmer}, {Baker}, {Baum}, {Bhatawdekar}, {Boyett}, {Chen}, {Chevallard}, {Circosta}, {Curti}, {Danhaive}, {DeCoursey}, {de Graaff}, {Dressler}, {Egami}, {Helton}, {Hviding}, {Ji}, {Jones}, {Kumari}, {L{\"u}tzgendorf}, {Laseter}, {Looser}, {Lyu}, {Maseda}, {Nelson}, {Parlanti}, {Perna}, {Pusk{\'a}s}, {Rawle}, {Rodr{\'\i}guez Del Pino}, {Sandles}, {Saxena}, {Scholtz}, {Sharpe}, {Shivaei}, {Silcock}, {Simmonds}, {Skarbinski}, {Smit}, {Stone}, {Suess}, {Sun}, {Tang}, {Topping}, {{\"U}bler}, {Villanueva}, {Wallace}, {Whitler}, {Witstok}, \& {Woodrum}}]{eisenstein23}
{Eisenstein}, D.~J., {Willott}, C., {Alberts}, S., {et~al.} 2023{\natexlab{a}}, arXiv e-prints, arXiv:2306.02465

\bibitem[{{Eisenstein} {et~al.}(2023{\natexlab{b}}){Eisenstein}, {Johnson}, {Robertson}, {Tacchella}, {Hainline}, {Jakobsen}, {Maiolino}, {Bonaventura}, {Bunker}, {Cameron}, {Cargile}, {Curtis-Lake}, {Hausen}, {Pusk{\'a}s}, {Rieke}, {Sun}, {Willmer}, {Willott}, {Alberts}, {Arribas}, {Baker}, {Baum}, {Bhatawdekar}, {Carniani}, {Charlot}, {Chen}, {Chevallard}, {Curti}, {DeCoursey}, {D'Eugenio}, {de Graaff}, {Egami}, {Helton}, {Ji}, {Jones}, {Kumari}, {L{\"u}tzgendorf}, {Laseter}, {Looser}, {Lyu}, {Maseda}, {Nelson}, {Parlanti}, {Rauscher}, {Rawle}, {Rieke}, {Rix}, {Rujopakarn}, {Sandles}, {Saxena}, {Scholtz}, {Sharpe}, {Shivaei}, {Simmonds}, {Smit}, {Topping}, {{\"U}bler}, {Venturi}, {Williams}, {Witstok}, \& {Woodrum}}]{eisenstein23b}
{Eisenstein}, D.~J., {Johnson}, B.~D., {Robertson}, B., {et~al.} 2023{\natexlab{b}}, arXiv e-prints, arXiv:2310.12340

\bibitem[{{Elbaz} {et~al.}(2011){Elbaz}, {Dickinson}, {Hwang}, {D{\'\i}az-Santos}, {Magdis}, {Magnelli}, {Le Borgne}, {Galliano}, {Pannella}, {Chanial}, {Armus}, {Charmandaris}, {Daddi}, {Aussel}, {Popesso}, {Kartaltepe}, {Altieri}, {Valtchanov}, {Coia}, {Dannerbauer}, {Dasyra}, {Leiton}, {Mazzarella}, {Alexander}, {Buat}, {Burgarella}, {Chary}, {Gilli}, {Ivison}, {Juneau}, {Le Floc'h}, {Lutz}, {Morrison}, {Mullaney}, {Murphy}, {Pope}, {Scott}, {Brodwin}, {Calzetti}, {Cesarsky}, {Charlot}, {Dole}, {Eisenhardt}, {Ferguson}, {F{\"o}rster Schreiber}, {Frayer}, {Giavalisco}, {Huynh}, {Koekemoer}, {Papovich}, {Reddy}, {Surace}, {Teplitz}, {Yun}, \& {Wilson}}]{elbaz11}
{Elbaz}, D., {Dickinson}, M., {Hwang}, H.~S., {et~al.} 2011, \aap, 533, A119

\bibitem[{{Elbaz} {et~al.}(2018){Elbaz}, {Leiton}, {Nagar}, {Okumura}, {Franco}, {Schreiber}, {Pannella}, {Wang}, {Dickinson}, {D{\'\i}az-Santos}, {Ciesla}, {Daddi}, {Bournaud}, {Magdis}, {Zhou}, \& {Rujopakarn}}]{elbaz17}
{Elbaz}, D., {Leiton}, R., {Nagar}, N., {et~al.} 2018, \aap, 616, A110

\bibitem[{{Ferland} {et~al.}(2017){Ferland}, {Chatzikos}, {Guzm{\'a}n}, {Lykins}, {van Hoof}, {Williams}, {Abel}, {Badnell}, {Keenan}, {Porter}, \& {Stancil}}]{ferland17}
{Ferland}, G.~J., {Chatzikos}, M., {Guzm{\'a}n}, F., {et~al.} 2017, \rmxaa, 53, 385

\bibitem[{{Franco} {et~al.}(2018){Franco}, {Elbaz}, {B{\'e}thermin}, {Magnelli}, {Schreiber}, {Ciesla}, {Dickinson}, {Nagar}, {Silverman}, {Daddi}, {Alexander}, {Wang}, {Pannella}, {Le Floc'h}, {Pope}, {Giavalisco}, {Maury}, {Bournaud}, {Chary}, {Demarco}, {Ferguson}, {Finkelstein}, {Inami}, {Iono}, {Juneau}, {Lagache}, {Leiton}, {Lin}, {Magdis}, {Messias}, {Motohara}, {Mullaney}, {Okumura}, {Papovich}, {Pforr}, {Rujopakarn}, {Sargent}, {Shu}, \& {Zhou}}]{franco18}
{Franco}, M., {Elbaz}, D., {B{\'e}thermin}, M., {et~al.} 2018, \aap, 620, A152

\bibitem[{{Fujimoto} {et~al.}(2023){Fujimoto}, {Bezanson}, {Labbe}, {Brammer}, {Price}, {Wang}, {Weaver}, {Fudamoto}, {Oesch}, {Williams}, {Dayal}, {Feldmann}, {Greene}, {Leja}, {Whitaker}, {Zitrin}, {Cutler}, {Furtak}, {Pan}, {Chemerynska}, {Kokorev}, {Miller}, {Atek}, {van Dokkum}, {Juneau}, {Kassin}, {Khullar}, {Marchesini}, {Maseda}, {Nelson}, {Setton}, \& {Smit}}]{fujimoto25}
{Fujimoto}, S., {Bezanson}, R., {Labbe}, I., {et~al.} 2023, arXiv e-prints, arXiv:2309.07834

\bibitem[{{Fujimoto} {et~al.}(2024){Fujimoto}, {Kohno}, {Ouchi}, {Oguri}, {Kokorev}, {Brammer}, {Sun}, {Gonz{\'a}lez-L{\'o}pez}, {Bauer}, {Caminha}, {Hatsukade}, {Richard}, {Smail}, {Tsujita}, {Ueda}, {Uematsu}, {Zitrin}, {Coe}, {Kneib}, {Postman}, {Umetsu}, {Lagos}, {Popping}, {Ao}, {Bradley}, {Caputi}, {Dessauges-Zavadsky}, {Egami}, {Espada}, {Ivison}, {Jauzac}, {Knudsen}, {Koekemoer}, {Magdis}, {Mahler}, {Mu{\~n}oz Arancibia}, {Rawle}, {Shimasaku}, {Toft}, {Umehata}, {Valentino}, {Wang}, \& {Wang}}]{fujimoto23a}
{Fujimoto}, S., {Kohno}, K., {Ouchi}, M., {et~al.} 2024, \apjs, 275, 36

\bibitem[{{Furtak} {et~al.}(2023{\natexlab{a}}){Furtak}, {Zitrin}, {Weaver}, {Atek}, {Bezanson}, {Labb{\'e}}, {Whitaker}, {Leja}, {Price}, {Brammer}, {Wang}, {Marchesini}, {Pan}, {Dayal}, {van Dokkum}, {Feldmann}, {Fujimoto}, {Franx}, {Khullar}, {Nelson}, \& {Mowla}}]{furtak23b}
{Furtak}, L.~J., {Zitrin}, A., {Weaver}, J.~R., {et~al.} 2023{\natexlab{a}}, \mnras, 523, 4568

\bibitem[{{Furtak} {et~al.}(2023{\natexlab{b}}){Furtak}, {Zitrin}, {Plat}, {Fujimoto}, {Wang}, {Nelson}, {Labb{\'e}}, {Bezanson}, {Brammer}, {van Dokkum}, {Endsley}, {Glazebrook}, {Greene}, {Leja}, {Price}, {Smit}, {Stark}, {Weaver}, {Whitaker}, {Atek}, {Chevallard}, {Curtis-Lake}, {Dayal}, {Feltre}, {Franx}, {Fudamoto}, {Marchesini}, {Mowla}, {Pan}, {Suess}, {Vidal-Garc{\'\i}a}, \& {Williams}}]{furtak23a}
{Furtak}, L.~J., {Zitrin}, A., {Plat}, A., {et~al.} 2023{\natexlab{b}}, \apj, 952, 142

\bibitem[{{Garilli} {et~al.}(2021){Garilli}, {McLure}, {Pentericci}, {Franzetti}, {Gargiulo}, {Carnall}, {Cucciati}, {Iovino}, {Amorin}, {Bolzonella}, {Bongiorno}, {Castellano}, {Cimatti}, {Cirasuolo}, {Cullen}, {Dunlop}, {Elbaz}, {Finkelstein}, {Fontana}, {Fontanot}, {Fumana}, {Guaita}, {Hartley}, {Jarvis}, {Juneau}, {Maccagni}, {McLeod}, {Nandra}, {Pompei}, {Pozzetti}, {Scodeggio}, {Talia}, {Calabr{\`o}}, {Cresci}, {Fynbo}, {Hathi}, {Hibon}, {Koekemoer}, {Magliocchetti}, {Salvato}, {Vietri}, {Zamorani}, {Almaini}, {Balestra}, {Bardelli}, {Begley}, {Brammer}, {Bell}, {Bowler}, {Brusa}, {Buitrago}, {Caputi}, {Cassata}, {Charlot}, {Citro}, {Cristiani}, {Curtis-Lake}, {Dickinson}, {Fazio}, {Ferguson}, {Fiore}, {Franco}, {Georgakakis}, {Giavalisco}, {Grazian}, {Hamadouche}, {Jung}, {Kim}, {Khusanova}, {Le F{\`e}vre}, {Longhetti}, {Lotz}, {Mannucci}, {Maltby}, {Matsuoka}, {Mendez-Hernandez}, {Mendez-Abreu}, {Mignoli}, {Moresco}, {Nonino}, {Pannella}, {Papovich}, {Popesso}, {Roberts-Borsani}, {Rosario},
  {Saldana-Lopez}, {Santini}, {Saxena}, {Schaerer}, {Schreiber}, {Stark}, {Tasca}, {Thomas}, {Vanzella}, {Wild}, {Williams}, \& {Zucca}}]{garilli21}
{Garilli}, B., {McLure}, R., {Pentericci}, L., {et~al.} 2021, \aap, 647, A150

\bibitem[{{Gibson} {et~al.}(2024){Gibson}, {Nelson}, {Williams}, {Price}, {Whitaker}, {Suess}, {de Graaff}, {Johnson}, {Bunker}, {Baker}, {Bhatawdekar}, {Boyett}, {Charlot}, {Curtis-Lake}, {Eisenstein}, {Hainline}, {Hausen}, {Maiolino}, {Rieke}, {Rieke}, {Robertson}, {Tacchella}, \& {Willott}}]{gibson24}
{Gibson}, J.~L., {Nelson}, E., {Williams}, C.~C., {et~al.} 2024, \apj, 974, 48

\bibitem[{{Gillman} {et~al.}(2023){Gillman}, {Gullberg}, {Brammer}, {Vijayan}, {Lee}, {Bl{\'a}nquez}, {Brinch}, {Greve}, {Jermann}, {Jin}, {Kokorev}, {Liu}, {Magdis}, {Rizzo}, \& {Valentino}}]{gillman23}
{Gillman}, S., {Gullberg}, B., {Brammer}, G., {et~al.} 2023, \aap, 676, A26

\bibitem[{{Gillman} {et~al.}(2024){Gillman}, {Smail}, {Gullberg}, {Swinbank}, {Vijayan}, {Lee}, {Brammer}, {Dudzevi{\v{c}}i{\={u}}t{\.{e}}}, {Greve}, {Almaini}, {Brinch}, {Chapman}, {Chen}, {Ikarashi}, {Matsuda}, {Wang}, {Walter}, \& {van der Werf}}]{gillman24}
{Gillman}, S., {Smail}, I., {Gullberg}, B., {et~al.} 2024, \aap, 691, A299

\bibitem[{{G{\'o}mez-Guijarro} {et~al.}(2022){G{\'o}mez-Guijarro}, {Elbaz}, {Xiao}, {B{\'e}thermin}, {Franco}, {Magnelli}, {Daddi}, {Dickinson}, {Demarco}, {Inami}, {Rujopakarn}, {Magdis}, {Shu}, {Chary}, {Zhou}, {Alexander}, {Bournaud}, {Ciesla}, {Ferguson}, {Finkelstein}, {Giavalisco}, {Iono}, {Juneau}, {Kartaltepe}, {Lagache}, {Le Floc'h}, {Leiton}, {Lin}, {Motohara}, {Mullaney}, {Okumura}, {Pannella}, {Papovich}, {Pope}, {Sargent}, {Silverman}, {Treister}, \& {Wang}}]{gomez-guijarro22}
{G{\'o}mez-Guijarro}, C., {Elbaz}, D., {Xiao}, M., {et~al.} 2022, \aap, 658, A43

\bibitem[{{Gonz{\'a}lez-L{\'o}pez} {et~al.}(2017){Gonz{\'a}lez-L{\'o}pez}, {Bauer}, {Romero-Ca{\~n}izales}, {Kneissl}, {Villard}, {Carvajal}, {Kim}, {Laporte}, {Anguita}, {Aravena}, {Bouwens}, {Bradley}, {Carrasco}, {Demarco}, {Ford}, {Ibar}, {Infante}, {Messias}, {Mu{\~n}oz Arancibia}, {Nagar}, {Padilla}, {Treister}, {Troncoso}, \& {Zitrin}}]{gonzalez-lopez17}
{Gonz{\'a}lez-L{\'o}pez}, J., {Bauer}, F.~E., {Romero-Ca{\~n}izales}, C., {et~al.} 2017, \aap, 597, A41

\bibitem[{{Gonz{\'a}lez-L{\'o}pez} {et~al.}(2019){Gonz{\'a}lez-L{\'o}pez}, {Decarli}, {Pavesi}, {Walter}, {Aravena}, {Carilli}, {Boogaard}, {Popping}, {Weiss}, {Assef}, {Bauer}, {Bertoldi}, {Bouwens}, {Contini}, {Cortes}, {Cox}, {da Cunha}, {Daddi}, {D{\'\i}az-Santos}, {Inami}, {Hodge}, {Ivison}, {Le F{\`e}vre}, {Magnelli}, {Oesch}, {Riechers}, {Rix}, {Smail}, {Swinbank}, {Somerville}, {Uzgil}, \& {van der Werf}}]{gonzalez-lopez19}
{Gonz{\'a}lez-L{\'o}pez}, J., {Decarli}, R., {Pavesi}, R., {et~al.} 2019, \apj, 882, 139

\bibitem[{{Gottumukkala} {et~al.}(2024){Gottumukkala}, {Barrufet}, {Oesch}, {Weibel}, {Allen}, {Alcalde Pampliega}, {Nelson}, {Williams}, {Brammer}, {Fudamoto}, {Gonz{\'a}lez}, {Heintz}, {Illingworth}, {Magee}, {Naidu}, {Shuntov}, {Stefanon}, {Toft}, {Valentino}, \& {Xiao}}]{gottumukkala23}
{Gottumukkala}, R., {Barrufet}, L., {Oesch}, P.~A., {et~al.} 2024, \mnras, 530, 966

\bibitem[{{Greene} {et~al.}(2024){Greene}, {Labbe}, {Goulding}, {Furtak}, {Chemerynska}, {Kokorev}, {Dayal}, {Volonteri}, {Williams}, {Wang}, {Setton}, {Burgasser}, {Bezanson}, {Atek}, {Brammer}, {Cutler}, {Feldmann}, {Fujimoto}, {Glazebrook}, {de Graaff}, {Khullar}, {Leja}, {Marchesini}, {Maseda}, {Matthee}, {Miller}, {Naidu}, {Nanayakkara}, {Oesch}, {Pan}, {Papovich}, {Price}, {van Dokkum}, {Weaver}, {Whitaker}, \& {Zitrin}}]{greene24}
{Greene}, J.~E., {Labbe}, I., {Goulding}, A.~D., {et~al.} 2024, \apj, 964, 39

\bibitem[{{Hainline} {et~al.}(2011){Hainline}, {Blain}, {Smail}, {Alexander}, {Armus}, {Chapman}, \& {Ivison}}]{hainline11}
{Hainline}, L.~J., {Blain}, A.~W., {Smail}, I., {et~al.} 2011, \apj, 740, 96

\bibitem[{{Hatsukade} {et~al.}(2018){Hatsukade}, {Kohno}, {Yamaguchi}, {Umehata}, {Ao}, {Aretxaga}, {Caputi}, {Dunlop}, {Egami}, {Espada}, {Fujimoto}, {Hayatsu}, {Hughes}, {Ikarashi}, {Iono}, {Ivison}, {Kawabe}, {Kodama}, {Lee}, {Matsuda}, {Nakanishi}, {Ohta}, {Ouchi}, {Rujopakarn}, {Suzuki}, {Tamura}, {Ueda}, {Wang}, {Wang}, {Wilson}, {Yoshimura}, \& {Yun}}]{hatsukade18}
{Hatsukade}, B., {Kohno}, K., {Yamaguchi}, Y., {et~al.} 2018, \pasj, 70, 105

\bibitem[{{Hodge} {et~al.}(2016){Hodge}, {Swinbank}, {Simpson}, {Smail}, {Walter}, {Alexander}, {Bertoldi}, {Biggs}, {Brandt}, {Chapman}, {Chen}, {Coppin}, {Cox}, {Dannerbauer}, {Edge}, {Greve}, {Ivison}, {Karim}, {Knudsen}, {Menten}, {Rix}, {Schinnerer}, {Wardlow}, {Weiss}, \& {van der Werf}}]{hodge16}
{Hodge}, J.~A., {Swinbank}, A.~M., {Simpson}, J.~M., {et~al.} 2016, \apj, 833, 103

\bibitem[{{Hodge} {et~al.}(2025){Hodge}, {Cunha}, {Kendrew}, {Li}, {Smail}, {Westoby}, {Nayak}, {Swinbank}, {Chen}, {Walter}, {van der Werf}, {Cracraft}, {Battisti}, {Brandt}, {Calistro Rivera}, {Chapman}, {Cox}, {Dannerbauer}, {Decarli}, {Frias Castillo}, {Greve}, {Knudsen}, {Leslie}, {Menten}, {Rybak}, {Schinnerer}, {Wardlow}, \& {Weiss}}]{hodge24}
{Hodge}, J.~A., {Cunha}, E.~d., {Kendrew}, S., {et~al.} 2025, \apj, 978, 165

\bibitem[{{Hopkins} {et~al.}(2008){Hopkins}, {Hernquist}, {Cox}, \& {Kere{\v{s}}}}]{hopkins08}
{Hopkins}, P.~F., {Hernquist}, L., {Cox}, T.~J., \& {Kere{\v{s}}}, D. 2008, \apjs, 175, 356

\bibitem[{{Hsu} {et~al.}(2016){Hsu}, {Cowie}, {Chen}, {Barger}, \& {Wang}}]{hsu16}
{Hsu}, L.-Y., {Cowie}, L.~L., {Chen}, C.-C., {Barger}, A.~J., \& {Wang}, W.-H. 2016, \apj, 829, 25

\bibitem[{{Hughes} {et~al.}(1998){Hughes}, {Serjeant}, {Dunlop}, {Rowan-Robinson}, {Blain}, {Mann}, {Ivison}, {Peacock}, {Efstathiou}, {Gear}, {Oliver}, {Lawrence}, {Longair}, {Goldschmidt}, \& {Jenness}}]{hughes98}
{Hughes}, D.~H., {Serjeant}, S., {Dunlop}, J., {et~al.} 1998, \nat, 394, 241

\bibitem[{{Illingworth} {et~al.}(2016){Illingworth}, {Magee}, {Bouwens}, {Oesch}, {Labbe}, {van Dokkum}, {Whitaker}, {Holden}, {Franx}, \& {Gonzalez}}]{illingworth16}
{Illingworth}, G., {Magee}, D., {Bouwens}, R., {et~al.} 2016, arXiv e-prints, arXiv:1606.00841

\bibitem[{{Inami} {et~al.}(2017){Inami}, {Bacon}, {Brinchmann}, {Richard}, {Contini}, {Conseil}, {Hamer}, {Akhlaghi}, {Bouch{\'e}}, {Cl{\'e}ment}, {Desprez}, {Drake}, {Hashimoto}, {Leclercq}, {Maseda}, {Michel-Dansac}, {Paalvast}, {Tresse}, {Ventou}, {Kollatschny}, {Boogaard}, {Finley}, {Marino}, {Schaye}, \& {Wisotzki}}]{inami17}
{Inami}, H., {Bacon}, R., {Brinchmann}, J., {et~al.} 2017, \aap, 608, A2

\bibitem[{{Ivison} {et~al.}(2002){Ivison}, {Greve}, {Smail}, {Dunlop}, {Roche}, {Scott}, {Page}, {Stevens}, {Almaini}, {Blain}, {Willott}, {Fox}, {Gilbank}, {Serjeant}, \& {Hughes}}]{ivison02}
{Ivison}, R.~J., {Greve}, T.~R., {Smail}, I., {et~al.} 2002, \mnras, 337, 1

\bibitem[{{Jain} \& {Wadadekar}(2024)}]{jain24}
{Jain}, R., \& {Wadadekar}, Y. 2024, arXiv e-prints, arXiv:2412.04834

\bibitem[{{Kamieneski} {et~al.}(2023){Kamieneski}, {Frye}, {Pascale}, {Cohen}, {Windhorst}, {Jansen}, {Yun}, {Cheng}, {Summers}, {Carleton}, {Harrington}, {Diego}, {Yan}, {Koekemoer}, {Willmer}, {Petric}, {Furtak}, {Foo}, {Conselice}, {Coe}, {Driver}, {Grogin}, {Marshall}, {Nonino}, {Pirzkal}, {Robotham}, {Ryan}, \& {Tompkins}}]{kamieneski23}
{Kamieneski}, P.~S., {Frye}, B.~L., {Pascale}, M., {et~al.} 2023, \apj, 955, 91

\bibitem[{{Kartaltepe} {et~al.}(2023){Kartaltepe}, {Rose}, {Vanderhoof}, {McGrath}, {Costantin}, {Cox}, {Yung}, {Kocevski}, {Wuyts}, {Ferguson}, {Bagley}, {Finkelstein}, {Amor{\'\i}n}, {Andrews}, {Arrabal Haro}, {Backhaus}, {Behroozi}, {Bisigello}, {Calabr{\`o}}, {Casey}, {Coogan}, {Cooper}, {Croton}, {de la Vega}, {Dickinson}, {Fontana}, {Franco}, {Grazian}, {Grogin}, {Hathi}, {Holwerda}, {Huertas-Company}, {Iyer}, {Jogee}, {Jung}, {Kewley}, {Kirkpatrick}, {Koekemoer}, {Liu}, {Lotz}, {Lucas}, {Newman}, {Pacifici}, {Pandya}, {Papovich}, {Pentericci}, {P{\'e}rez-Gonz{\'a}lez}, {Petersen}, {Pirzkal}, {Rafelski}, {Ravindranath}, {Simons}, {Snyder}, {Somerville}, {Stanway}, {Straughn}, {Tacchella}, {Trump}, {Vega-Ferrero}, {Wilkins}, {Yang}, \& {Zavala}}]{kartaltepe23}
{Kartaltepe}, J.~S., {Rose}, C., {Vanderhoof}, B.~N., {et~al.} 2023, \apjl, 946, L15

\bibitem[{{Kennicutt}(1998)}]{kennicutt98}
{Kennicutt}, Robert~C., J. 1998, \araa, 36, 189

\bibitem[{{Knudsen} {et~al.}(2008){Knudsen}, {van der Werf}, \& {Kneib}}]{knudsen08}
{Knudsen}, K.~K., {van der Werf}, P.~P., \& {Kneib}, J.~P. 2008, \mnras, 384, 1611

\bibitem[{{Kocevski} {et~al.}(2024){Kocevski}, {Finkelstein}, {Barro}, {Taylor}, {Calabr{\`o}}, {Laloux}, {Buchner}, {Trump}, {Leung}, {Yang}, {Dickinson}, {P{\'e}rez-Gonz{\'a}lez}, {Pacucci}, {Inayoshi}, {Somerville}, {McGrath}, {Akins}, {Bagley}, {Bisigello}, {Bowler}, {Carnall}, {Casey}, {Cheng}, {Cleri}, {Costantin}, {Cullen}, {Davis}, {Donnan}, {Dunlop}, {Ellis}, {Ferguson}, {Fujimoto}, {Fontana}, {Giavalisco}, {Grazian}, {Grogin}, {Hathi}, {Hirschmann}, {Huertas-Company}, {Holwerda}, {Illingworth}, {Juneau}, {Kartaltepe}, {Koekemoer}, {Li}, {Lucas}, {Magee}, {Mason}, {McLeod}, {McLure}, {Napolitano}, {Papovich}, {Pirzkal}, {Rodighiero}, {Santini}, {Wilkins}, \& {Yung}}]{kocevski24}
{Kocevski}, D.~D., {Finkelstein}, S.~L., {Barro}, G., {et~al.} 2024, arXiv e-prints, arXiv:2404.03576

\bibitem[{{Kokorev} {et~al.}(2023){Kokorev}, {Fujimoto}, {Labbe}, {Greene}, {Bezanson}, {Dayal}, {Nelson}, {Atek}, {Brammer}, {Caputi}, {Chemerynska}, {Cutler}, {Feldmann}, {Fudamoto}, {Furtak}, {Goulding}, {de Graaff}, {Leja}, {Marchesini}, {Miller}, {Nanayakkara}, {Oesch}, {Pan}, {Price}, {Setton}, {Smit}, {Stefanon}, {Wang}, {Weaver}, {Whitaker}, {Williams}, \& {Zitrin}}]{kokorev23}
{Kokorev}, V., {Fujimoto}, S., {Labbe}, I., {et~al.} 2023, \apjl, 957, L7

\bibitem[{{Kriek} {et~al.}(2015){Kriek}, {Shapley}, {Reddy}, {Siana}, {Coil}, {Mobasher}, {Freeman}, {de Groot}, {Price}, {Sanders}, {Shivaei}, {Brammer}, {Momcheva}, {Skelton}, {van Dokkum}, {Whitaker}, {Aird}, {Azadi}, {Kassis}, {Bullock}, {Conroy}, {Dav{\'e}}, {Kere{\v{s}}}, \& {Krumholz}}]{kriek15}
{Kriek}, M., {Shapley}, A.~E., {Reddy}, N.~A., {et~al.} 2015, \apjs, 218, 15

\bibitem[{{Kroupa}(2001)}]{kroupa01}
{Kroupa}, P. 2001, \mnras, 322, 231

\bibitem[{{Kurk} {et~al.}(2013){Kurk}, {Cimatti}, {Daddi}, {Mignoli}, {Pozzetti}, {Dickinson}, {Bolzonella}, {Zamorani}, {Cassata}, {Rodighiero}, {Franceschini}, {Renzini}, {Rosati}, {Halliday}, \& {Berta}}]{kurk12}
{Kurk}, J., {Cimatti}, A., {Daddi}, E., {et~al.} 2013, \aap, 549, A63

\bibitem[{{Labbe} {et~al.}(2025){Labbe}, {Greene}, {Bezanson}, {Fujimoto}, {Furtak}, {Goulding}, {Matthee}, {Naidu}, {Oesch}, {Atek}, {Brammer}, {Chemerynska}, {Coe}, {Cutler}, {Dayal}, {Feldmann}, {Franx}, {Glazebrook}, {Leja}, {Maseda}, {Marchesini}, {Nanayakkara}, {Nelson}, {Pan}, {Papovich}, {Price}, {Suess}, {Wang}, {Weaver}, {Whitaker}, {Williams}, \& {Zitrin}}]{labbe23}
{Labbe}, I., {Greene}, J.~E., {Bezanson}, R., {et~al.} 2025, \apj, 978, 92

\bibitem[{{Le Bail} {et~al.}(2024){Le Bail}, {Daddi}, {Elbaz}, {Dickinson}, {Giavalisco}, {Magnelli}, {G{\'o}mez-Guijarro}, {Kalita}, {Koekemoer}, {Holwerda}, {Bournaud}, {de la Vega}, {Calabr{\`o}}, {Dekel}, {Cheng}, {Bisigello}, {Franco}, {Costantin}, {Lucas}, {P{\'e}rez-Gonz{\'a}lez}, {Lu}, {Wilkins}, {Arrabal Haro}, {Bagley}, {Finkelstein}, {Kartaltepe}, {Papovich}, {Pirzkal}, \& {Yung}}]{lebail24}
{Le Bail}, A., {Daddi}, E., {Elbaz}, D., {et~al.} 2024, \aap, 688, A53

\bibitem[{{Le F{\`e}vre} {et~al.}(2015){Le F{\`e}vre}, {Tasca}, {Cassata}, {Garilli}, {Le Brun}, {Maccagni}, {Pentericci}, {Thomas}, {Vanzella}, {Zamorani}, {Zucca}, {Amorin}, {Bardelli}, {Capak}, {Cassar{\`a}}, {Castellano}, {Cimatti}, {Cuby}, {Cucciati}, {de la Torre}, {Durkalec}, {Fontana}, {Giavalisco}, {Grazian}, {Hathi}, {Ilbert}, {Lemaux}, {Moreau}, {Paltani}, {Ribeiro}, {Salvato}, {Schaerer}, {Scodeggio}, {Sommariva}, {Talia}, {Taniguchi}, {Tresse}, {Vergani}, {Wang}, {Charlot}, {Contini}, {Fotopoulou}, {L{\'o}pez-Sanjuan}, {Mellier}, \& {Scoville}}]{lefevre15}
{Le F{\`e}vre}, O., {Tasca}, L.~A.~M., {Cassata}, P., {et~al.} 2015, \aap, 576, A79

\bibitem[{{Lim} {et~al.}(2020){Lim}, {Wang}, {Smail}, {Scott}, {Chen}, {Chang}, {Simpson}, {Toba}, {Shu}, {Clements}, {Greenslade}, {Ao}, {Babul}, {Birkin}, {Chapman}, {Cheng}, {Cho}, {Dannerbauer}, {Dudzevi{\v{c}}i{\={u}}t{\.{e}}}, {Dunlop}, {Gao}, {Goto}, {Ho}, {Hsu}, {Hwang}, {Jeong}, {Koprowski}, {Lee}, {Lin}, {Lin}, {Micha{\l}owski}, {Parsons}, {Sawicki}, {Shirley}, {Shim}, {Urquhart}, {Wang}, \& {Wang}}]{lim20}
{Lim}, C.-F., {Wang}, W.-H., {Smail}, I., {et~al.} 2020, \apj, 889, 80

\bibitem[{{Lo Faro} {et~al.}(2017){Lo Faro}, {Buat}, {Roehlly}, {Alvarez-Marquez}, {Burgarella}, {Silva}, \& {Efstathiou}}]{lofaro17}
{Lo Faro}, B., {Buat}, V., {Roehlly}, Y., {et~al.} 2017, \mnras, 472, 1372

\bibitem[{{Manning} {et~al.}(2022){Manning}, {Casey}, {Zavala}, {Magdis}, {Drew}, {Champagne}, {Aravena}, {B{\'e}thermin}, {Clements}, {Finkelstein}, {Fujimoto}, {Hayward}, {Hodge}, {Ilbert}, {Kartaltepe}, {Knudsen}, {Koekemoer}, {Man}, {Sanders}, {Sheth}, {Spilker}, {Staguhn}, {Talia}, {Treister}, \& {Yun}}]{manning22}
{Manning}, S.~M., {Casey}, C.~M., {Zavala}, J.~A., {et~al.} 2022, \apj, 925, 23

\bibitem[{{Matthee} {et~al.}(2024){Matthee}, {Naidu}, {Brammer}, {Chisholm}, {Eilers}, {Goulding}, {Greene}, {Kashino}, {Labbe}, {Lilly}, {Mackenzie}, {Oesch}, {Weibel}, {Wuyts}, {Xiao}, {Bordoloi}, {Bouwens}, {van Dokkum}, {Illingworth}, {Kramarenko}, {Maseda}, {Mason}, {Meyer}, {Nelson}, {Reddy}, {Shivaei}, {Simcoe}, \& {Yue}}]{matthee24}
{Matthee}, J., {Naidu}, R.~P., {Brammer}, G., {et~al.} 2024, \apj, 963, 129

\bibitem[{{McAlpine} {et~al.}(2019){McAlpine}, {Smail}, {Bower}, {Swinbank}, {Trayford}, {Theuns}, {Baes}, {Camps}, {Crain}, \& {Schaye}}]{mcalpine19}
{McAlpine}, S., {Smail}, I., {Bower}, R.~G., {et~al.} 2019, \mnras, 488, 2440

\bibitem[{{McKay} {et~al.}(2024){McKay}, {Barger}, \& {Cowie}}]{mckay24}
{McKay}, S.~J., {Barger}, A.~J., \& {Cowie}, L.~L. 2024, \apj, 962, 128

\bibitem[{{McKay} {et~al.}(2023){McKay}, {Barger}, {Cowie}, {Bauer}, \& {Rosenthal}}]{mckay23}
{McKay}, S.~J., {Barger}, A.~J., {Cowie}, L.~L., {Bauer}, F.~E., \& {Rosenthal}, M.~J.~N. 2023, \apj, 951, 48

\bibitem[{{McKinney} {et~al.}(2025){McKinney}, {Casey}, {Long}, {Cooper}, {Manning}, {Franco}, {Akins}, {Lambrides}, {Gammon}, {Silva}, {Gentile}, {Zavala}, {Amvrosiadis}, {Andika}, {Brinch}, {Champagne}, {Chartab}, {Drakos}, {Faisst}, {Fujimoto}, {Gillman}, {Gozaliasl}, {Greve}, {Harish}, {Hayward}, {Hirschmann}, {Ilbert}, {Kalita}, {Kartaltepe}, {Koekemoer}, {Kokorev}, {Liu}, {Magdis}, {McCracken}, {Rhodes}, {Robertson}, {Talia}, {Valentino}, \& {Vijayan}}]{mckinney24}
{McKinney}, J., {Casey}, C.~M., {Long}, A.~S., {et~al.} 2025, \apj, 979, 229

\bibitem[{{McLure} {et~al.}(2018){McLure}, {Pentericci}, {Cimatti}, {Dunlop}, {Elbaz}, {Fontana}, {Nandra}, {Amorin}, {Bolzonella}, {Bongiorno}, {Carnall}, {Castellano}, {Cirasuolo}, {Cucciati}, {Cullen}, {De Barros}, {Finkelstein}, {Fontanot}, {Franzetti}, {Fumana}, {Gargiulo}, {Garilli}, {Guaita}, {Hartley}, {Iovino}, {Jarvis}, {Juneau}, {Karman}, {Maccagni}, {Marchi}, {M{\'a}rmol-Queralt{\'o}}, {Pompei}, {Pozzetti}, {Scodeggio}, {Sommariva}, {Talia}, {Almaini}, {Balestra}, {Bardelli}, {Bell}, {Bourne}, {Bowler}, {Brusa}, {Buitrago}, {Caputi}, {Cassata}, {Charlot}, {Citro}, {Cresci}, {Cristiani}, {Curtis-Lake}, {Dickinson}, {Fazio}, {Ferguson}, {Fiore}, {Franco}, {Fynbo}, {Galametz}, {Georgakakis}, {Giavalisco}, {Grazian}, {Hathi}, {Jung}, {Kim}, {Koekemoer}, {Khusanova}, {Le F{\`e}vre}, {Lotz}, {Mannucci}, {Maltby}, {Matsuoka}, {McLeod}, {Mendez-Hernandez}, {Mendez-Abreu}, {Mignoli}, {Moresco}, {Mortlock}, {Nonino}, {Pannella}, {Papovich}, {Popesso}, {Rosario}, {Salvato}, {Santini}, {Schaerer},
  {Schreiber}, {Stark}, {Tasca}, {Thomas}, {Treu}, {Vanzella}, {Wild}, {Williams}, {Zamorani}, \& {Zucca}}]{mclure18}
{McLure}, R.~J., {Pentericci}, L., {Cimatti}, A., {et~al.} 2018, \mnras, 479, 25

\bibitem[{{Meyer} {et~al.}(2024){Meyer}, {Barrufet}, {Boogaard}, {Naidu}, {Oesch}, \& {Walter}}]{meyer24}
{Meyer}, R.~A., {Barrufet}, L., {Boogaard}, L.~A., {et~al.} 2024, \aap, 681, L3

\bibitem[{{Micha{\l}owski} {et~al.}(2014){Micha{\l}owski}, {Hayward}, {Dunlop}, {Bruce}, {Cirasuolo}, {Cullen}, \& {Hernquist}}]{michalowski14}
{Micha{\l}owski}, M.~J., {Hayward}, C.~C., {Dunlop}, J.~S., {et~al.} 2014, \aap, 571, A75

\bibitem[{{Mignoli} {et~al.}(2005){Mignoli}, {Cimatti}, {Zamorani}, {Pozzetti}, {Daddi}, {Renzini}, {Broadhurst}, {Cristiani}, {D'Odorico}, {Fontana}, {Giallongo}, {Gilmozzi}, {Menci}, \& {Saracco}}]{mignoli05}
{Mignoli}, M., {Cimatti}, A., {Zamorani}, G., {et~al.} 2005, \aap, 437, 883

\bibitem[{{Mihos} \& {Hernquist}(1996)}]{mihos96}
{Mihos}, J.~C., \& {Hernquist}, L. 1996, \apj, 464, 641

\bibitem[{{Momcheva} {et~al.}(2016){Momcheva}, {Brammer}, {van Dokkum}, {Skelton}, {Whitaker}, {Nelson}, {Fumagalli}, {Maseda}, {Leja}, {Franx}, {Rix}, {Bezanson}, {Da Cunha}, {Dickey}, {F{\"o}rster Schreiber}, {Illingworth}, {Kriek}, {Labb{\'e}}, {Ulf Lange}, {Lundgren}, {Magee}, {Marchesini}, {Oesch}, {Pacifici}, {Patel}, {Price}, {Tal}, {Wake}, {van der Wel}, \& {Wuyts}}]{momcheva16}
{Momcheva}, I.~G., {Brammer}, G.~B., {van Dokkum}, P.~G., {et~al.} 2016, \apjs, 225, 27

\bibitem[{{Mu{\~n}oz Arancibia} {et~al.}(2023){Mu{\~n}oz Arancibia}, {Gonz{\'a}lez-L{\'o}pez}, {Ibar}, {Bauer}, {Anguita}, {Aravena}, {Demarco}, {Kneissl}, {Koekemoer}, {Troncoso-Iribarren}, \& {Zitrin}}]{munoz-arancibia23}
{Mu{\~n}oz Arancibia}, A.~M., {Gonz{\'a}lez-L{\'o}pez}, J., {Ibar}, E., {et~al.} 2023, \aap, 675, A85

\bibitem[{{Naidu} {et~al.}(2022){Naidu}, {Oesch}, {Setton}, {Matthee}, {Conroy}, {Johnson}, {Weaver}, {Bouwens}, {Brammer}, {Dayal}, {Illingworth}, {Barrufet}, {Belli}, {Bezanson}, {Bose}, {Heintz}, {Leja}, {Leonova}, {Marques-Chaves}, {Stefanon}, {Toft}, {van der Wel}, {van Dokkum}, {Weibel}, \& {Whitaker}}]{naidu22}
{Naidu}, R.~P., {Oesch}, P.~A., {Setton}, D.~J., {et~al.} 2022, arXiv e-prints, arXiv:2208.02794

\bibitem[{{Naidu} {et~al.}(2024){Naidu}, {Matthee}, {Kramarenko}, {Weibel}, {Brammer}, {Oesch}, {Lechner}, {Furtak}, {Di Cesare}, {Torralba}, {Kotiwale}, {Bezanson}, {Bouwens}, {Chandra}, {Claeyssens}, {Danhaive}, {Frebel}, {de Graaff}, {Greene}, {Heintz}, {Ji}, {Kashino}, {Katz}, {Labbe}, {Leja}, {Li}, {Maseda}, {Richard}, {Shivaei}, {Simcoe}, {Sobral}, {Suess}, {Tacchella}, \& {Williams}}]{naidu24}
{Naidu}, R.~P., {Matthee}, J., {Kramarenko}, I., {et~al.} 2024, arXiv e-prints, arXiv:2410.01874

\bibitem[{{Narayanan} {et~al.}(2010){Narayanan}, {Hayward}, {Cox}, {Hernquist}, {Jonsson}, {Younger}, \& {Groves}}]{narayanan10}
{Narayanan}, D., {Hayward}, C.~C., {Cox}, T.~J., {et~al.} 2010, \mnras, 401, 1613

\bibitem[{{Nardiello} {et~al.}(2022){Nardiello}, {Bedin}, {Burgasser}, {Salaris}, {Cassisi}, {Griggio}, \& {Scalco}}]{nardiello22}
{Nardiello}, D., {Bedin}, L.~R., {Burgasser}, A., {et~al.} 2022, \mnras, 517, 484

\bibitem[{{Nedkova} {et~al.}(2024){Nedkova}, {Rafelski}, {Teplitz}, {Mehta}, {Degroot}, {Ravindranath}, {Alavi}, {Beckett}, {Grogin}, {H{\"a}u{\ss}ler}, {Koekemoer}, {Oyarz{\'u}n}, {Prichard}, {Revalski}, {Snyder}, {Sunnquist}, {Wang}, {Windhorst}, {Chartab}, {Conselice}, {Guo}, {Hathi}, {Hayes}, {Ji}, {Kim}, {Lucas}, {Mobasher}, {O'Connell}, {Sattari}, {Smith}, {Taamoli}, {Yung}, \& {The Uvcandels Team}}]{nedkova24}
{Nedkova}, K.~V., {Rafelski}, M., {Teplitz}, H.~I., {et~al.} 2024, \apj, 970, 188

\bibitem[{{Nelson} {et~al.}(2023){Nelson}, {Suess}, {Bezanson}, {Price}, {van Dokkum}, {Leja}, {Wang}, {Whitaker}, {Labb{\'e}}, {Barrufet}, {Brammer}, {Eisenstein}, {Gibson}, {Hartley}, {Johnson}, {Heintz}, {Mathews}, {Miller}, {Oesch}, {Sandles}, {Setton}, {Speagle}, {Tacchella}, {Tadaki}, {{\"U}bler}, \& {Weaver}}]{nelson23}
{Nelson}, E.~J., {Suess}, K.~A., {Bezanson}, R., {et~al.} 2023, \apjl, 948, L18

\bibitem[{{Oesch} {et~al.}(2023){Oesch}, {Brammer}, {Naidu}, {Bouwens}, {Chisholm}, {Illingworth}, {Matthee}, {Nelson}, {Qin}, {Reddy}, {Shapley}, {Shivaei}, {van Dokkum}, {Weibel}, {Whitaker}, {Wuyts}, {Covelo-Paz}, {Endsley}, {Fudamoto}, {Giovinazzo}, {Herard-Demanche}, {Kerutt}, {Kramarenko}, {Labbe}, {Leonova}, {Lin}, {Magee}, {Marchesini}, {Maseda}, {Mason}, {Matharu}, {Meyer}, {Neufeld}, {Prieto Lyon}, {Schaerer}, {Sharma}, {Shuntov}, {Smit}, {Stefanon}, {Wyithe}, \& {Xiao}}]{oesch23}
{Oesch}, P.~A., {Brammer}, G., {Naidu}, R.~P., {et~al.} 2023, \mnras, 525, 2864

\bibitem[{{Oke} \& {Gunn}(1983)}]{oke83}
{Oke}, J.~B., \& {Gunn}, J.~E. 1983, \apj, 266, 713

\bibitem[{{Oliver} {et~al.}(2012){Oliver}, {Bock}, {Altieri}, {Amblard}, {Arumugam}, {Aussel}, {Babbedge}, {Beelen}, {B{\'e}thermin}, {Blain}, {Boselli}, {Bridge}, {Brisbin}, {Buat}, {Burgarella}, {Castro-Rodr{\'\i}guez}, {Cava}, {Chanial}, {Cirasuolo}, {Clements}, {Conley}, {Conversi}, {Cooray}, {Dowell}, {Dubois}, {Dwek}, {Dye}, {Eales}, {Elbaz}, {Farrah}, {Feltre}, {Ferrero}, {Fiolet}, {Fox}, {Franceschini}, {Gear}, {Giovannoli}, {Glenn}, {Gong}, {Gonz{\'a}lez Solares}, {Griffin}, {Halpern}, {Harwit}, {Hatziminaoglou}, {Heinis}, {Hurley}, {Hwang}, {Hyde}, {Ibar}, {Ilbert}, {Isaak}, {Ivison}, {Lagache}, {Le Floc'h}, {Levenson}, {Faro}, {Lu}, {Madden}, {Maffei}, {Magdis}, {Mainetti}, {Marchetti}, {Marsden}, {Marshall}, {Mortier}, {Nguyen}, {O'Halloran}, {Omont}, {Page}, {Panuzzo}, {Papageorgiou}, {Patel}, {Pearson}, {P{\'e}rez-Fournon}, {Pohlen}, {Rawlings}, {Raymond}, {Rigopoulou}, {Riguccini}, {Rizzo}, {Rodighiero}, {Roseboom}, {Rowan-Robinson}, {S{\'a}nchez Portal}, {Schulz}, {Scott}, {Seymour}, {Shupe},
  {Smith}, {Stevens}, {Symeonidis}, {Trichas}, {Tugwell}, {Vaccari}, {Valtchanov}, {Vieira}, {Viero}, {Vigroux}, {Wang}, {Ward}, {Wardlow}, {Wright}, {Xu}, \& {Zemcov}}]{oliver12}
{Oliver}, S.~J., {Bock}, J., {Altieri}, B., {et~al.} 2012, \mnras, 424, 1614

\bibitem[{{Ormerod} {et~al.}(2024){Ormerod}, {Conselice}, {Adams}, {Harvey}, {Austin}, {Trussler}, {Ferreira}, {Caruana}, {Lucatelli}, {Li}, \& {Roper}}]{ormerod24}
{Ormerod}, K., {Conselice}, C.~J., {Adams}, N.~J., {et~al.} 2024, \mnras, 527, 6110

\bibitem[{{Pacifici} {et~al.}(2023){Pacifici}, {Iyer}, {Mobasher}, {da Cunha}, {Acquaviva}, {Burgarella}, {Calistro Rivera}, {Carnall}, {Chang}, {Chartab}, {Cooke}, {Fairhurst}, {Kartaltepe}, {Leja}, {Ma{\l}ek}, {Salmon}, {Torelli}, {Vidal-Garc{\'\i}a}, {Boquien}, {Brammer}, {Brown}, {Capak}, {Chevallard}, {Circosta}, {Croton}, {Davidzon}, {Dickinson}, {Duncan}, {Faber}, {Ferguson}, {Fontana}, {Guo}, {Haeussler}, {Hemmati}, {Jafariyazani}, {Kassin}, {Larson}, {Lee}, {Mantha}, {Marchi}, {Nayyeri}, {Newman}, {Pandya}, {Pforr}, {Reddy}, {Sanders}, {Shah}, {Shahidi}, {Stevans}, {Triani}, {Tyler}, {Vanderhoof}, {de la Vega}, {Wang}, \& {Weston}}]{pacifici23}
{Pacifici}, C., {Iyer}, K.~G., {Mobasher}, B., {et~al.} 2023, \apj, 944, 141

\bibitem[{{Padilla} \& {Strauss}(2008)}]{padilla08}
{Padilla}, N.~D., \& {Strauss}, M.~A. 2008, \mnras, 388, 1321

\bibitem[{{Paris} {et~al.}(2023){Paris}, {Merlin}, {Fontana}, {Bonchi}, {Brammer}, {Correnti}, {Treu}, {Boyett}, {Calabr{\`o}}, {Castellano}, {Chen}, {Yang}, {Glazebrook}, {Kelly}, {Koekemoer}, {Leethochawalit}, {Mascia}, {Mason}, {Morishita}, {Nonino}, {Pentericci}, {Polenta}, {Roberts-Borsani}, {Santini}, {Trenti}, {Vanzella}, {Vulcani}, {Windhorst}, {Nanayakkara}, \& {Wang}}]{paris23}
{Paris}, D., {Merlin}, E., {Fontana}, A., {et~al.} 2023, \apj, 952, 20

\bibitem[{{Peng} {et~al.}(2002){Peng}, {Ho}, {Impey}, \& {Rix}}]{peng02}
{Peng}, C.~Y., {Ho}, L.~C., {Impey}, C.~D., \& {Rix}, H.-W. 2002, \aj, 124, 266

\bibitem[{{Peng} {et~al.}(2010){Peng}, {Ho}, {Impey}, \& {Rix}}]{peng10}
---. 2010, \aj, 139, 2097

\bibitem[{{Pentericci} {et~al.}(2018){Pentericci}, {McLure}, {Garilli}, {Cucciati}, {Franzetti}, {Iovino}, {Amorin}, {Bolzonella}, {Bongiorno}, {Carnall}, {Castellano}, {Cimatti}, {Cirasuolo}, {Cullen}, {De Barros}, {Dunlop}, {Elbaz}, {Finkelstein}, {Fontana}, {Fontanot}, {Fumana}, {Gargiulo}, {Guaita}, {Hartley}, {Jarvis}, {Juneau}, {Karman}, {Maccagni}, {Marchi}, {Marmol-Queralto}, {Nandra}, {Pompei}, {Pozzetti}, {Scodeggio}, {Sommariva}, {Talia}, {Almaini}, {Balestra}, {Bardelli}, {Bell}, {Bourne}, {Bowler}, {Brusa}, {Buitrago}, {Caputi}, {Cassata}, {Charlot}, {Citro}, {Cresci}, {Cristiani}, {Curtis-Lake}, {Dickinson}, {Fazio}, {Ferguson}, {Fiore}, {Franco}, {Fynbo}, {Galametz}, {Georgakakis}, {Giavalisco}, {Grazian}, {Hathi}, {Jung}, {Kim}, {Koekemoer}, {Khusanova}, {Le F{\`e}vre}, {Lotz}, {Mannucci}, {Maltby}, {Matsuoka}, {McLeod}, {Mendez-Hernandez}, {Mendez-Abreu}, {Mignoli}, {Moresco}, {Mortlock}, {Nonino}, {Pannella}, {Papovich}, {Popesso}, {Rosario}, {Salvato}, {Santini}, {Schaerer}, {Schreiber},
  {Stark}, {Tasca}, {Thomas}, {Treu}, {Vanzella}, {Wild}, {Williams}, {Zamorani}, \& {Zucca}}]{pentericci18}
{Pentericci}, L., {McLure}, R.~J., {Garilli}, B., {et~al.} 2018, \aap, 616, A174

\bibitem[{{P{\'e}rez-Gonz{\'a}lez} {et~al.}(2023){P{\'e}rez-Gonz{\'a}lez}, {Barro}, {Annunziatella}, {Costantin}, {Garc{\'\i}a-Argum{\'a}nez}, {McGrath}, {M{\'e}rida}, {Zavala}, {Arrabal Haro}, {Bagley}, {Backhaus}, {Behroozi}, {Bell}, {Bisigello}, {Buat}, {Calabr{\`o}}, {Casey}, {Cleri}, {Coogan}, {Cooper}, {Cooray}, {Dekel}, {Dickinson}, {Elbaz}, {Ferguson}, {Finkelstein}, {Fontana}, {Franco}, {Gardner}, {Giavalisco}, {G{\'o}mez-Guijarro}, {Grazian}, {Grogin}, {Guo}, {Huertas-Company}, {Jogee}, {Kartaltepe}, {Kewley}, {Kirkpatrick}, {Kocevski}, {Koekemoer}, {Long}, {Lotz}, {Lucas}, {Papovich}, {Pirzkal}, {Ravindranath}, {Somerville}, {Tacchella}, {Trump}, {Wang}, {Wilkins}, {Wuyts}, {Yang}, \& {Yung}}]{perez-gonzalez23}
{P{\'e}rez-Gonz{\'a}lez}, P.~G., {Barro}, G., {Annunziatella}, M., {et~al.} 2023, \apjl, 946, L16

\bibitem[{{P{\'e}rez-Gonz{\'a}lez} {et~al.}(2024){P{\'e}rez-Gonz{\'a}lez}, {Rinaldi}, {Caputi}, {{\'A}lvarez-M{\'a}rquez}, {Annunziatella}, {Langeroodi}, {Moutard}, {Boogaard}, {Iani}, {Melinder}, {Costantin}, {{\"O}stlin}, {Colina}, {Greve}, {Wright}, {Alonso-Herrero}, {Bik}, {Bosman}, {Crespo G{\'o}mez}, {Dicken}, {Eckart}, {Garc{\'\i}a-Mar{\'\i}n}, {Gillman}, {G{\"u}del}, {Henning}, {Hjorth}, {Jermann}, {Labiano}, {Meyer}, {Pei{\ensuremath{\beta}}ker}, {Pye}, {Ray}, {Tikkanen}, {Walter}, \& {van der Werf}}]{perez-gonzalez24}
{P{\'e}rez-Gonz{\'a}lez}, P.~G., {Rinaldi}, P., {Caputi}, K.~I., {et~al.} 2024, \apjl, 969, L10

\bibitem[{{Polletta} {et~al.}(2024){Polletta}, {Frye}, {Garuda}, {Willner}, {Berta}, {Kneissl}, {Dole}, {Jansen}, {Lehnert}, {Cohen}, {Summers}, {Windhorst}, {D'Silva}, {Koekemoer}, {Coe}, {Conselice}, {Driver}, {Grogin}, {Marshall}, {Nonino}, {Ortiz}, {Pirzkal}, {Robotham}, {Ryan}, {Willmer}, {Yan}, {Arumugam}, {Cheng}, {Gim}, {Hathi}, {Holwerda}, {Kamieneski}, {Keel}, {Li}, {Pascale}, {Rottgering}, {Smith}, \& {Yun}}]{polletta24}
{Polletta}, M., {Frye}, B.~L., {Garuda}, N., {et~al.} 2024, \aap, 690, A285

\bibitem[{{Pope} {et~al.}(2005){Pope}, {Borys}, {Scott}, {Conselice}, {Dickinson}, \& {Mobasher}}]{pope05}
{Pope}, A., {Borys}, C., {Scott}, D., {et~al.} 2005, \mnras, 358, 149

\bibitem[{{Price} {et~al.}(2025{\natexlab{a}}){Price}, {Bezanson}, {Labbe}, {Furtak}, {de Graaff}, {Greene}, {Kokorev}, {Setton}, {Suess}, {Brammer}, {Cutler}, {Leja}, {Pan}, {Wang}, {Weaver}, {Whitaker}, {Atek}, {Burgasser}, {Chemerynska}, {Dayal}, {Feldmann}, {F{\"o}rster Schreiber}, {Fudamoto}, {Fujimoto}, {Glazebrook}, {Goulding}, {Khullar}, {Kriek}, {Marchesini}, {Maseda}, {Miller}, {Muzzin}, {Nanayakkara}, {Nelson}, {Oesch}, {Shipley}, {Smit}, {Taylor}, {Dokkum}, {Williams}, \& {Zitrin}}]{price24}
{Price}, S.~H., {Bezanson}, R., {Labbe}, I., {et~al.} 2025{\natexlab{a}}, \apj, 982, 51

\bibitem[{{Price} {et~al.}(2025{\natexlab{b}}){Price}, {Suess}, {Williams}, {Bezanson}, {Khullar}, {Nelson}, {Wang}, {Weaver}, {Fujimoto}, {Kokorev}, {Greene}, {Brammer}, {Cutler}, {Dayal}, {Furtak}, {Labbe}, {Leja}, {Miller}, {Nanayakkara}, {Pan}, \& {Whitaker}}]{price23}
{Price}, S.~H., {Suess}, K.~A., {Williams}, C.~C., {et~al.} 2025{\natexlab{b}}, \apj, 980, 11

\bibitem[{{Ren} {et~al.}(2025){Ren}, {Liu}, {Li}, {Zhao}, {Cui}, {Song}, {Li}, {Mo}, {Yesuf}, {Wang}, {An}, \& {Zheng}}]{ren25}
{Ren}, J., {Liu}, F.~S., {Li}, N., {et~al.} 2025, \apj, 982, 200

\bibitem[{{Rieke, Marcia} {et~al.}(2023){Rieke, Marcia}, {Robertson, Brant}, {Tacchella, Sandro}, {Willmer, Christopher}, {Johnson, Ben}, {Carniani, Stefano}, {Bunker, Andy}, \& {Willott, Chris}}]{10.17909/8tdj-8n28}
{Rieke, Marcia}, {Robertson, Brant}, {Tacchella, Sandro}, {et~al.} 2023, Data from the JWST Advanced Deep Extragalactic Survey (JADES),  STScI/MAST, \dodoi{10.17909/8TDJ-8N28}

\bibitem[{{Roper} {et~al.}(2022){Roper}, {Lovell}, {Vijayan}, {Marshall}, {Irodotou}, {Kuusisto}, {Thomas}, \& {Wilkins}}]{roper22}
{Roper}, W.~J., {Lovell}, C.~C., {Vijayan}, A.~P., {et~al.} 2022, \mnras, 514, 1921

\bibitem[{{Sanders} \& {Mirabel}(1996)}]{sanders96}
{Sanders}, D.~B., \& {Mirabel}, I.~F. 1996, \araa, 34, 749

\bibitem[{{Smail} {et~al.}(1997){Smail}, {Ivison}, \& {Blain}}]{smail97}
{Smail}, I., {Ivison}, R.~J., \& {Blain}, A.~W. 1997, \apjl, 490, L5

\bibitem[{{Smol{\v{c}}i{\'c}} {et~al.}(2012){Smol{\v{c}}i{\'c}}, {Aravena}, {Navarrete}, {Schinnerer}, {Riechers}, {Bertoldi}, {Feruglio}, {Finoguenov}, {Salvato}, {Sargent}, {McCracken}, {Albrecht}, {Karim}, {Capak}, {Carilli}, {Cappelluti}, {Elvis}, {Ilbert}, {Kartaltepe}, {Lilly}, {Sanders}, {Sheth}, {Scoville}, \& {Taniguchi}}]{smolcic12}
{Smol{\v{c}}i{\'c}}, V., {Aravena}, M., {Navarrete}, F., {et~al.} 2012, \aap, 548, A4

\bibitem[{{Stach} {et~al.}(2019){Stach}, {Dudzevi{\v{c}}i{\={u}}t{\.{e}}}, {Smail}, {Swinbank}, {Geach}, {Simpson}, {An}, {Almaini}, {Arumugam}, {Blain}, {Chapman}, {Chen}, {Conselice}, {Cooke}, {Coppin}, {da Cunha}, {Dunlop}, {Farrah}, {Gullberg}, {Hodge}, {Ivison}, {Kocevski}, {Micha{\l}owski}, {Miyaji}, {Scott}, {Thomson}, {Wardlow}, {Weiss}, \& {van der Werf}}]{stach19}
{Stach}, S.~M., {Dudzevi{\v{c}}i{\={u}}t{\.{e}}}, U., {Smail}, I., {et~al.} 2019, \mnras, 487, 4648

\bibitem[{{Sun} {et~al.}(2022){Sun}, {Egami}, {Fujimoto}, {Rawle}, {Bauer}, {Kohno}, {Smail}, {P{\'e}rez-Gonz{\'a}lez}, {Ao}, {Chapman}, {Combes}, {Dessauges-Zavadsky}, {Espada}, {Gonz{\'a}lez-L{\'o}pez}, {Koekemoer}, {Kokorev}, {Lee}, {Morokuma-Matsui}, {Mu{\~n}oz Arancibia}, {Oguri}, {Pell{\'o}}, {Ueda}, {Uematsu}, {Valentino}, {Van der Werf}, {Walth}, {Zemcov}, \& {Zitrin}}]{sun22}
{Sun}, F., {Egami}, E., {Fujimoto}, S., {et~al.} 2022, \apj, 932, 77

\bibitem[{{Sun} {et~al.}(2024{\natexlab{a}}){Sun}, {Helton}, {Egami}, {Hainline}, {Rieke}, {Willmer}, {Eisenstein}, {Johnson}, {Rieke}, {Robertson}, {Tacchella}, {Alberts}, {Baker}, {Bhatawdekar}, {Boyett}, {Bunker}, {Charlot}, {Chen}, {Chevallard}, {Curtis-Lake}, {Danhaive}, {DeCoursey}, {Ji}, {Lyu}, {Maiolino}, {Rujopakarn}, {Sandles}, {Shivaei}, {{\"U}bler}, {Willott}, \& {Witstok}}]{sun24}
{Sun}, F., {Helton}, J.~M., {Egami}, E., {et~al.} 2024{\natexlab{a}}, \apj, 961, 69

\bibitem[{{Sun} {et~al.}(2025){Sun}, {Wang}, {Yang}, {Champagne}, {Decarli}, {Fan}, {Ba{\~n}ados}, {Cai}, {Colina}, {Egami}, {Hennawi}, {Jin}, {Jun}, {Khusanova}, {Li}, {Li}, {Lin}, {Liu}, {Meyer}, {Pudoka}, {Rieke}, {Shen}, {Tee}, {Venemans}, {Walter}, {Wu}, {Zhang}, \& {Zou}}]{sun24c}
{Sun}, F., {Wang}, F., {Yang}, J., {et~al.} 2025, \apj, 980, 12

\bibitem[{{Sun} {et~al.}(2024{\natexlab{b}}){Sun}, {Ho}, {Zhuang}, {Ma}, {Chen}, \& {Li}}]{sun24a}
{Sun}, W., {Ho}, L.~C., {Zhuang}, M.-Y., {et~al.} 2024{\natexlab{b}}, \apj, 960, 104

\bibitem[{{Suzuki} {et~al.}(2023){Suzuki}, {van Mierlo}, \& {Caputi}}]{suzuki24}
{Suzuki}, T.~L., {van Mierlo}, S.~E., \& {Caputi}, K.~I. 2023, \apj, 959, 82

\bibitem[{{Swinbank} {et~al.}(2010){Swinbank}, {Smail}, {Chapman}, {Borys}, {Alexander}, {Blain}, {Conselice}, {Hainline}, \& {Ivison}}]{swinbank10}
{Swinbank}, A.~M., {Smail}, I., {Chapman}, S.~C., {et~al.} 2010, \mnras, 405, 234

\bibitem[{{Tan} {et~al.}(2024){Tan}, {Daddi}, {Magnelli}, {Correa}, {Bournaud}, {Adscheid}, {Zhang}, {Elbaz}, {G{\'o}mez-Guijarro}, {Kalita}, {Liu}, {Liu}, {Pety}, {Puglisi}, {Schinnerer}, {Silverman}, \& {Valentino}}]{tan24}
{Tan}, Q.-H., {Daddi}, E., {Magnelli}, B., {et~al.} 2024, \nat, 636, 69

\bibitem[{{Targett} {et~al.}(2013){Targett}, {Dunlop}, {Cirasuolo}, {McLure}, {Bruce}, {Fontana}, {Galametz}, {Paris}, {Dav{\'e}}, {Dekel}, {Faber}, {Ferguson}, {Grogin}, {Kartaltepe}, {Kocevski}, {Koekemoer}, {Kurczynski}, {Lai}, \& {Lotz}}]{targett13}
{Targett}, T.~A., {Dunlop}, J.~S., {Cirasuolo}, M., {et~al.} 2013, \mnras, 432, 2012

\bibitem[{{Taylor} {et~al.}(2024){Taylor}, {Finkelstein}, {Kocevski}, {Jeon}, {Bromm}, {Amorin}, {Arrabal Haro}, {Backhaus}, {Bagley}, {Ba{\~n}ados}, {Bhatawdekar}, {Brooks}, {Calabro}, {Chavez Ortiz}, {Cheng}, {Cleri}, {Cole}, {Davis}, {Dickinson}, {Donnan}, {Dunlop}, {Ellis}, {Fernandez}, {Fontana}, {Fujimoto}, {Giavalisco}, {Grazian}, {Guo}, {Hathi}, {Holwerda}, {Hirschmann}, {Inayoshi}, {Kartaltepe}, {Khusanova}, {Koekemoer}, {Kokorev}, {Larson}, {Leung}, {Lucas}, {McLeod}, {Napolitano}, {Onoue}, {Pacucci}, {Papovich}, {P{\'e}rez-Gonz{\'a}lez}, {Pirzkal}, {Somerville}, {Trump}, {Wilkins}, {Yung}, \& {Zhang}}]{taylor24}
{Taylor}, A.~J., {Finkelstein}, S.~L., {Kocevski}, D.~D., {et~al.} 2024, arXiv e-prints, arXiv:2409.06772

\bibitem[{{Treu} {et~al.}(2022){Treu}, {Roberts-Borsani}, {Bradac}, {Brammer}, {Fontana}, {Henry}, {Mason}, {Morishita}, {Pentericci}, {Wang}, {Acebron}, {Bagley}, {Bergamini}, {Belfiori}, {Bonchi}, {Boyett}, {Boutsia}, {Calabr{\'o}}, {Caminha}, {Castellano}, {Dressler}, {Glazebrook}, {Grillo}, {Jacobs}, {Jones}, {Kelly}, {Leethochawalit}, {Malkan}, {Marchesini}, {Mascia}, {Mercurio}, {Merlin}, {Nanayakkara}, {Nonino}, {Paris}, {Poggianti}, {Rosati}, {Santini}, {Scarlata}, {Shipley}, {Strait}, {Trenti}, {Tubthong}, {Vanzella}, {Vulcani}, \& {Yang}}]{treu22}
{Treu}, T., {Roberts-Borsani}, G., {Bradac}, M., {et~al.} 2022, \apj, 935, 110

\bibitem[{{Treu, Tommaso} \& {Paris, Diego}(2023)}]{10.17909/kw3c-n857}
{Treu, Tommaso}, \& {Paris, Diego}. 2023, High-level science products produced by the GLASS-JWST team, as described in Paris, et al. 2024 (ADS bibcode: 2023ApJ...952...20P), and in Mascia, et al. 2024 (ADS bibcode: 2024A\&A...690A...2M).,  STScI/MAST, \dodoi{10.17909/KW3C-N857}

\bibitem[{{Uematsu} {et~al.}(2024){Uematsu}, {Ueda}, {Kohno}, {Toba}, {Yamada}, {Smail}, {Umehata}, {Fujimoto}, {Hatsukade}, {Ao}, {Bauer}, {Brammer}, {Dessauges-Zavadsky}, {Espada}, {Jolly}, {Koekemoer}, {Kokorev}, {Magdis}, {Oguri}, \& {Sun}}]{uematsu24}
{Uematsu}, R., {Ueda}, Y., {Kohno}, K., {et~al.} 2024, \apj, 965, 108

\bibitem[{{van der Wel} {et~al.}(2012){van der Wel}, {Bell}, {H{\"a}ussler}, {McGrath}, {Chang}, {Guo}, {McIntosh}, {Rix}, {Barden}, {Cheung}, {Faber}, {Ferguson}, {Galametz}, {Grogin}, {Hartley}, {Kartaltepe}, {Kocevski}, {Koekemoer}, {Lotz}, {Mozena}, {Peth}, \& {Peng}}]{vanderwel12}
{van der Wel}, A., {Bell}, E.~F., {H{\"a}ussler}, B., {et~al.} 2012, \apjs, 203, 24

\bibitem[{{van der Wel} {et~al.}(2014){van der Wel}, {Chang}, {Bell}, {Holden}, {Ferguson}, {Giavalisco}, {Rix}, {Skelton}, {Whitaker}, {Momcheva}, {Brammer}, {Kassin}, {Martig}, {Dekel}, {Ceverino}, {Koo}, {Mozena}, {van Dokkum}, {Franx}, {Faber}, \& {Primack}}]{vanderwel14}
{van der Wel}, A., {Chang}, Y.-Y., {Bell}, E.~F., {et~al.} 2014, \apjl, 792, L6

\bibitem[{{Vanzella} {et~al.}(2008){Vanzella}, {Cristiani}, {Dickinson}, {Giavalisco}, {Kuntschner}, {Haase}, {Nonino}, {Rosati}, {Cesarsky}, {Ferguson}, {Fosbury}, {Grazian}, {Moustakas}, {Rettura}, {Popesso}, {Renzini}, {Stern}, \& {GOODS Team}}]{vanzella08}
{Vanzella}, E., {Cristiani}, S., {Dickinson}, M., {et~al.} 2008, \aap, 478, 83

\bibitem[{Virtanen {et~al.}(2020)Virtanen, Gommers, Oliphant, Haberland, Reddy, Cournapeau, Burovski, Peterson, Weckesser, Bright, {van der Walt}, Brett, Wilson, Millman, Mayorov, Nelson, Jones, Kern, Larson, Carey, Polat, Feng, Moore, {VanderPlas}, Laxalde, Perktold, Cimrman, Henriksen, Quintero, Harris, Archibald, Ribeiro, Pedregosa, {van Mulbregt}, \& {SciPy 1.0 Contributors}}]{scipy20}
Virtanen, P., Gommers, R., Oliphant, T.~E., {et~al.} 2020, Nature Methods, 17, 261

\bibitem[{{Walter} {et~al.}(2016){Walter}, {Decarli}, {Aravena}, {Carilli}, {Bouwens}, {da Cunha}, {Daddi}, {Ivison}, {Riechers}, {Smail}, {Swinbank}, {Weiss}, {Anguita}, {Assef}, {Bacon}, {Bauer}, {Bell}, {Bertoldi}, {Chapman}, {Colina}, {Cortes}, {Cox}, {Dickinson}, {Elbaz}, {G{\'o}nzalez-L{\'o}pez}, {Ibar}, {Inami}, {Infante}, {Hodge}, {Karim}, {Le Fevre}, {Magnelli}, {Neri}, {Oesch}, {Ota}, {Popping}, {Rix}, {Sargent}, {Sheth}, {van der Wel}, {van der Werf}, \& {Wagg}}]{walter16}
{Walter}, F., {Decarli}, R., {Aravena}, M., {et~al.} 2016, \apj, 833, 67

\bibitem[{{Wang} {et~al.}(2016){Wang}, {Elbaz}, {Schreiber}, {Pannella}, {Shu}, {Willner}, {Ashby}, {Huang}, {Fontana}, {Dekel}, {Daddi}, {Ferguson}, {Dunlop}, {Ciesla}, {Koekemoer}, {Giavalisco}, {Boutsia}, {Finkelstein}, {Juneau}, {Barro}, {Koo}, {Micha{\l}owski}, {Orellana}, {Lu}, {Castellano}, {Bourne}, {Buitrago}, {Santini}, {Faber}, {Hathi}, {Lucas}, \& {P{\'e}rez-Gonz{\'a}lez}}]{wang16}
{Wang}, T., {Elbaz}, D., {Schreiber}, C., {et~al.} 2016, \apj, 816, 84

\bibitem[{{Wang} {et~al.}(2019){Wang}, {Schreiber}, {Elbaz}, {Yoshimura}, {Kohno}, {Shu}, {Yamaguchi}, {Pannella}, {Franco}, {Huang}, {Lim}, \& {Wang}}]{wang19}
{Wang}, T., {Schreiber}, C., {Elbaz}, D., {et~al.} 2019, \nat, 572, 211

\bibitem[{{Wang} {et~al.}(2012){Wang}, {Barger}, \& {Cowie}}]{wang12}
{Wang}, W.-H., {Barger}, A.~J., \& {Cowie}, L.~L. 2012, \apj, 744, 155

\bibitem[{{Weaver} {et~al.}(2024){Weaver}, {Cutler}, {Pan}, {Whitaker}, {Labb{\'e}}, {Price}, {Bezanson}, {Brammer}, {Marchesini}, {Leja}, {Wang}, {Furtak}, {Zitrin}, {Atek}, {Chemerynska}, {Coe}, {Dayal}, {van Dokkum}, {Feldmann}, {F{\"o}rster Schreiber}, {Franx}, {Fujimoto}, {Fudamoto}, {Glazebrook}, {de Graaff}, {Greene}, {Juneau}, {Kassin}, {Kriek}, {Khullar}, {Maseda}, {Mowla}, {Muzzin}, {Nanayakkara}, {Nelson}, {Oesch}, {Pacifici}, {Papovich}, {Setton}, {Shapley}, {Shipley}, {Smit}, {Stefanon}, {Taylor}, {Weibel}, \& {Williams}}]{weaver24}
{Weaver}, J.~R., {Cutler}, S.~E., {Pan}, R., {et~al.} 2024, \apjs, 270, 7

\bibitem[{{Whitaker} {et~al.}(2019){Whitaker}, {Ashas}, {Illingworth}, {Magee}, {Leja}, {Oesch}, {van Dokkum}, {Mowla}, {Bouwens}, {Franx}, {Holden}, {Labb{\'e}}, {Rafelski}, {Teplitz}, \& {Gonzalez}}]{whitaker19}
{Whitaker}, K.~E., {Ashas}, M., {Illingworth}, G., {et~al.} 2019, \apjs, 244, 16

\bibitem[{{Williams} {et~al.}(2024){Williams}, {Alberts}, {Ji}, {Hainline}, {Lyu}, {Rieke}, {Endsley}, {Suess}, {Sun}, {Johnson}, {Florian}, {Shivaei}, {Rujopakarn}, {Baker}, {Bhatawdekar}, {Boyett}, {Bunker}, {Cameron}, {Carniani}, {Charlot}, {Curtis-Lake}, {DeCoursey}, {de Graaff}, {Egami}, {Eisenstein}, {Gibson}, {Hausen}, {Helton}, {Maiolino}, {Maseda}, {Nelson}, {P{\'e}rez-Gonz{\'a}lez}, {Rieke}, {Robertson}, {Saxena}, {Tacchella}, {Willmer}, \& {Willott}}]{williams23}
{Williams}, C.~C., {Alberts}, S., {Ji}, Z., {et~al.} 2024, \apj, 968, 34

\bibitem[{{Wuyts} {et~al.}(2011){Wuyts}, {F{\"o}rster Schreiber}, {van der Wel}, {Magnelli}, {Guo}, {Genzel}, {Lutz}, {Aussel}, {Barro}, {Berta}, {Cava}, {Graci{\'a}-Carpio}, {Hathi}, {Huang}, {Kocevski}, {Koekemoer}, {Lee}, {Le Floc'h}, {McGrath}, {Nordon}, {Popesso}, {Pozzi}, {Riguccini}, {Rodighiero}, {Saintonge}, \& {Tacconi}}]{wuyts11}
{Wuyts}, S., {F{\"o}rster Schreiber}, N.~M., {van der Wel}, A., {et~al.} 2011, \apj, 742, 96

\bibitem[{{Xiao} {et~al.}(2024){Xiao}, {Oesch}, {Elbaz}, {Bing}, {Nelson}, {Weibel}, {Illingworth}, {van Dokkum}, {Naidu}, {Daddi}, {Bouwens}, {Matthee}, {Wuyts}, {Chisholm}, {Brammer}, {Dickinson}, {Magnelli}, {Leroy}, {Schaerer}, {Herard-Demanche}, {Lim}, {Barrufet}, {Endsley}, {Fudamoto}, {G{\'o}mez-Guijarro}, {Gottumukkala}, {Labb{\'e}}, {Magee}, {Marchesini}, {Maseda}, {Qin}, {Reddy}, {Shapley}, {Shivaei}, {Shuntov}, {Stefanon}, {Whitaker}, \& {Wyithe}}]{xiao23}
{Xiao}, M., {Oesch}, P.~A., {Elbaz}, D., {et~al.} 2024, \nat, 635, 311

\bibitem[{{Xiao} {et~al.}(2023){Xiao}, {Elbaz}, {G{\'o}mez-Guijarro}, {Leroy}, {Bing}, {Daddi}, {Magnelli}, {Franco}, {Zhou}, {Dickinson}, {Wang}, {Rujopakarn}, {Magdis}, {Treister}, {Inami}, {Demarco}, {Sargent}, {Shu}, {Kartaltepe}, {Alexander}, {B{\'e}thermin}, {Bournaud}, {Ciesla}, {Ferguson}, {Finkelstein}, {Giavalisco}, {Gu}, {Iono}, {Juneau}, {Lagache}, {Leiton}, {Messias}, {Motohara}, {Mullaney}, {Nagar}, {Pannella}, {Papovich}, {Pope}, {Schreiber}, \& {Silverman}}]{xiao23a}
{Xiao}, M.~Y., {Elbaz}, D., {G{\'o}mez-Guijarro}, C., {et~al.} 2023, \aap, 672, A18

\bibitem[{{Yamaguchi} {et~al.}(2019){Yamaguchi}, {Kohno}, {Hatsukade}, {Wang}, {Yoshimura}, {Ao}, {Caputi}, {Dunlop}, {Egami}, {Espada}, {Fujimoto}, {Hayatsu}, {Ivison}, {Kodama}, {Kusakabe}, {Nagao}, {Ouchi}, {Rujopakarn}, {Tadaki}, {Tamura}, {Ueda}, {Umehata}, {Wang}, \& {Yun}}]{yamaguchi19}
{Yamaguchi}, Y., {Kohno}, K., {Hatsukade}, B., {et~al.} 2019, \apj, 878, 73

\bibitem[{{Yan} {et~al.}(2024){Yan}, {Sun}, \& {Ling}}]{yan24}
{Yan}, H., {Sun}, B., \& {Ling}, C. 2024, \apj, 975, 44

\bibitem[{{Zavala} {et~al.}(2021){Zavala}, {Casey}, {Manning}, {Aravena}, {Bethermin}, {Caputi}, {Clements}, {Cunha}, {Drew}, {Finkelstein}, {Fujimoto}, {Hayward}, {Hodge}, {Kartaltepe}, {Knudsen}, {Koekemoer}, {Long}, {Magdis}, {Man}, {Popping}, {Sanders}, {Scoville}, {Sheth}, {Staguhn}, {Toft}, {Treister}, {Vieira}, \& {Yun}}]{zavala21}
{Zavala}, J.~A., {Casey}, C.~M., {Manning}, S.~M., {et~al.} 2021, \apj, 909, 165

\bibitem[{{Zavala} {et~al.}(2023){Zavala}, {Buat}, {Casey}, {Finkelstein}, {Burgarella}, {Bagley}, {Ciesla}, {Daddi}, {Dickinson}, {Ferguson}, {Franco}, {Jim{\'e}nez-Andrade}, {Kartaltepe}, {Koekemoer}, {Le Bail}, {Murphy}, {Papovich}, {Tacchella}, {Wilkins}, {Aretxaga}, {Behroozi}, {Champagne}, {Fontana}, {Giavalisco}, {Grazian}, {Grogin}, {Kewley}, {Kocevski}, {Kirkpatrick}, {Lotz}, {Pentericci}, {P{\'e}rez-Gonz{\'a}lez}, {Pirzkal}, {Ravindranath}, {Somerville}, {Trump}, {Yang}, {Yung}, {Almaini}, {Amor{\'\i}n}, {Annunziatella}, {Arrabal Haro}, {Backhaus}, {Barro}, {Bell}, {Bhatawdekar}, {Bisigello}, {Buitrago}, {Calabr{\`o}}, {Castellano}, {Ch{\'a}vez Ortiz}, {Chworowsky}, {Cleri}, {Cohen}, {Cole}, {Cooke}, {Cooper}, {Cooray}, {Costantin}, {Cox}, {Croton}, {Dav{\'e}}, {de La Vega}, {Dekel}, {Elbaz}, {Estrada-Carpenter}, {Fern{\'a}ndez}, {Finkelstein}, {Freundlich}, {Fujimoto}, {Garc{\'\i}a-Argum{\'a}nez}, {Gardner}, {Gawiser}, {G{\'o}mez-Guijarro}, {Guo}, {Hamilton}, {Hathi}, {Holwerda}, {Hirschmann},
  {Huertas-Company}, {Hutchison}, {Iyer}, {Jaskot}, {Jha}, {Jogee}, {Juneau}, {Jung}, {Kassin}, {Kurczynski}, {Larson}, {Leung}, {Long}, {Lucas}, {Magnelli}, {Mantha}, {Matharu}, {McGrath}, {McIntosh}, {Medrano}, {Merlin}, {Mobasher}, {Morales}, {Newman}, {Nicholls}, {Pandya}, {Rafelski}, {Ronayne}, {Rose}, {Ryan}, {Santini}, {Seill{\'e}}, {Shah}, {Shen}, {Simons}, {Snyder}, {Stanway}, {Straughn}, {Teplitz}, {Vanderhoof}, {Vega-Ferrero}, {Wang}, {Weiner}, {Willmer}, {Wuyts}, \& {Ceers Team}}]{zavala23}
{Zavala}, J.~A., {Buat}, V., {Casey}, C.~M., {et~al.} 2023, \apjl, 943, L9

\bibitem[{{Zhuang} \& {Shen}(2024)}]{zhuang24}
{Zhuang}, M.-Y., \& {Shen}, Y. 2024, \apj, 962, 139

\end{thebibliography}
\bibliographystyle{aasjournal}

\end{document}